\documentclass[pdflatex,sn-mathphys-num]{sn-jnl}

\usepackage{graphicx}%
\usepackage{multirow}%
\usepackage{amsmath,amssymb,amsfonts}%
\usepackage{amsthm}%
\usepackage{mathrsfs}%
\usepackage[title]{appendix}%
\usepackage{xcolor}%
\usepackage{textcomp}%
\usepackage{manyfoot}%
\usepackage{booktabs}%
\usepackage{algorithm}%
\usepackage{algorithmicx}%
\usepackage{algpseudocode}%
\usepackage{listings}%
\usepackage{subcaption}
\usepackage{float}

\theoremstyle{thmstyleone}%
\theoremstyle{thmstyletwo}%

\theoremstyle{thmstylethree}%

\usepackage[table]{xcolor}
\renewcommand{\arraystretch}{2.5}

\begin{document}

\title[]{Wigner–Husimi phase-space structure of quasi-exactly solvable sextic potential}


\author[1]{\fnm{Angelina N.} \sur{Mendoza Tavera}}\email{cbi2182800711@xanum.uam.mx}
\equalcont{These authors contributed equally to this work.}

\author[1]{\fnm{Adrian M.} \sur{Escobar Ruiz}}\email{admau@xanum.uam.mx}
\equalcont{These authors contributed equally to this work.}

\author[2]{\fnm{Robin P.} \sur{Sagar}}\email{sagar@xanum.uam.mx}
\equalcont{These authors contributed equally to this work.}

\equalcont{These authors contributed equally to this work.}

\affil*[1]{\orgdiv{Departamento de Física}, \orgname{Universidad Autónoma Metropolitana - Iztapalapa}, \orgaddress{\street{Rafael
Atlixco 186}, \city{CDMX}, \postcode{09340}, \state{Ciudad de México}, \country{México}}}

\affil*[2]{\orgdiv{Departamento de Química}, \orgname{Universidad Autónoma Metropolitana - Iztapalapa}, \orgaddress{\street{Rafael
Atlixco 186}, \city{CDMX}, \postcode{09340}, \state{Ciudad de México}, \country{México}}}


\abstract{
In this study, we compare the Wigner function $W$, its modulus, and the Husimi distribution $H$ in a one-dimensional quantum system exhibiting a transition from a single-well to a double-well configuration, using the quasi-exactly solvable sextic oscillator as a representative example. High-accuracy variational wavefunctions for the lowest states are used to compute two-dimensional phase-space structures, one-dimensional marginals, and the corresponding Shannon entropies, mutual information, and Cumulative Residual Jeffreys divergences. The analysis shows that the Wigner representation is uniquely responsive to interference effects and displays clear, nonmonotonic entropic behavior as the wells separate, whereas the modulus-Wigner and Husimi distributions account only for geometric splitting or coarse-grained delocalization. These findings establish a quantitative hierarchy in the ability of $W$, $|W|$, and $H$ to resolve structural changes in a quantum state and provide a general framework for assessing the descriptive power of different phase-space representations in systems with emerging bimodality or tunneling.
}





\maketitle

\section{Introduction}

Phase-space formulations of quantum mechanics provide a unified framework for
analyzing coherence, interference, and localization through quasiprobability
distributions such as the Wigner and Husimi functions. Since the original works of
Wigner~\cite{Wigner1932} and Husimi~\cite{Husimi1940}, these representations have
become indispensable in quantum optics, semiclassical dynamics, and information
theory. Their versatility has been demonstrated in AMO physics, including coherence
in Talbot--Lau interferometry~\cite{Imhof2016Atoms}, optical quantum communication
via stochastic and Wigner formalisms~\cite{Casado2019Atoms}, and phase-space
analyses of instabilities in Rydberg complexes~\cite{Dimitrijevic2019Atoms}. Broader
developments—such as generalized Wigner families for finite-level systems~\cite{Abgaryan2021Atoms}
and conceptual studies highlighting the atom as a central building block of modern
physics~\cite{Connerade2023Atoms}—further underscore the importance of phase-space
methods.

The Wigner function encodes the full quantum structure of a state, including its
regions of negativity associated with nonclassical interference~\cite{Laguna2010},
whereas the Husimi distribution provides a positive-definite, Gaussian-smeared
semiclassical portrait~\cite{Wehrl1979,Calixto2012,Floerchinger2021}. Related approaches
such as the Wigner--Kirkwood expansion~\cite{Kazandjian2022Atoms} incorporate
systematic semiclassical corrections. Recently, the modulus $|W|$ has attracted
renewed interest as an intermediate representation preserving geometric phase-space
features while eliminating sign oscillations~\cite{Salazar2023}. Applications include the
study of coherence and classical inhomogeneities~\cite{Smith2022Atoms} and nonlinear
atom--light interactions~\cite{Yamada2025Atoms}.

Beyond structural aspects, these phase-space distributions admit a rich
information-theoretic characterization through Shannon, Rényi, Tsallis, and Fisher
entropies, as well as correlation measures. Shannon entropies quantify global
delocalization~\cite{Gadre1985}, while the entropic uncertainty relation provides a
sharper bound than the variance-based Heisenberg relation~\cite{Bialynicki2006}. These
concepts have been applied from AMO to condensed-matter physics
\cite{Nagy2010,Angulo2008,Anteneodo1996,Nagy2023}, including complexity
\cite{Nagy2023}, quantum-phase transitions~\cite{Calixto2012}, and entanglement
studies~\cite{Floerchinger2021}. Tsallis entropies further illuminate electron correlation
and nonlinear structure~\cite{Tsallis1988,FloresGallegos2018}.

A broad set of measures also exists for quantifying statistical correlations. These
include correlation coefficients~\cite{Kutzelnigg1968,Thakkar1981}, mutual information in
position and momentum spaces~\cite{Laguna2011A,Laguna2011B}, and higher-order
interaction information~\cite{Matsuda2000,Salazar2020,Salazar2024}. Such tools have been
applied to nuclear systems~\cite{Moustakidis2005}, Bose--Einstein condensates
\cite{Sriraman2017,Zhao2019,Chakrabarti2024}, chemical reactivity~\cite{FloresGomez2023}, and
wave-packet dynamics~\cite{Romera2007}. Their relevance to AMO-related phase-space
analysis is reflected in recent Wigner and Husimi studies
\cite{Imhof2016Atoms,Kazandjian2022Atoms,Smith2022Atoms}.

Despite these developments, a unified comparative study of the Wigner, $|W|$, and
Husimi quasiprobabilities within a single, continuously tunable quantum system—using
modern information-theoretic indicators—has been missing. While Wigner and Husimi
functions have been widely applied, no previous work has systematically compared
them, together with $|W|$, across their two-dimensional structure, marginals, entropies,
mutual information, and CRJ divergences. A natural setting for such an investigation is
the quasi-exactly solvable (QES) sextic potential, which admits analytic eigenstates for
part of its spectrum and interpolates smoothly between single- and double-well
regimes. Its algebraic foundations were developed in the pioneering works
\cite{Turbiner1988,Turbiner1987,Ushveridze1988} and later linked to
orthogonal-polynomial structures~\cite{Bender1996}. Recent SUSY and WKB developments
further confirm its role as a benchmark model~\cite{Contreras2025,Contreras2024}.

Our recent study~\cite{Tavera2025} showed that one-dimensional densities
$|\psi(x)|^2$ and $|\phi(p)|^2$ already reveal tunneling, quasi-degeneracy, and parity
effects through entropic and divergence-based indicators. However, that analysis
remained strictly in configuration and momentum space and did \emph{not} construct any
phase-space distributions nor examine Wigner negativity or smoothing effects.

In this work, we extend the informational analysis of~\cite{Tavera2025} to the \emph{full
phase-space domain}, providing the first unified comparison of the Wigner function
$W(x,p)$, its modulus $|W(x,p)|$, and the Husimi distribution $H(x,p)$ for the QES
sextic oscillator. To our knowledge, no prior study has jointly compared (i) Wigner,
(ii) modulus-Wigner, and (iii) Husimi representations across: (a) their full phase-space
structure, (b) their marginals, (c) their entropies, (d) their mutual information, and
(e) their CRJ divergences—features inaccessible to purely one-dimensional approaches.

The central question guiding this work is:
\emph{How do $W(x,p)$, $|W(x,p)|$, and $H(x,p)$ differ in their structural and
information-theoretic response to the single- to double-well transition in the QES
sextic potential?}

The paper is organized as follows. Section~\ref{sec:QES} introduces the QES sextic
potential and summarizes the variational construction of low-lying states.
Section~\ref{SEC3} reviews the three quasiprobability representations. Section~\ref{SEC4}
presents the main phase-space results. Section~\ref{SEC5} discusses entropic and
correlation measures derived from the marginals, while Section~\ref{SEC6} incorporates
mutual information and CRJ divergences. Section~\ref{SEC7} summarizes the main
insights, and Section~\ref{sec:conclusions} outlines possible directions for future work.
Appendices~A and B provide supporting algebraic and numerical details.

\section{The model: QES sextic potential}
\label{sec:QES}

Quasi-exactly solvable (QES) models
form a distinguished class of quantum systems in which a finite portion of the spectrum
can be determined in closed algebraic form, while higher states require approximation.
This intermediate status between exactly solvable and generic models makes them
particularly valuable: algebraic eigenstates provide exact benchmarks, whereas the
non-algebraic sector tests the reliability of variational, numerical, and
information-theoretic methods. The sextic QES potential is particularly rich, as tuning
its control parameter $\lambda$ drives the system from a single-well regime into a
symmetric double well, giving rise to localization-delocalization transitions,
quasi-degenerate tunneling doublets, and nonperturbative instanton effects. These
features render QES systems natural testbeds for studying the interplay between
algebraic solvability, semiclassical physics, and entropic measures of coherence and
correlation.

So, we focus on the one-dimensional spectral problem of non-relativistic quantum mechanics,  
\begin{equation}
\label{Hp}
    {\cal H}\,\psi(x) \;=\; E\,\psi(x)\,, 
    \qquad \psi(x) \in {\cal L}^2(\mathbb{R}) \,,
\end{equation}
defined on the full real line $x \in (-\infty,\infty)$.  
The Hamiltonian operator is of the form 
\begin{equation}
\label{HQES}
    {\cal H} \;=\; -\frac{\hbar^2}{2m}\,\frac{d^2}{dx^2} \;+\; V^{\rm QES}(x;\lambda)\,,
\end{equation}
with particle mass $m$ and the quasi-exactly solvable sextic potential  
\begin{equation}
\label{vqes}
    V^{\rm QES}(x;\lambda) \;=\; \tfrac{1}{2}\,\Big(x^{6} + 2x^{4} - 2(2\lambda+1)\,x^{2}\Big)\,,
\end{equation}
where $\lambda \in \mathbb{R}$ is the control parameter.  

This potential belongs to the class of quasi-exactly solvable (QES) models: for special values of $\lambda$, a finite subset of eigenstates can be obtained in closed algebraic form, while the remainder requires approximation. Hereafter we adopt atomic units, $\hbar = 1$ and $m = 1$. Equation~\eqref{Hp} with potential~\eqref{vqes} defines the central problem addressed in this work.

\begin{figure}[h]
\begin{center}
\includegraphics[width=10.5
cm]{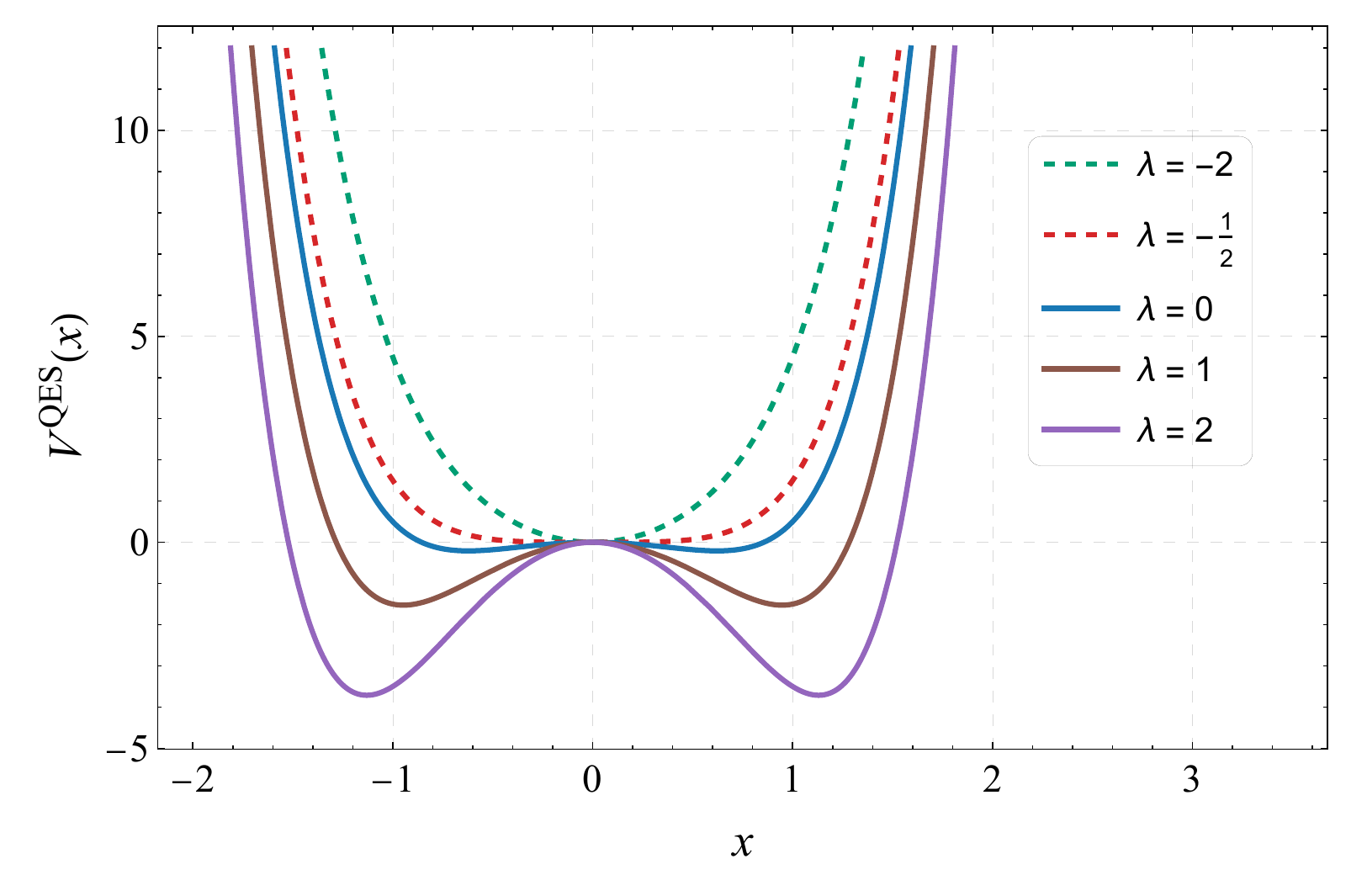}
\caption{\small QES confining sextic potential $V^{\rm QES}(x;\lambda)$ in (\ref{vqes}) for different values of the parameter $\lambda$. For $\lambda>-\frac{1}{2}$ it develops two symmetric degenerate minima, and the $\lambda-$independent maximum located at $x = 0$ corresponds to a potential energy $V = 0$.
}
\label{F1}
\end{center}
\end{figure}

\subsection{Variational calculations of low-lying states}

To solve the spectral problem~\eqref{HQES}, we employ a variational approach. For the explicit calculations, we follow the methodology presented in~\cite{Tavera2025}. Since the full construction of the variational ansatz and the details and results of its optimization are thoroughly discussed in that reference, we do not reproduce them here, and instead summarize only the essential elements relevant for the present work. Details on the convergence and accuracy of calculation are presented in Appendix \ref{A2}. 

The real trial wavefunction for the $n$th state is chosen in the form  
\begin{equation}
\psi_n^{(\rm var)}(x;\lambda)\ =\ Q_{k}^{(n)}(x;\lambda)\, e^{-x^4 / 4},\qquad n=0,1,2,\ldots \, ,
\end{equation}
where the exponential factor guarantees the correct asymptotic decay as $|x| \to \infty$, and the polynomial $Q_k^{(n)}(x;\lambda)$ enforces the exact parity symmetry of the Hamiltonian under the reflection $x \to -x$.

Explicitly, for the ground state ($n=0$), the polynomial factor 
\begin{equation}
Q_k^{(0)}(x;\lambda)=\sum_{i=0}^{k} c_i\, x^{2i} \ ,
\end{equation}
is of degree $k=k(n)$ in the variable $x^2$. Thus $Q_k^{(0)}(x)=Q_k^{(0)}(-x)$, and it has no real zeros. The linear coefficients $c_i$ are determined by minimizing the variational energy,
\begin{equation}
    E^{(\rm{var})}
    = 
    \frac{\int_{-\infty}^{\infty} \psi_n^{(\rm var)}\,\hat H\,\psi_n^{(\rm var)}\,dx}
         {\int_{-\infty}^{\infty} {\big(\psi_n^{(\rm var)}\big)}^2\,dx}\quad ;
    \qquad 
    \frac{\partial E^{(\rm{var})}}{\partial c_i}=0 \ ,\quad i=1,2,\ldots,k \, .
\end{equation}

The first excited state $n=1$ is, instead, described by an odd polynomial factor $Q_k^{(1)}(x;\lambda)$, which possesses a single real root at $x=0$, as required by parity.

For excited states ($n>0$), several coefficients in the polynomials $Q_k^{(n)}(x;\lambda)$ can be fixed by enforcing orthogonality and normalization
\begin{equation}
\langle \psi_n \,|\, \psi_{\tilde n} \rangle = \delta_{n,\tilde n} \ ,
\end{equation}
while the remaining parameters are optimized by minimizing the corresponding energy expectation value. The polynomial order $k=k(n)$ is chosen minimally to ensure that the resulting variational energies reach a relative accuracy of $\sim 10^{-8}$ when compared to Lagrange-mesh ~\cite{Valle2024} calculations. This ansatz also admits closed-form Fourier transforms of the wavefunctions, enabling accurate momentum-space and entropy analyses.

That way, the resulting variational energies agree with exact solutions and with Lagrange-mesh results to within relative errors of order $10^{-8}$, confirming the reliability of the trial functions across all relevant values of the control parameter $\lambda$.

\section{Phase-space representations}
\label{SEC3}

Phase-space quasiprobability functions provide complementary viewpoints on the 
structure of a quantum state. In this work, we focus on three widely used 
representations—the Wigner distribution $W(x,p)$, its (normalized) absolute value $|W(x,p)|$, and the Husimi function $H(x,p)$—and study their similarities and differences for 
the low-lying states of the QES sextic oscillator. Although all three originate 
from the density operator $\hat{\rho}=|\psi\rangle\langle\psi|$, each representation encodes coherence, interference, and smoothing in a distinct manner, making a 
systematic comparison essential.

\clearpage

\subsection{The Wigner function}

For a fixed quantum state $n$ of the QES Hamiltonian (\ref{HQES}), the corresponding quasiprobability distribution Wigner function $W_n(x,p)$, defined in the phase-space $(x,y)$, is given by
\begin{equation}
W_n(x,p) = \frac{1}{\pi} \int_{-\infty}^{\infty}
\psi_n^{*}\!\left(x+y\right)\,
\psi_n\!\left(x-y\right)\, e^{2ipy}\, dy .
\label{eq:wigner}
\end{equation}
It is a real-valued function normalized according to
\begin{equation}
\int_{-\infty}^{\infty}\int_{-\infty}^{\infty} W_n(x,p)\, dx\, dp = 1,
\end{equation}
and its marginals yield the  probability quantum densities, namely
\begin{equation}
\int_{-\infty}^{\infty} W_n(x,p)\, dp = |\psi_n(x)|^{2}, \qquad
\int_{-\infty}^{\infty} W_n(x,p)\, dx = |\phi_n(p)|^{2}.
\end{equation}

Despite its normalization properties, $W_n(x,p)$ is not guaranteed to be non-negative. 
Its negative lobes play an important role: they are widely regarded as signatures of
quantum interference, nonclassicality, and phase--space coherence. 

In the sextic
oscillator studied here, the evolution of these negative regions as the coupling $\lambda$
varies provides detailed information about the emergence of tunneling, lobe separation,
and parity-dependent interference. In this work, we will focus exclusively on the two
lowest states, namely the ground state $(n=0)$ and the first excited state $(n=1)$.

\subsection{The modulus of the Wigner function}

For each eigenstate, we consider the modulus of the Wigner distribution,
$|W_n(x,p)|$, as a strictly non–negative phase–space representation. Although it
does not correspond to any operator ordering and is therefore not a true
quasiprobability in the Weyl–Heisenberg sense, $|W_n|$ has proven useful as an
interference–suppressed diagnostic: by removing the sign oscillations responsible for
Wigner negativity, it filters out fine interference fringes while preserving the global
geometric structure of the state.

Since $|W_n|$ is not normalized, we always use its normalized version when
computing entropies or marginals, i.e.,
\begin{equation}
|W_n| \;\rightarrow\;
\frac{|W_n(x,p)|}{\displaystyle \int_{-\infty}^{\infty} \int_{-\infty}^{\infty} |W_n(x,p)|\,dx\,dp}\ .    
\end{equation}
This ensures a consistent comparison with the normalized Wigner and Husimi
distributions.

In practice, $|W_n|$ inherits the main phase–space features of $W_n$ (peak
locations, lobe separation, elongation) while eliminating its alternating sign pattern,
and remains more structured than the Gaussian–smoothed Husimi function, making it an informative intermediate description.

\subsection{The Husimi function}

The Husimi function associated with the $n$th eigenstate is defined as the squared 
overlap of $\psi_n$ with the minimum-uncertainty coherent states $|x,p\rangle$,
\begin{equation}
H_n(x,p) = \frac{1}{\pi}\, 
\big|\langle x,p \,|\, \psi_n \rangle\big|^{2},
\label{eq:husimi}
\end{equation}
which is equivalent to a Gaussian smearing of the corresponding Wigner function,
\begin{equation}
H_n(x,p) = 
\frac{1}{\pi}
\int_{-\infty}^{\infty} \int_{-\infty}^{\infty} W_n(x',p')\, 
\exp\!\left[-(x-x')^{2}-(p-p')^{2}\right] 
\, dx'\, dp' .
\end{equation}

The Husimi function $H_n$ is smooth, positive definite, and normalized to unity.
Consequently, it lacks the negativity that characterizes $W_n$, but retains information 
about the spatial and momentum extent of the state. Its smoothing kernel has unit minimal
uncertainty, and therefore washes out rapid oscillations and interference fringes.
Comparing $H_n$ with $W_n$ and $|W_n|$ provides a clear picture of which features
in the Wigner function are robust under coarse graining and which arise solely from
nonclassical oscillatory structure.

\subsection{Rationale for systematic comparison}

The three quasiprobability functions considered here for each quantum state $n=0,1$ 
represent a hierarchy of differing degrees of coarse graining:
\[
W_n \;\longrightarrow\; |W_n| \;\longrightarrow\; H_n .
\]
The Wigner function retains full quantum coherence and interference; 
$|W_n|$ removes sign information but keeps much of the structural detail; 
and $H_n$ provides a non-negative, smoothed representation closely connected to
semiclassical phase--space descriptions.

A central goal of this work is to quantify how these representations differ as the
potential is tuned through single-well and double-well regimes. By analyzing 
two-dimensional plots, marginals, entropies, mutual information, and Cumulative
Residual Jeffreys divergences for $W_n$, $|W_n|$, and $H_n$ (with $n=0,1$), 
we obtain a unified framework for assessing their descriptive capacities and identifying 
the distinctive features each one contributes to the analysis of the sextic QES oscillator.

\section{2D phase-space structure of low-lying states}
\label{SEC4}

Before analyzing the phase--space distributions, we introduce the
state-dependent critical couplings $\lambda_{c}^{(n)}$. For each eigenstate $n$,
$\lambda_{c}^{(n)}$ denotes the value of $\lambda$ at which the energy $E_{n}$
reaches the height of the central maximum of the potential. At this point the
state becomes energetically comparable to the barrier top, and tunneling
effects begin to appear. These $\lambda_{c}^{(n)}$ serve as natural reference
markers in the phase-space and information-theoretic analysis that follows. In particular, for the two lowest states we have: 
\begin{equation}
\lambda_{0} = \lambda_{c}^{(0)} = 0.7329, 
\qquad
\lambda_{1} = \lambda_{c}^{(1)} = 1.4209.
\end{equation}

Now, the Wigner function, its modulus, and the Husimi distribution provide complementary
insights into how the quantum state is organized in phase space. In this section, we
analyze the two lowest states ($n=0,1$) of the QES sextic potential and examine how their phase-space
profiles evolve as the coupling parameter $\lambda$ drives the system from a shallow
single-well configuration to a well-developed double-well potential. Throughout, the
emphasis is placed on identifying the characteristic signatures that distinguish
$W$, $|W|$, and $H$.

It is worth mentioning that one-dimensional systems, such as the QES potential, allow a complete graphical description and analysis of phase-space distributions. In higher dimensions, such an analysis is restricted to projections in order to appreciate the behavior of these distributions.

\subsection{Ground state ($n=0$)}

\begin{figure}[H]
    \centering
    \begin{subfigure}[b]{0.45\textwidth}
        \includegraphics[width=\textwidth]{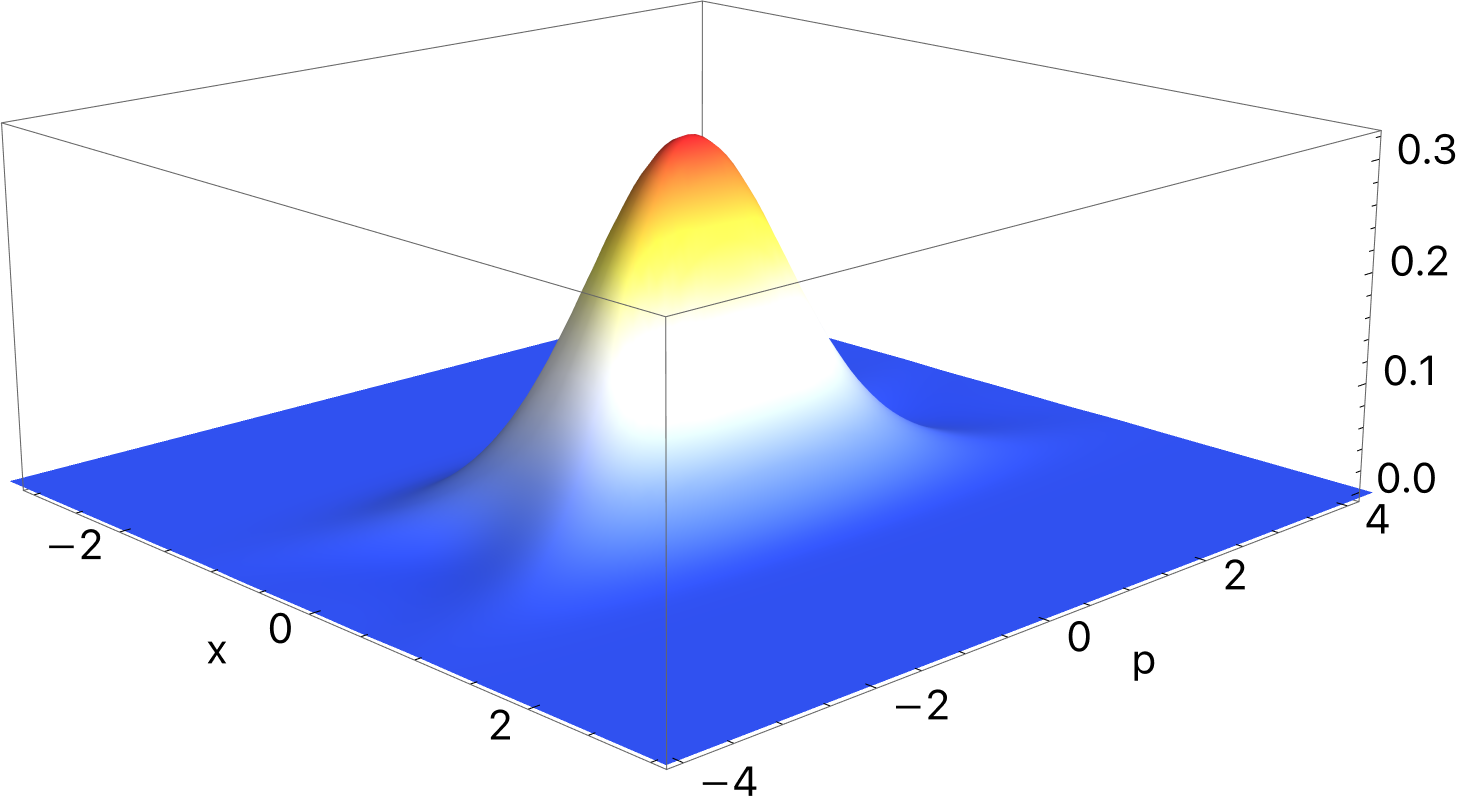}
        \caption{$\lambda=-\frac{3}{4}$}
    \end{subfigure}\hfill
    \begin{subfigure}[b]{0.45\textwidth}
        \includegraphics[width=\textwidth]{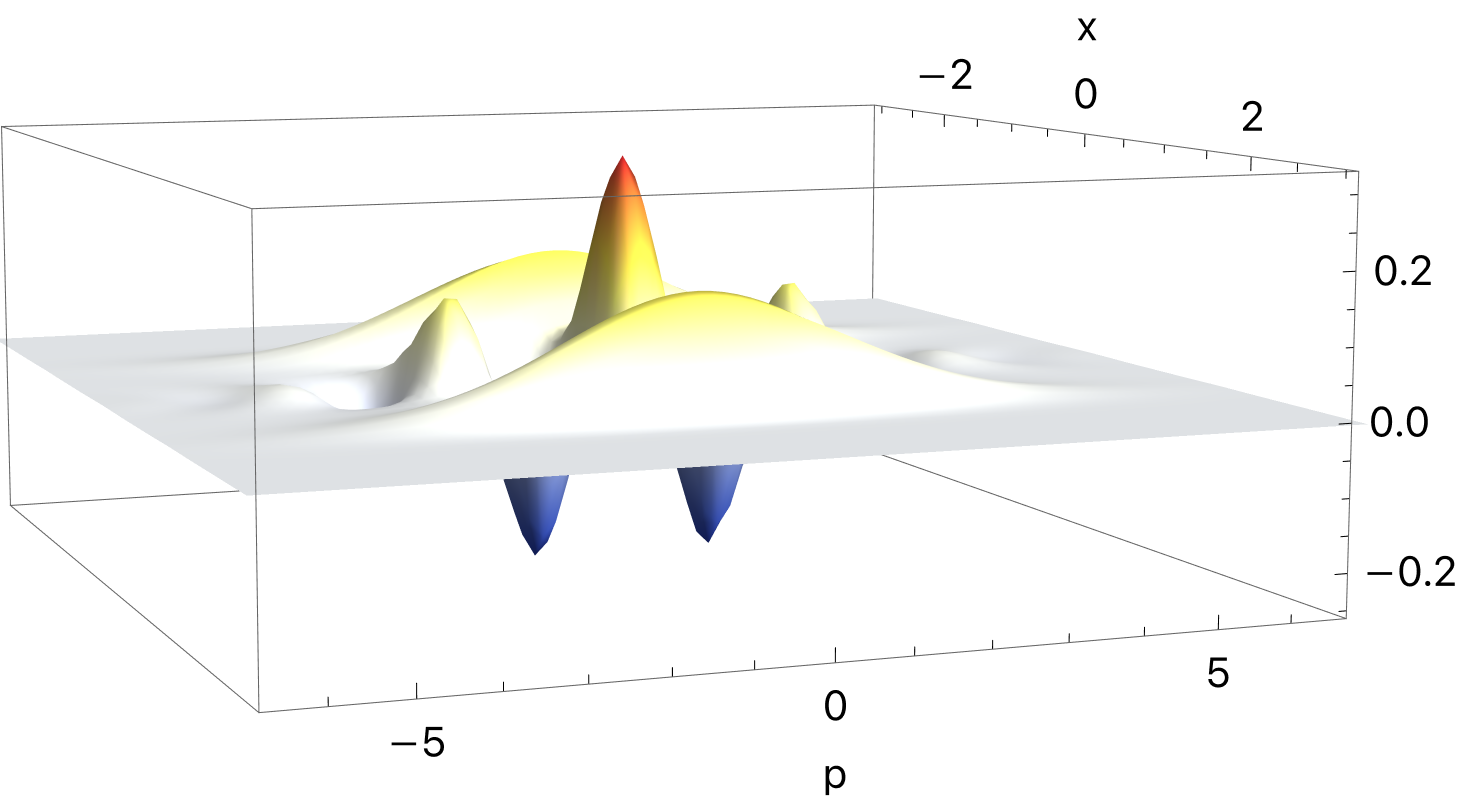}
        \caption{$\lambda=4$}
    \end{subfigure}

    \vspace{0.5cm} 

    \begin{subfigure}[b]{0.45\textwidth}
        \includegraphics[width=\textwidth]{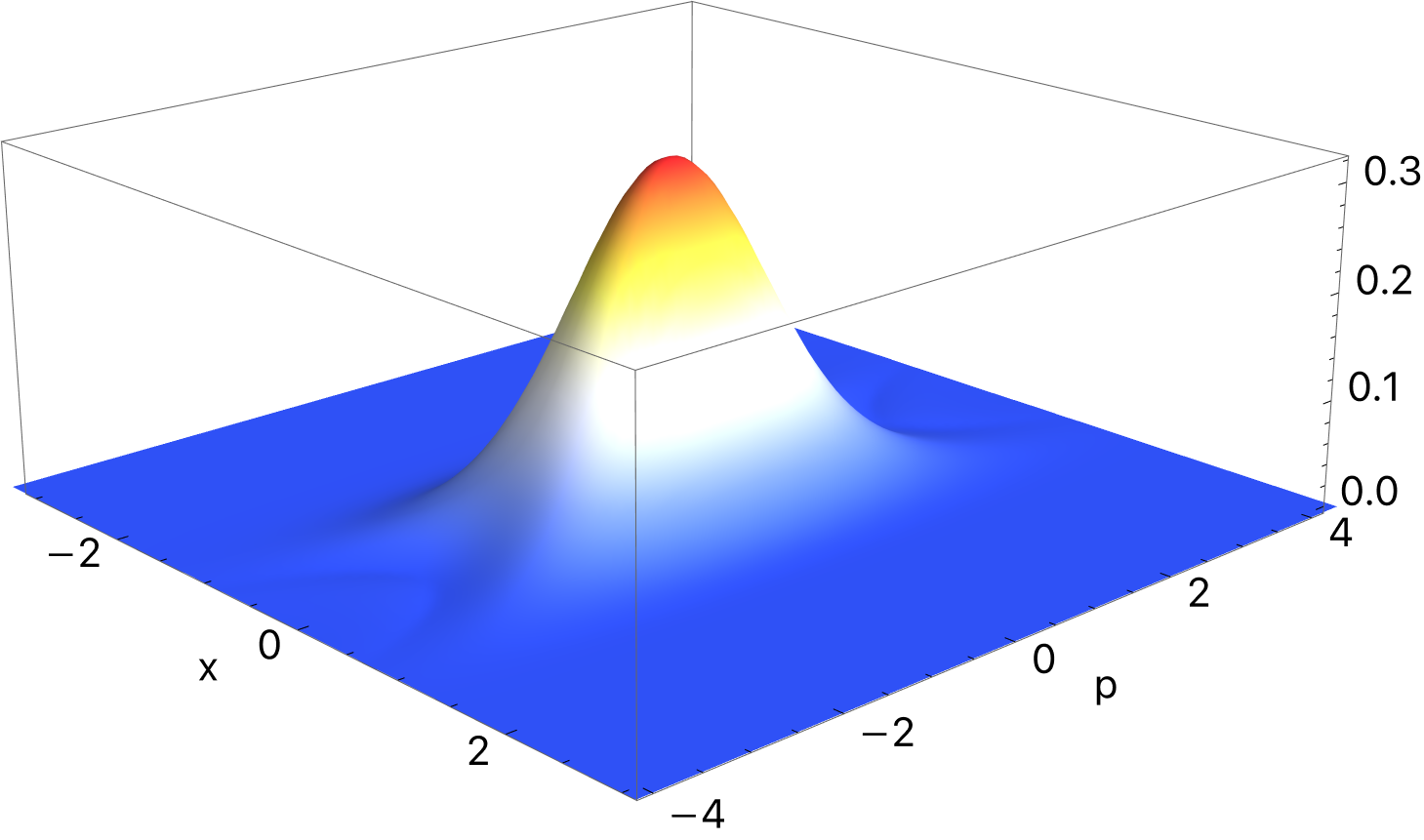}
        \caption{$\lambda=-\frac{3}{4}$}
    \end{subfigure}\hfill
    \begin{subfigure}[b]{0.45\textwidth}
        \includegraphics[width=\textwidth]{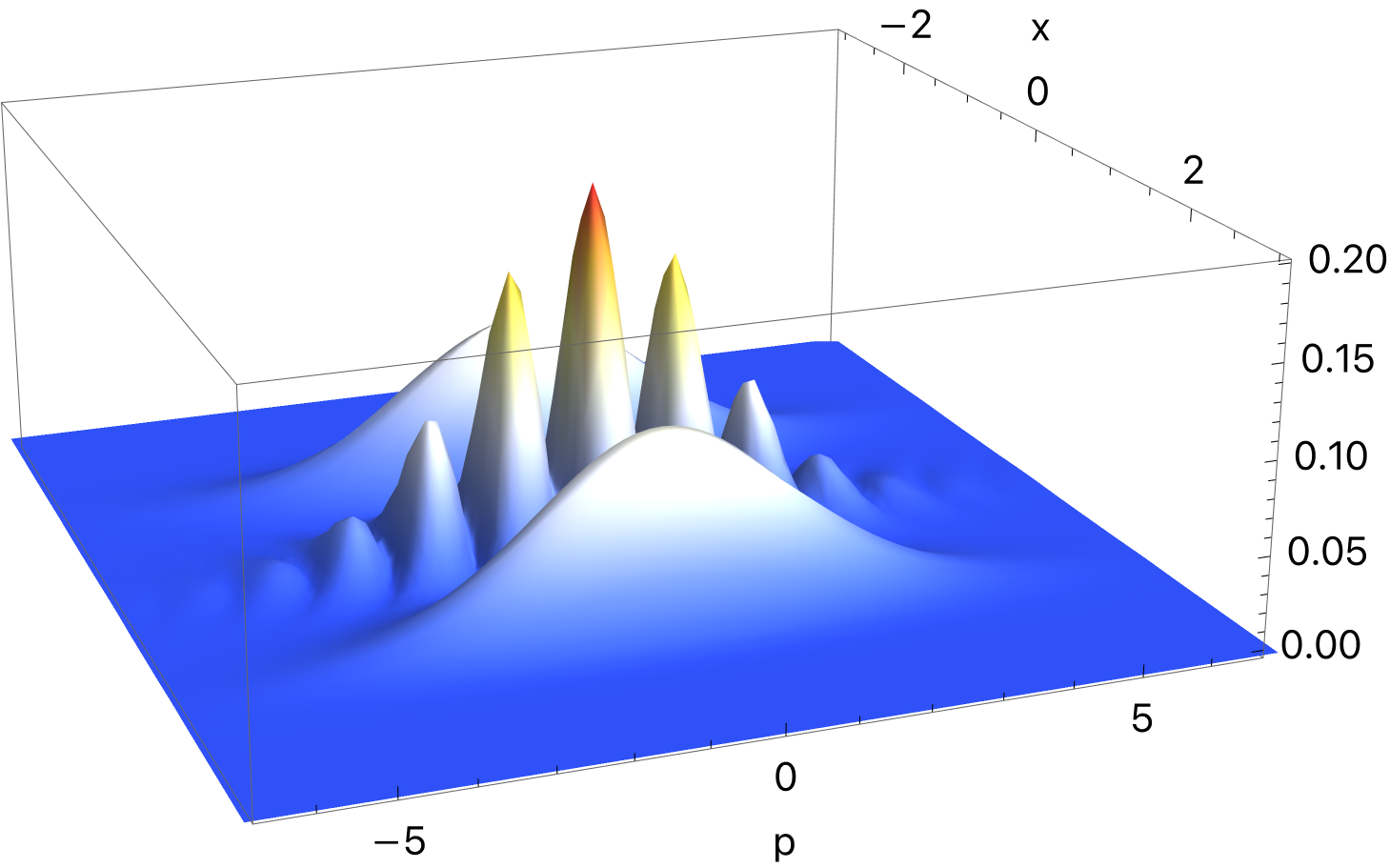}
        \caption{$\lambda=4$}
    \end{subfigure}

    \vspace{0.5cm} 

    \begin{subfigure}[b]{0.45\textwidth}
        \includegraphics[width=\textwidth]{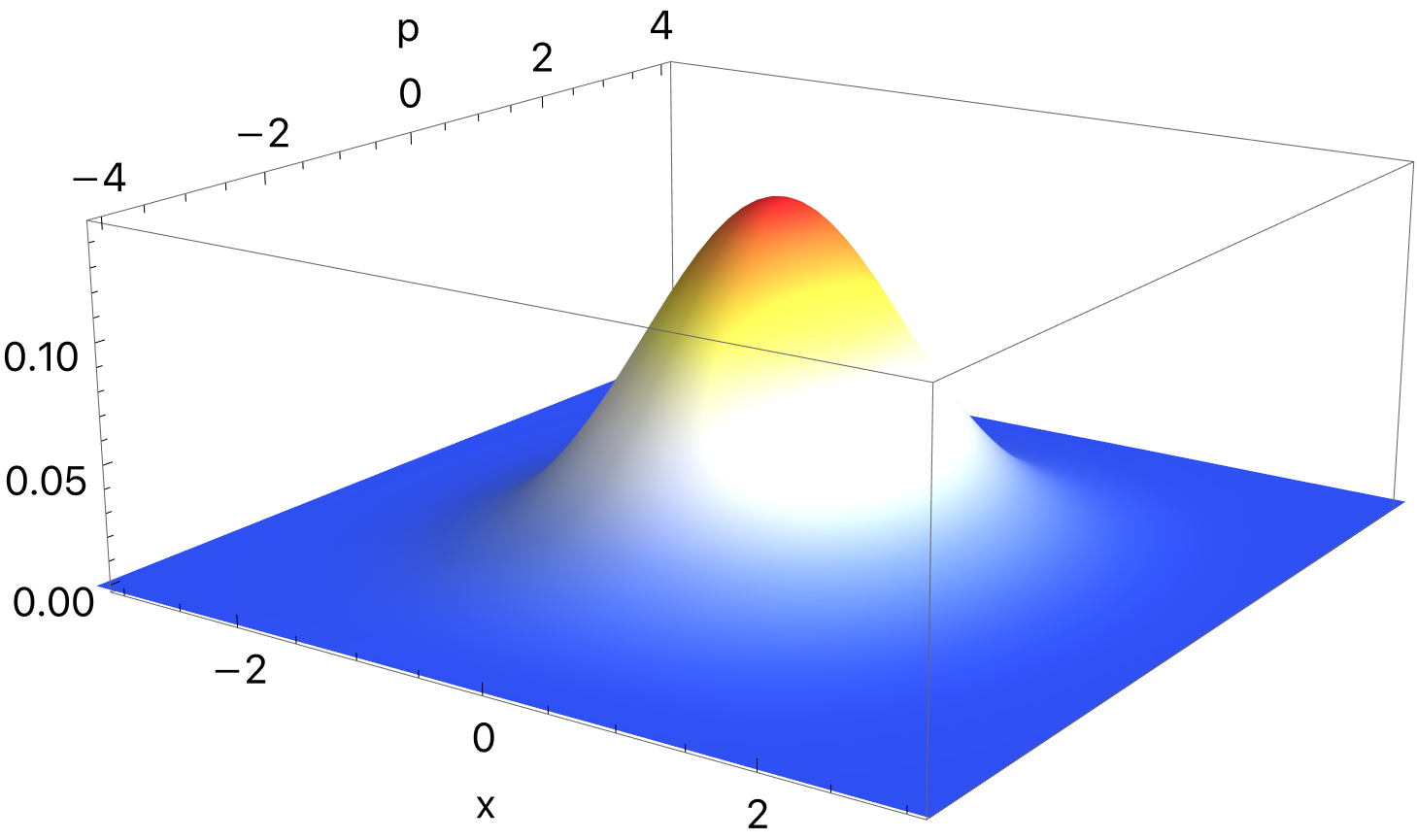}
        \caption{$\lambda=-\frac{3}{4}$}
    \end{subfigure}\hfill
    \begin{subfigure}[b]{0.45\textwidth}
        \includegraphics[width=\textwidth]{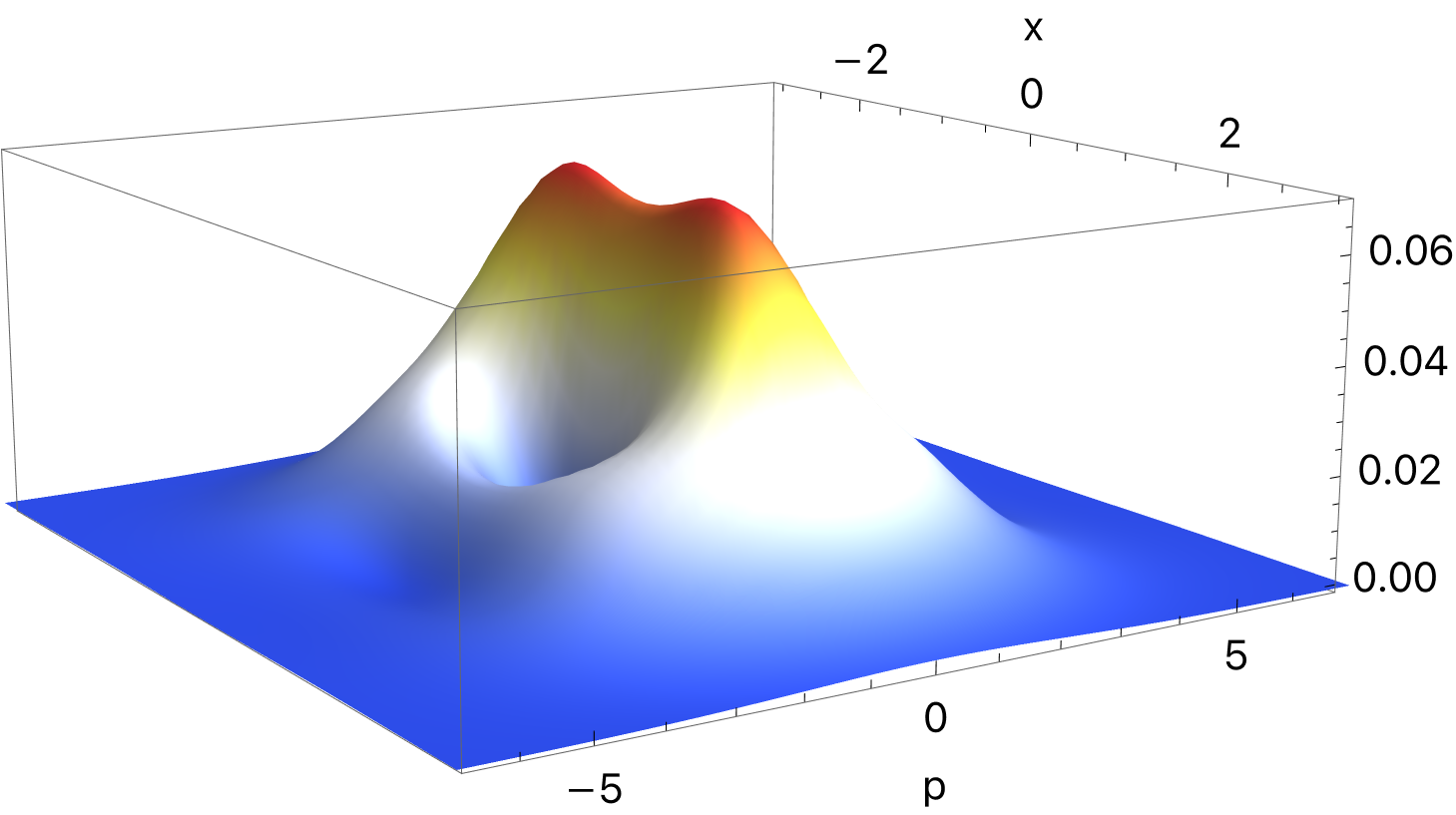}
        \caption{$\lambda=4$}
    \end{subfigure}
    \caption{\small Wigner quasiprobability distribution $W(x,p)$ (top), modulus of the Wigner function $|W(x,p)|$ (middle), and Husimi distribution $H(x,p)$ (bottom) for the ground state $n=0$ of the QES sextic oscillator at $\lambda=-\tfrac{3}{4}$ and $\lambda=4$.}
    \label{D-W-AbsW-Hn0}
\end{figure}

The Figure~\ref{D-W-AbsW-Hn0} shows that the Wigner function $W(x,p)$ for $n=0$ is initially
unimodal for small $\lambda$, with a single peak centered at $(x,p)=(0,0)$. As the
potential develops a central barrier, the phase--space density elongates along the $x$-axis
and eventually splits into two lobes located near the classical minima of the double well.
In this regime, $W(x,p)$ exhibits alternating positive and negative oscillations between the
two lobes, reflecting interference between amplitudes localized in opposite wells. In particular, at $\lambda = -\tfrac{3}{4}$ (single-well regime), $W(x,p)$ exhibits negative regions that are extremely small and not perceptible to the eye, whereas at $\lambda = 4$ (full double-well regime) the negative regions become clearly visible.

The modulus $|W(x,p)|$, removes the sign pattern
associated with interference while retaining the geometric structure of the distribution.
The region between the two phase--space lobes becomes uniformly positive, and the
fine interference ripples present in $W$ are suppressed without eliminating the overall
elongation or the separation of the main maxima. The resulting distribution lies midway
between the oscillatory Wigner profile and the smoothed Husimi function.

The Husimi distribution $H(x,p)$ presents a markedly different picture: its Gaussian
coarse graining smooths out the oscillations entirely and produces a strictly positive
phase--space density with broader contours. In the single-well region, $H$ is almost
circular near the origin; as $\lambda$ increases, it bifurcates into two smooth, separated
hills that track the underlying potential minima. The suppressed negativity and increased 
width of $H$ reflect the reduced resolution of coherent-state smoothing and provide a
more classical visual representation of the state.

\subsection{First excited state $(n=1)$}

For the first excited state ($n=1$) shown in Figure~\ref{D-W-AbsW-Hn1}, the Wigner function displays the characteristic
sign change associated with odd parity: a central nodal line at $x=0$ separates two
large lobes of opposite sign. In the single-well regime these lobes are close to each
other, producing a compact alternating pattern in $(x,p)$. As the potential becomes
double-welled, the lobes shift outward and become aligned with the classical turning
points, while strong interference fringes appear between them. This evolution is
particularly visible in the $p$-direction, where the oscillatory pattern becomes more
structured as the state develops distinct left- and right-localized components. Even at $\lambda = -\tfrac{3}{4}$ (single-well regime), the Wigner function exhibits pronounced negative regions. This behavior is markedly different from that of the ground state.

The modulus $|W(x,p)|$ again removes the nodal sign structure, resulting in two
equally-weighted positive peaks separated by a ridge of smaller amplitude. The outward
motion of these peaks with increasing $\lambda$ mirrors the movement of the wavefunction
nodes in configuration space. Compared with $n=0$, the excited-state modulus
retains more residual fine structure, indicating that $|W|$ still encodes aspects of the
interference present in $W$, even when negativity is removed.

For the Husimi distribution $H(x,p)$, the antisymmetric character of the state is
expressed not through negativity (which $H$ cannot display) but through the presence
of a broad valley along the $x=0$ axis. This smooth depletion replaces the sharp nodal
line seen in $W$. Note that the Wigner function retains more of the nodal structure as compared to the Husimi distribution. As $\lambda$ increases and the two classical wells deepen, the Husimi
density separates into two wide, rounded maxima located near the classical minima,
each corresponding to a coarse-grained envelope of the left- and right-localized lobes.

\subsection{Comparison and structural trends}

\begin{figure}[H]
    \centering

    \begin{subfigure}[b]{0.45\textwidth}
        \includegraphics[width=\textwidth]{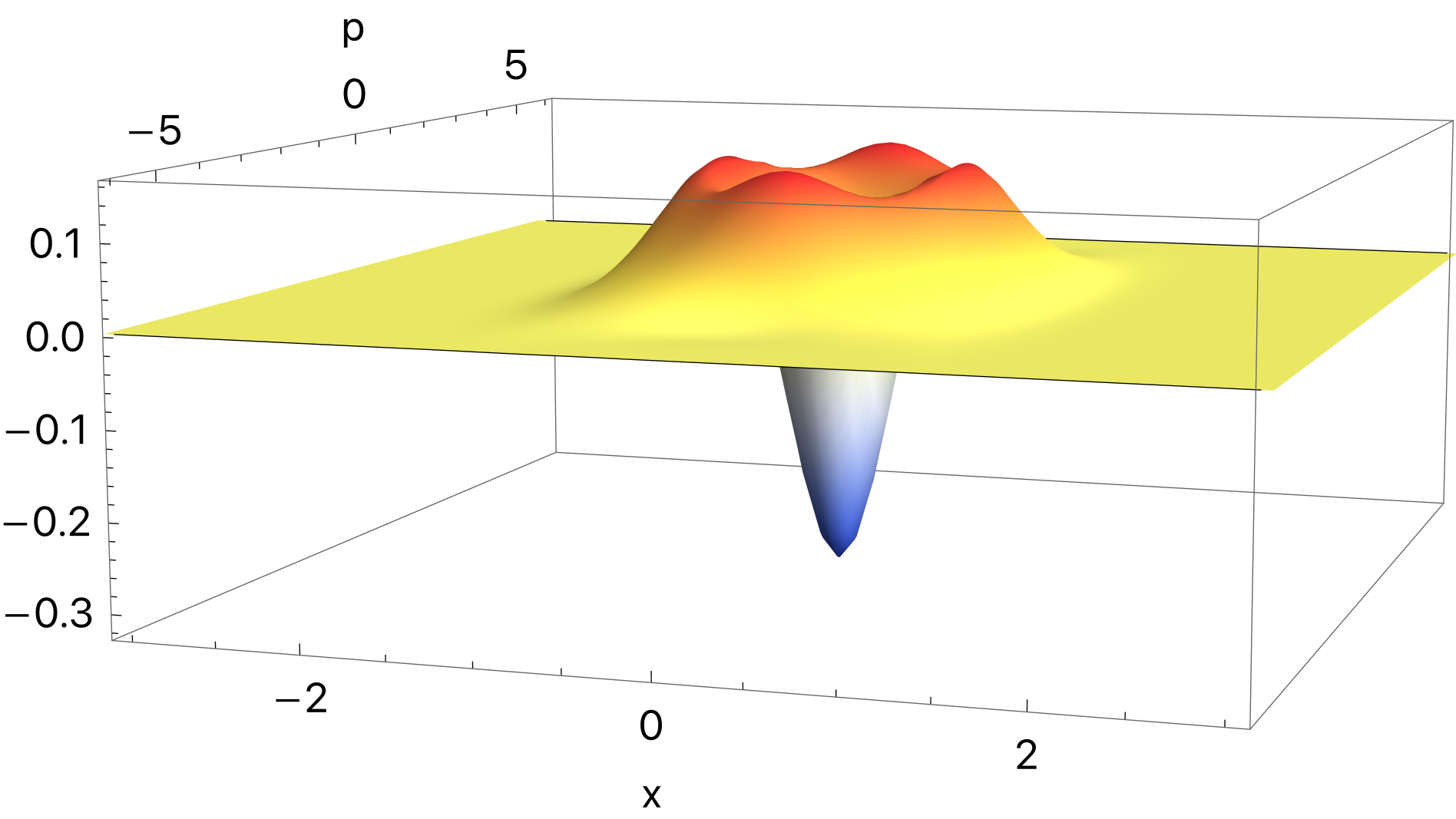}
        \caption{$\lambda=-\frac{3}{4}$}
    \end{subfigure}\hfill
    \begin{subfigure}[b]{0.45\textwidth}
        \includegraphics[width=\textwidth]{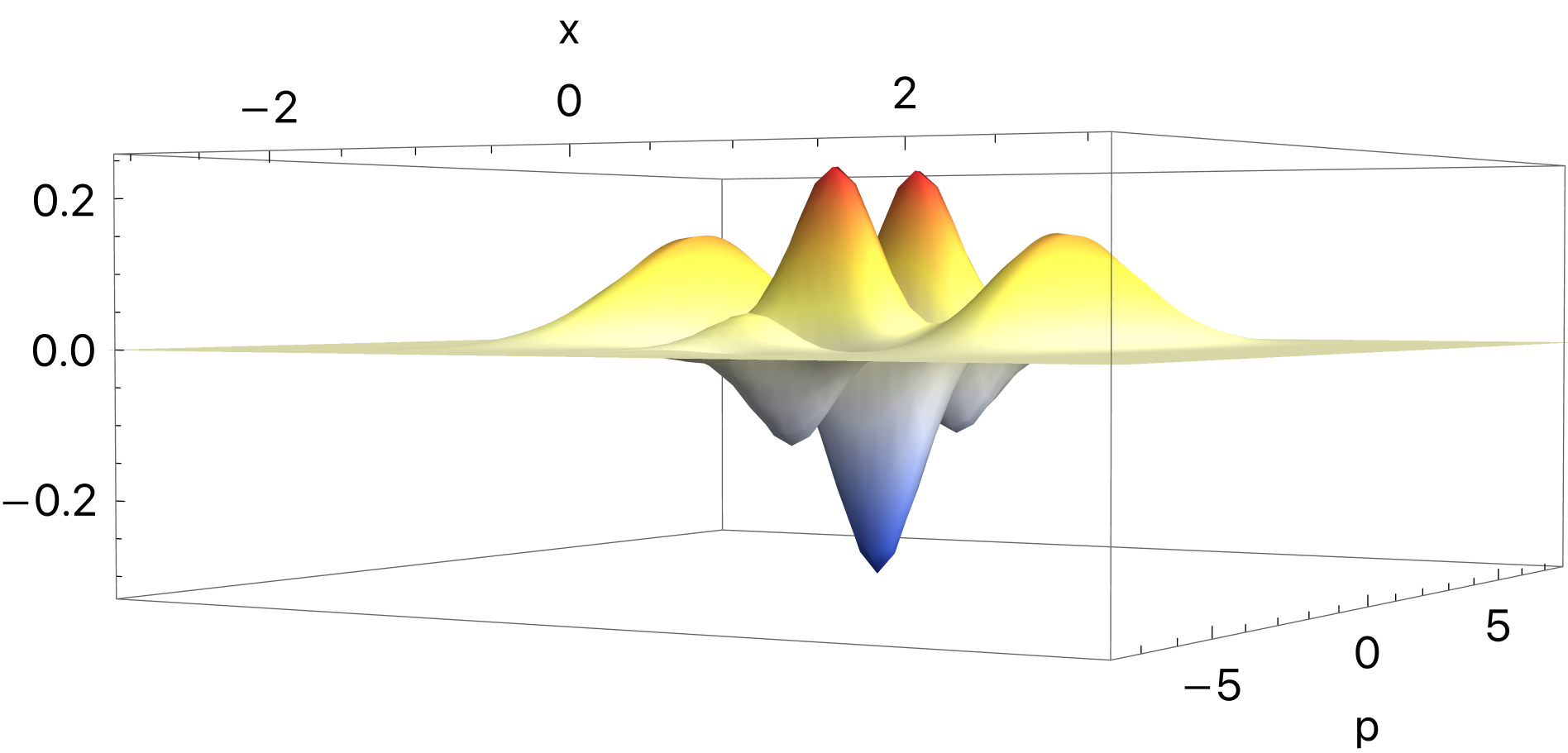}
        \caption{$\lambda=4$}
    \end{subfigure}

    \vspace{0.5cm} 

    \begin{subfigure}[b]{0.45\textwidth}
        \includegraphics[width=\textwidth]{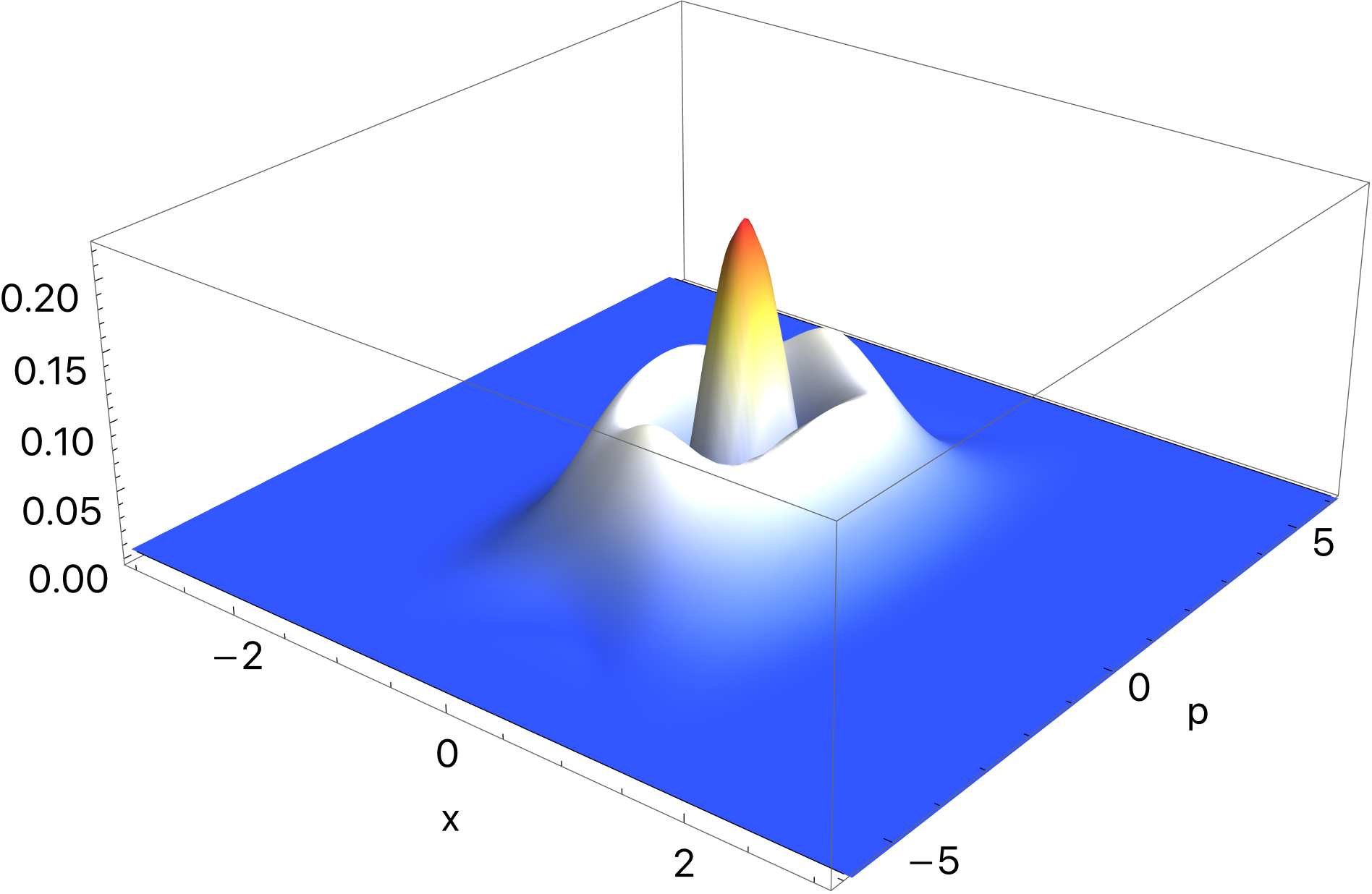}
        \caption{$\lambda=-\frac{3}{4}$}
    \end{subfigure}\hfill
    \begin{subfigure}[b]{0.45\textwidth}
        \includegraphics[width=\textwidth]{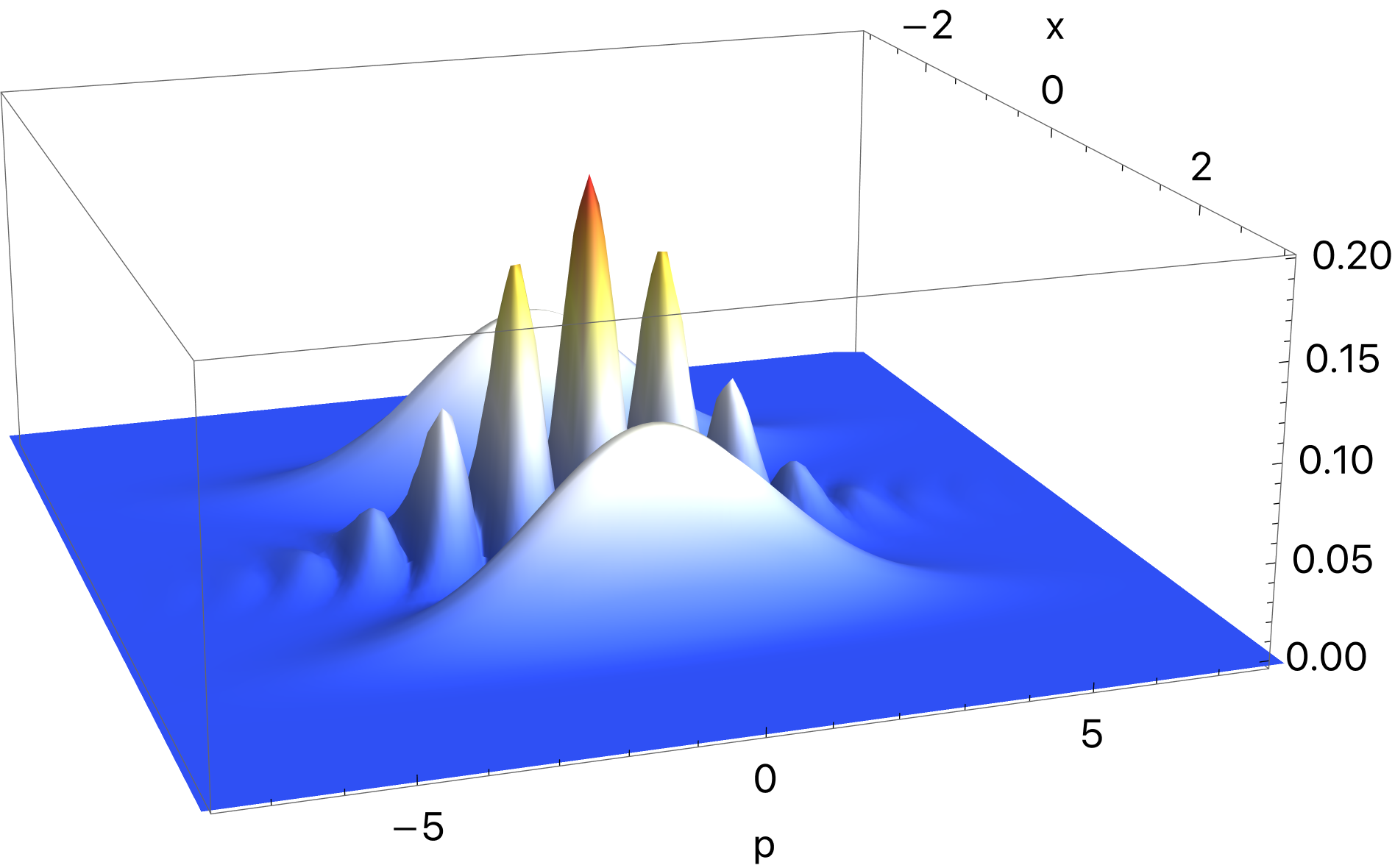}
        \caption{$\lambda=4$}
    \end{subfigure}

    \vspace{0.5cm} 

    \begin{subfigure}[b]{0.45\textwidth}
        \includegraphics[width=\textwidth]{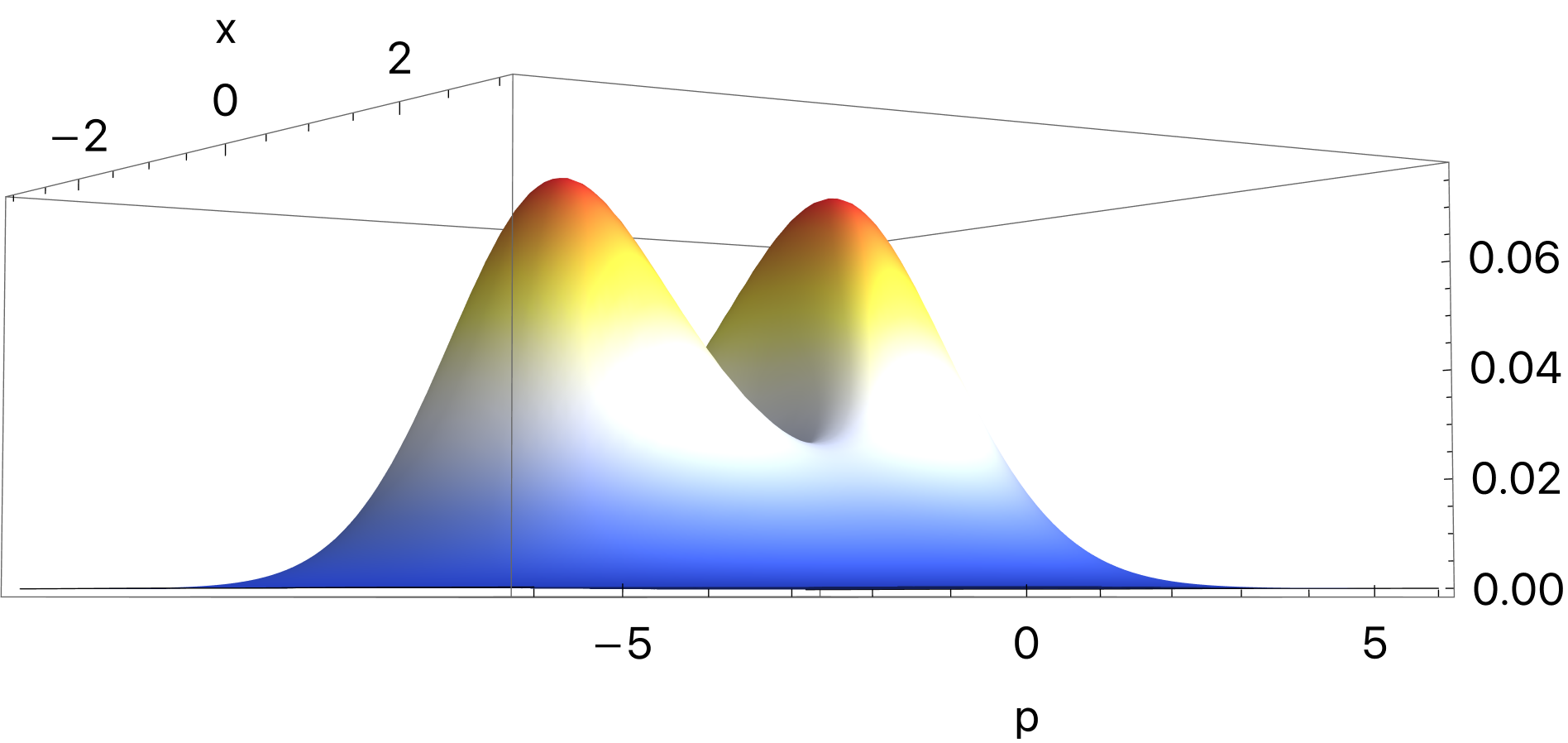}
        \caption{$\lambda=-\frac{3}{4}$}
    \end{subfigure}\hfill
    \begin{subfigure}[b]{0.45\textwidth}
        \includegraphics[width=\textwidth]{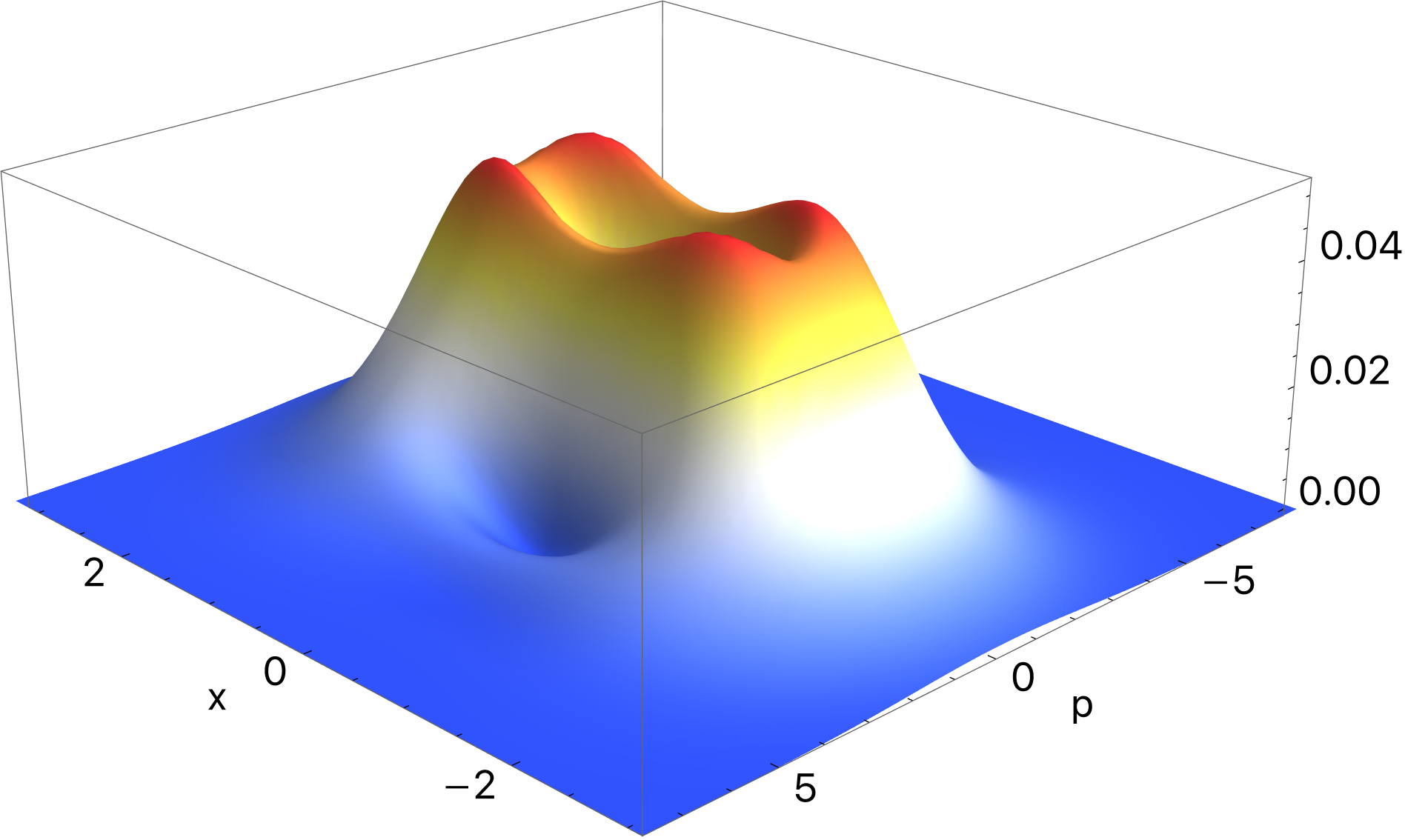}
        \caption{$\lambda=4$}
    \end{subfigure}
    \caption{\small  Wigner quasiprobability distribution $W(x,p)$ (top), modulus of the Wigner function $|W(x,p)|$ (middle), and Husimi distribution $H(x,p)$ (bottom) for the first excited state $n=1$ of the QES sextic oscillator at $\lambda=-\tfrac{3}{4}$ and $\lambda=4$.}
    \label{D-W-AbsW-Hn1}
\end{figure}

As $\lambda$ increases, the structural differences among \(W\), \(|W|\), and \(H\) observed in Figs.~\ref{D-W-AbsW-Hn0}-\ref{D-W-AbsW-Hn1} can be summarized as follows:

\begin{itemize}

\item \textbf{Before tunneling (\(\lambda \ll \lambda_c\)):}  
For \(n=0\), \(W\), \(|W|\), and \(H\) are nearly identical (Fig.~\ref{D-W-AbsW-Hn0}), all showing a single central lobe with minimal negativity, whereas for \(n=1\), these three representations already differ (Fig.~\ref{D-W-AbsW-Hn1}). In particular, \(W\) exhibits opposite-sign lobes separated by a nodal line, \(|W|\) retains a two-lobe geometry free of negativity, and \(H\) yields a broader, fully positive distribution with a smooth central valley.

\item \textbf{After tunneling (\(\lambda > \lambda_c\)):}  
Both states develop well-separated lobes, and the three representations diverge: \(W\) displays interference fringes, \(|W|\) retains the two-lobe geometry while eliminating all sign changes, but the interference fringes persist as positive oscillations, and \(H\) provides the smoothest, most classical-looking profile (Figs.~\ref{D-W-AbsW-Hn0}-\ref{D-W-AbsW-Hn1}).

\item \textbf{Cross-state comparison:}  
In the post-tunneling regime, \(|W|\) for \(n=0\) and \(n=1\) becomes remarkably similar, as sign removal suppresses parity-induced differences that remain evident in \(W\) and, to a lesser degree, in \(H\).

\item \textbf{Role of parity:} Even states produce a single symmetric lobe in $W$, while odd states exhibit two opposite-sign lobes separated by a nodal line. Taking the modulus $|W|$ removes negativity but preserves this one-peak (even) versus two-peak (odd) geometry, whereas the Husimi function $H$ smooths the structure into broad semiclassical peaks. These parity-imprinted features determine the distinct behavior of the $n=0$ and $n=1$ marginals, especially as $\lambda$ approaches the tunneling regime.

\end{itemize}

The phase–space plots above reveal the full two-dimensional structure of 
$W$, $|W|$, and $H$. We now turn to their position- and momentum-space 
marginals, which capture the projections of this structure onto the physical 
axes and serve as the fundamental inputs for the entropy, mutual-information, 
and CRJ analyses that follow.

\clearpage

\section{1D phase-space structure of low-lying states: marginal distributions}
\label{SEC5}
The position and momentum marginals derived from $W(x,p)$, $|W(x,p)|$, and
$H(x,p)$ provide simplified one-dimensional projections that retain the essential
localization and spreading properties of the phase-space distributions. These marginals
allow for a direct comparison between the three representations and clarify how
coarse graining, suppression of negativity, and smoothing influence the observable
profiles of the quantum state.

For each (quasi)probability distribution ${Q}(x,p) \in \{ W, |W|, H \}$, we define the 
position and momentum marginals by
\begin{equation}
{Q}_{x}(x) = \int_{-\infty}^{\infty} {Q}(x,p)\, dp,
\qquad
{Q}_{p}(p) = \int_{-\infty}^{\infty} {Q}(x,p)\, dx .
\end{equation}
For $W$, these marginals reproduce the exact quantum probability densities
$|\psi(x)|^{2}$ and $|\phi(p)|^{2}$, while for $|W|$ and $H$ they provide
coarse-grained and smoothed analogues of these quantities.

\subsection{Position Marginals}

\begin{figure}[H]
    \centering

    \begin{subfigure}[b]{0.23\textwidth}
        \includegraphics[width=\textwidth]{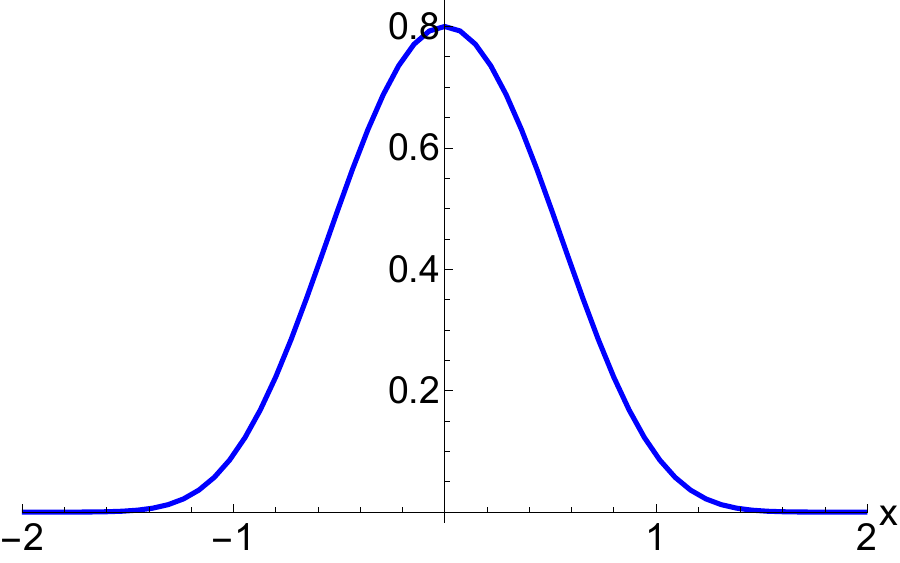}
        \caption{$\lambda=-\frac{3}{4}$}
    \end{subfigure}\hfill
    \begin{subfigure}[b]{0.23\textwidth}
        \includegraphics[width=\textwidth]{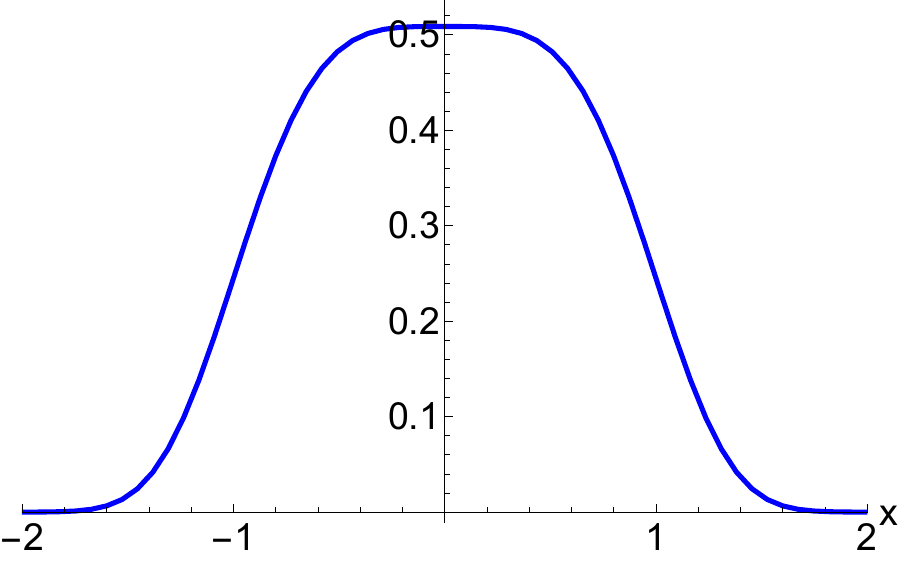}
        \caption{$\lambda_{c}^{n=0}=0.7329$}
    \end{subfigure}\hfill
    \begin{subfigure}[b]{0.23\textwidth}
        \includegraphics[width=\textwidth]{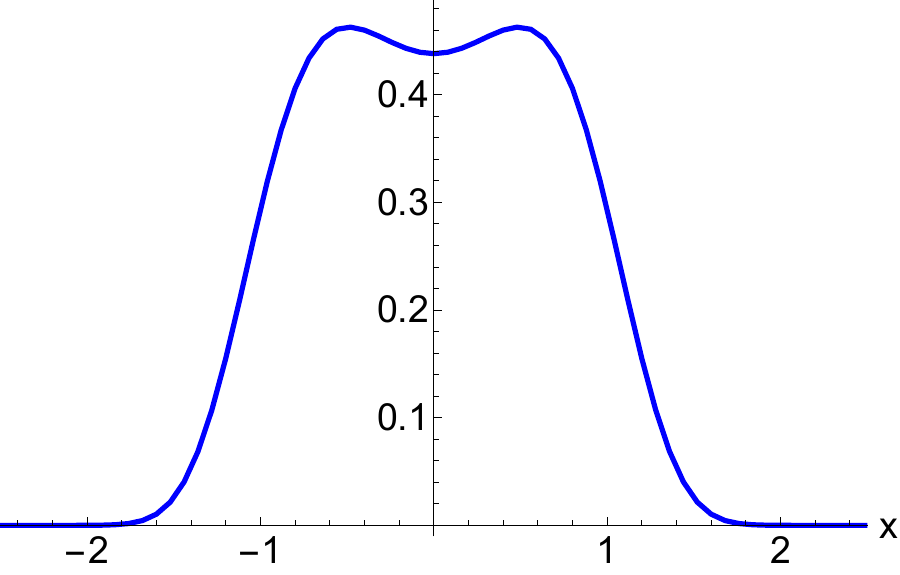}
        \caption{$\lambda=1$}
    \end{subfigure}\hfill
    \begin{subfigure}[b]{0.23\textwidth}
        \includegraphics[width=\textwidth]{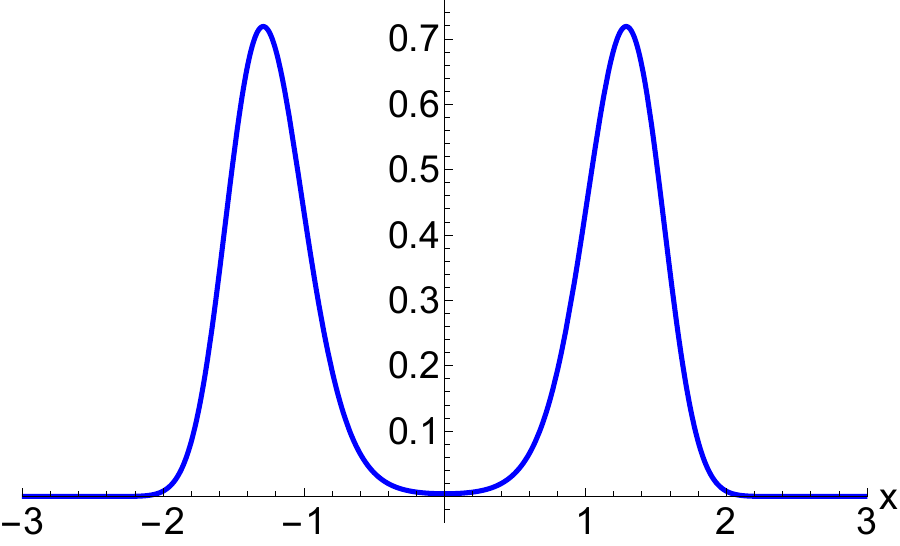}
        \caption{$\lambda=4$}
    \end{subfigure}

    \vspace{0.5cm} 

    \begin{subfigure}[b]{0.23\textwidth}
        \includegraphics[width=\textwidth]{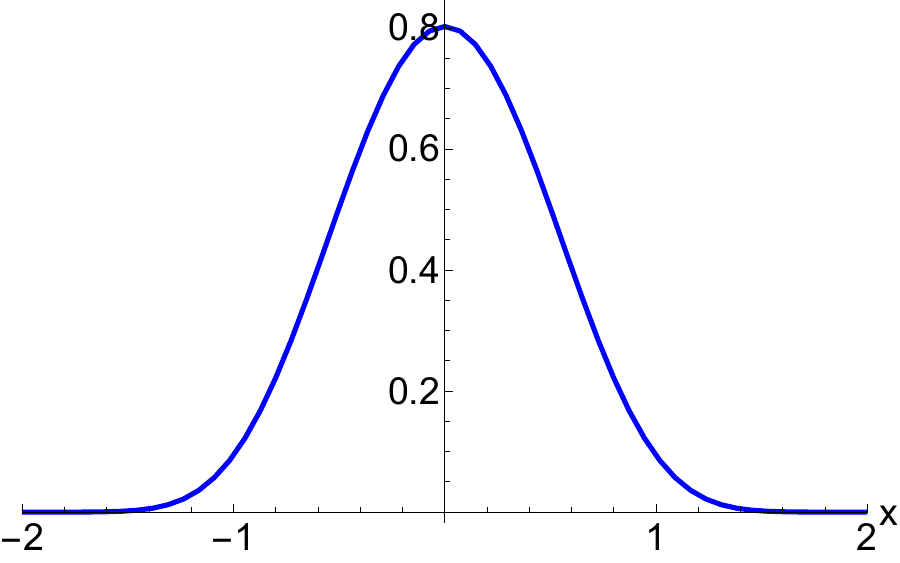}
        \caption{$\lambda=-\frac{3}{4}$}
    \end{subfigure}\hfill
    \begin{subfigure}[b]{0.23\textwidth}
        \includegraphics[width=\textwidth]{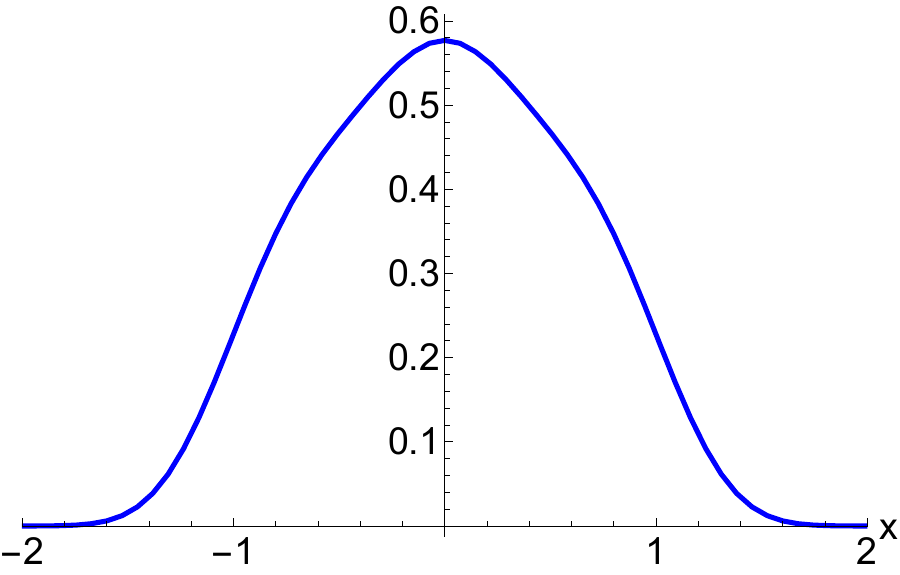}
        \caption{$\lambda_{c}^{n=0}=0.7329$}
    \end{subfigure}\hfill
    \begin{subfigure}[b]{0.23\textwidth}
        \includegraphics[width=\textwidth]{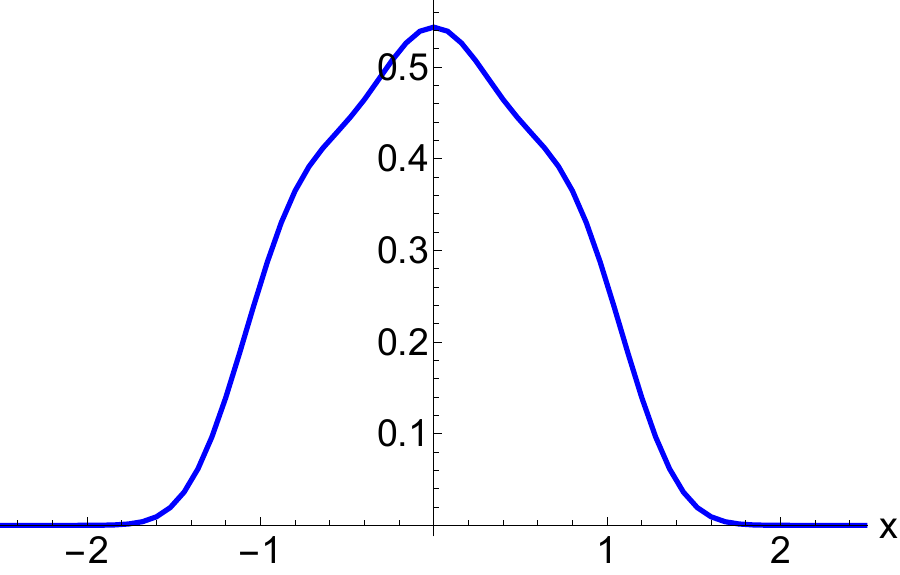}
        \caption{$\lambda=1$}
    \end{subfigure}\hfill
    \begin{subfigure}[b]{0.23\textwidth}
        \includegraphics[width=\textwidth]{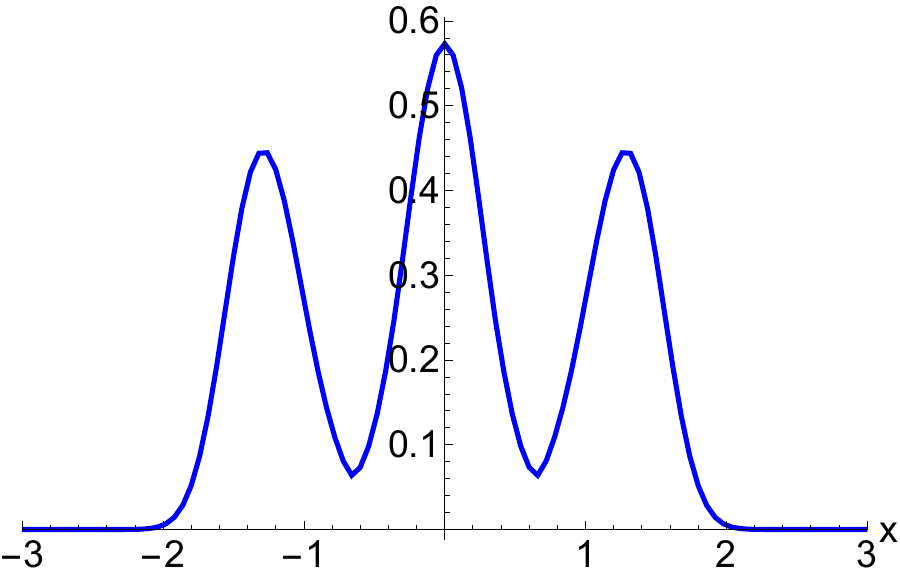}
        \caption{$\lambda=4$}
    \end{subfigure}

    \vspace{0.5cm} 

    \begin{subfigure}[b]{0.23\textwidth}
        \includegraphics[width=\textwidth]{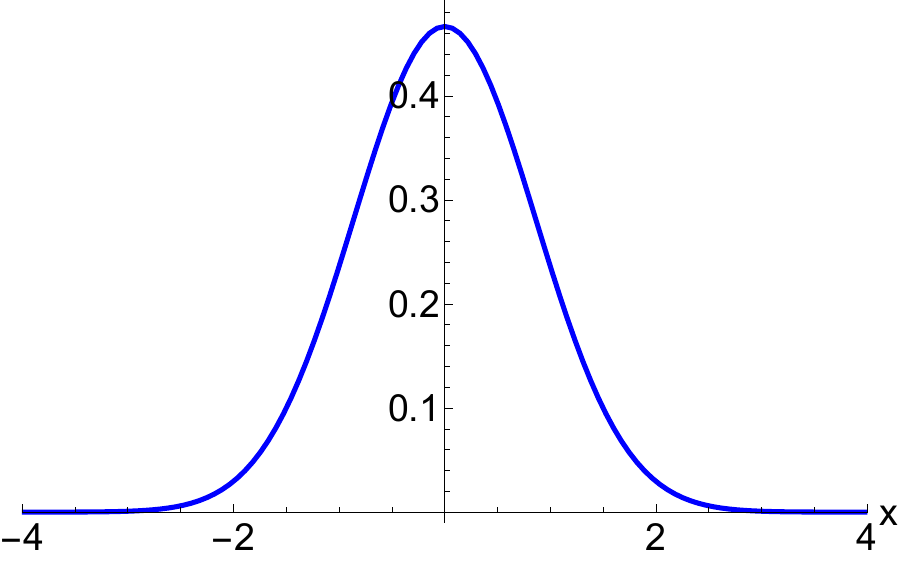}
        \caption{$\lambda=-\frac{3}{4}$}
    \end{subfigure}\hfill
    \begin{subfigure}[b]{0.23\textwidth}
        \includegraphics[width=\textwidth]{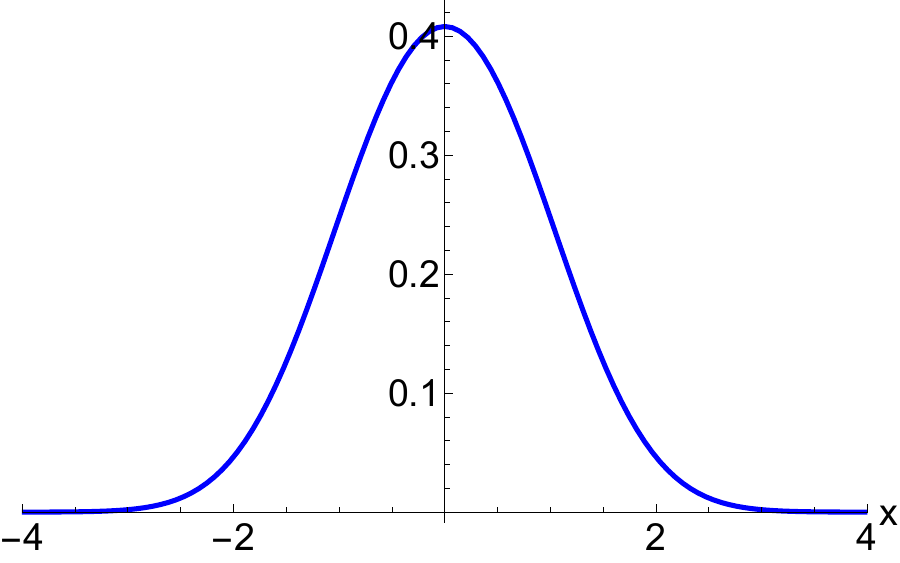}
        \caption{$\lambda_{c}^{n=0}=0.7329$}
    \end{subfigure}\hfill
    \begin{subfigure}[b]{0.23\textwidth}
        \includegraphics[width=\textwidth]{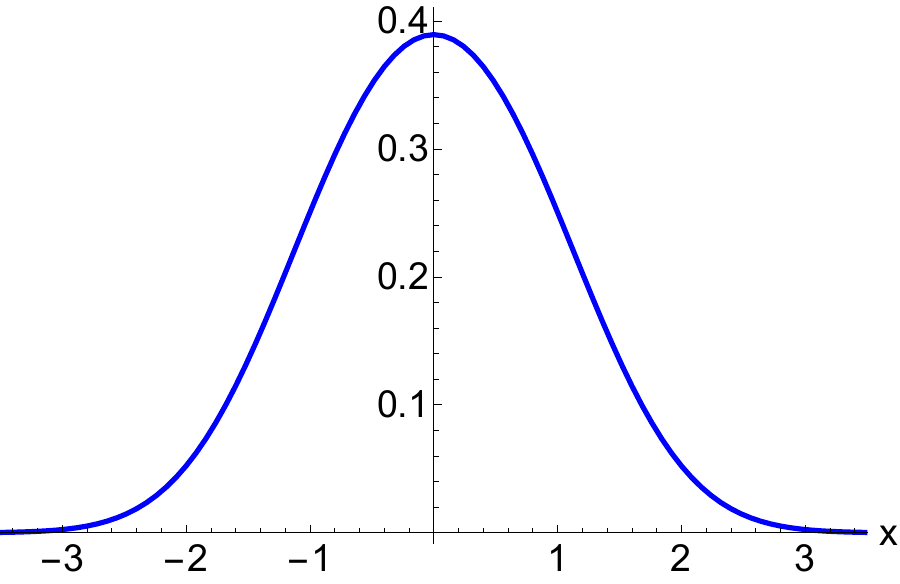}
        \caption{$\lambda=1$}
    \end{subfigure}\hfill
    \begin{subfigure}[b]{0.23\textwidth}
        \includegraphics[width=\textwidth]{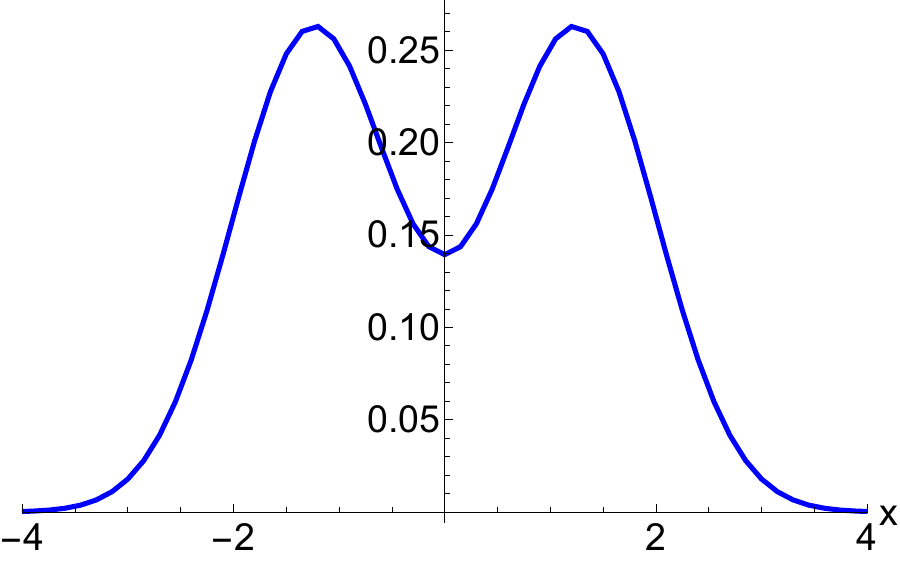}
        \caption{$\lambda=4$}
    \end{subfigure}

    \caption{Ground state $n=0$.  Position-space marginals $Q_x$ of the three distributions:  
(a)–(d) Wigner function $W$,  
(e)–(h) its modulus $|W|$, and  
(i)–(l) the Husimi distribution $H$,  
shown for different values of the parameter $\lambda$.
}
    \label{marginalesX0}
\end{figure}

\begin{figure}[H]
    \centering
    \begin{subfigure}[b]{0.48\textwidth}
        \centering
        \includegraphics[width=\textwidth]{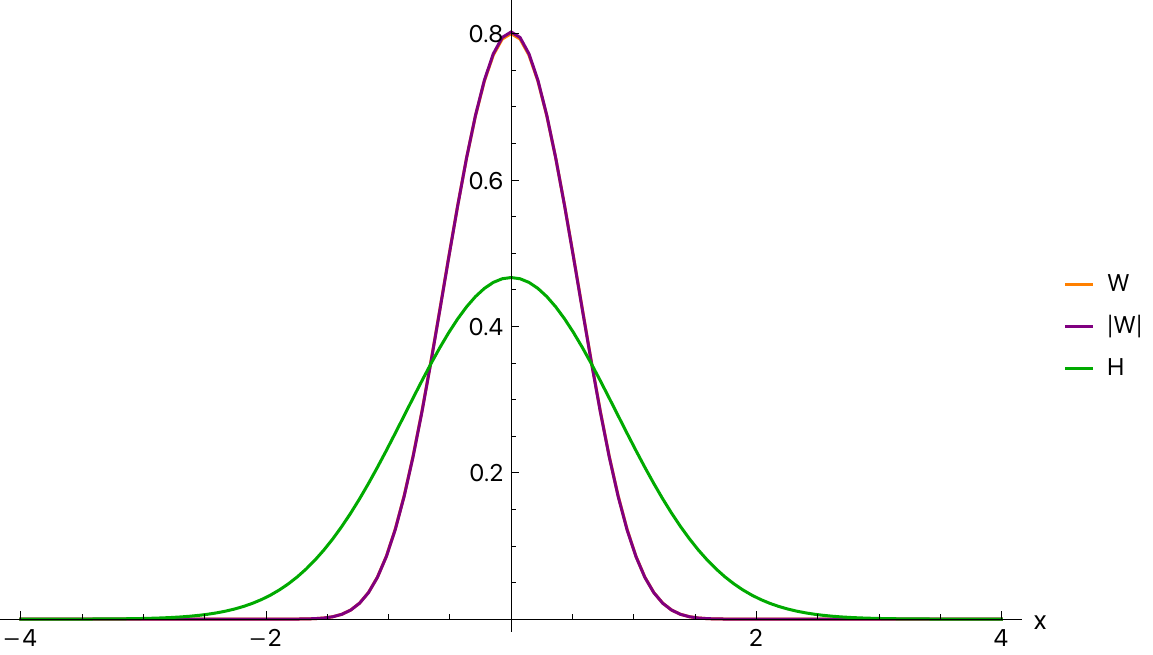}
        \caption{}
        \label{Den0m34x}
    \end{subfigure}
    \begin{subfigure}[b]{0.48\textwidth}
        \centering
        \includegraphics[width=\textwidth]{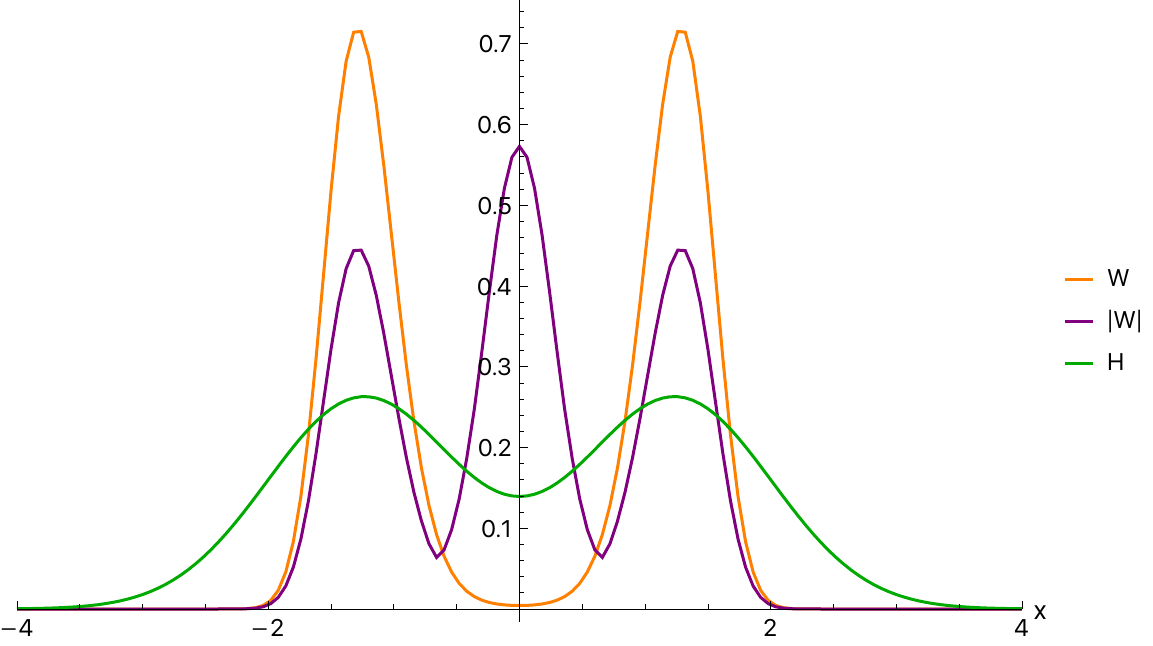}
        \caption{}
        \label{Den04x}
    \end{subfigure}
    \caption{Position-space marginals $Q_x$ of $W$, $|W|$, and $H$ for the ground state ($n=0$), shown on a common scale for (a) $\lambda = -\tfrac{3}{4}$ and (b) $\lambda = 4$.
}
    \label{densidadesjuntasX0}
\end{figure}

\begin{figure}[H]
    \centering

    \begin{subfigure}[b]{0.23\textwidth}
        \includegraphics[width=\textwidth]{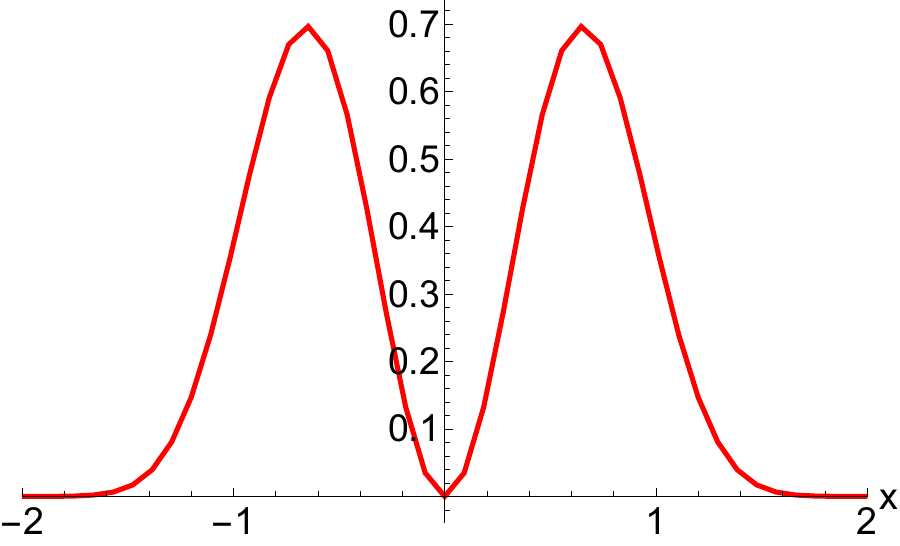}
        \caption{$\lambda=-\frac{3}{4}$}
    \end{subfigure}\hfill
    \begin{subfigure}[b]{0.23\textwidth}
        \includegraphics[width=\textwidth]{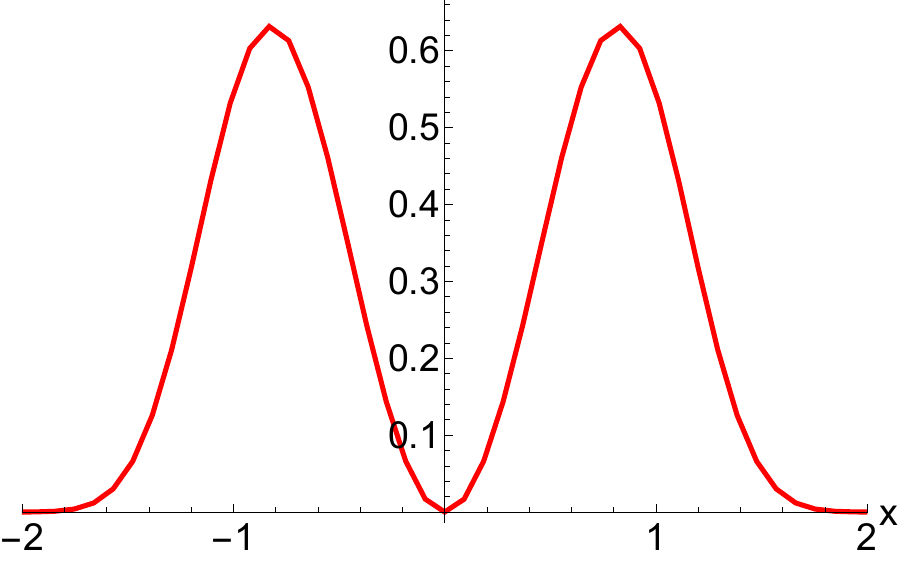}
        \caption{$\lambda=\frac{3}{4}$}
    \end{subfigure}\hfill
    \begin{subfigure}[b]{0.23\textwidth}
        \includegraphics[width=\textwidth]{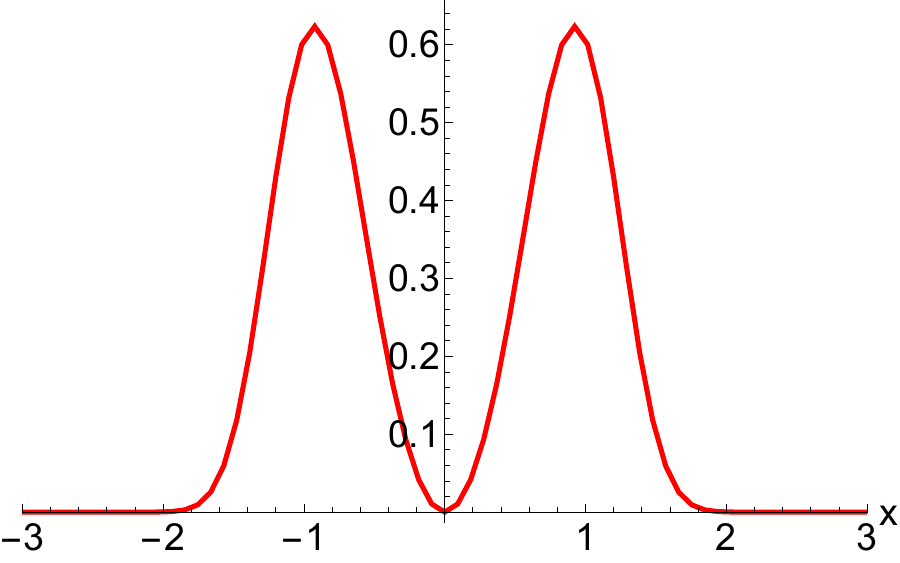}
        \caption{$\lambda_{c}^{n=1}=1.4209$}
    \end{subfigure}\hfill
    \begin{subfigure}[b]{0.23\textwidth}
        \includegraphics[width=\textwidth]{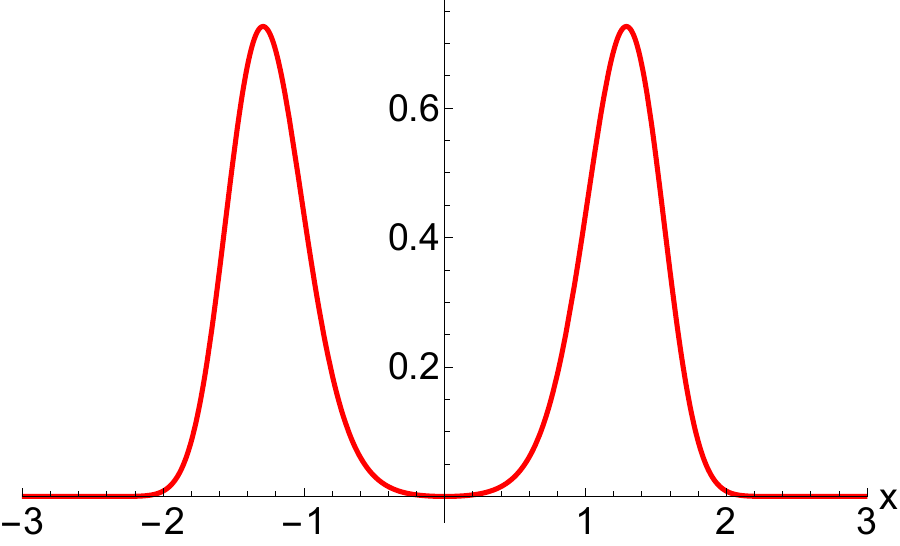}
        \caption{$\lambda=4$}
    \end{subfigure}

    \vspace{0.5cm} 

    \begin{subfigure}[b]{0.23\textwidth}
        \includegraphics[width=\textwidth]{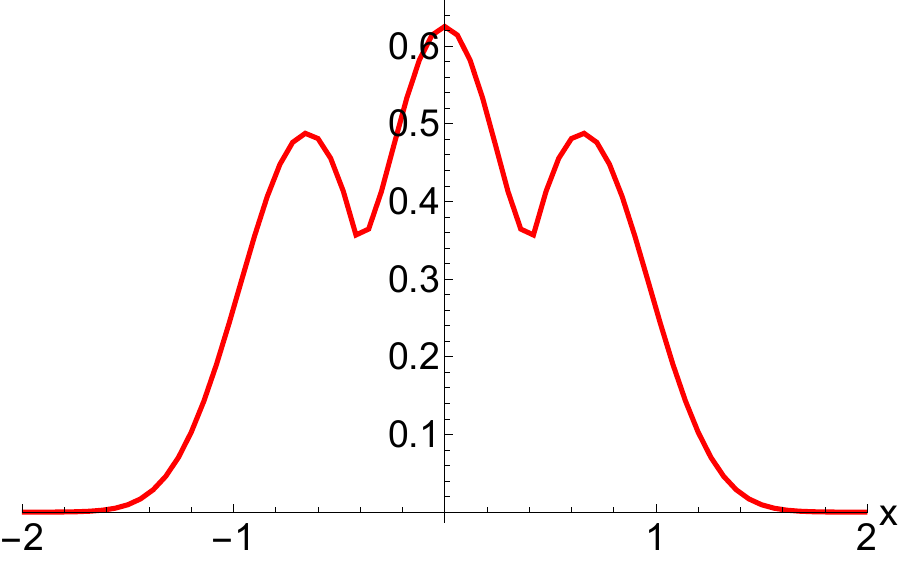}
        \caption{$\lambda=-\frac{3}{4}$}
    \end{subfigure}\hfill
    \begin{subfigure}[b]{0.23\textwidth}
        \includegraphics[width=\textwidth]{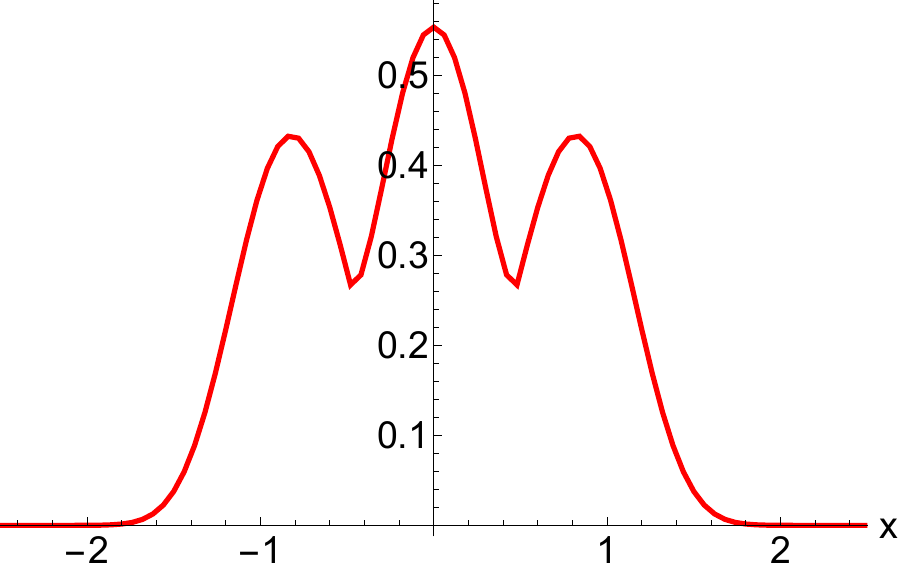}
        \caption{$\lambda=\frac{3}{4}$}
    \end{subfigure}\hfill
    \begin{subfigure}[b]{0.23\textwidth}
        \includegraphics[width=\textwidth]{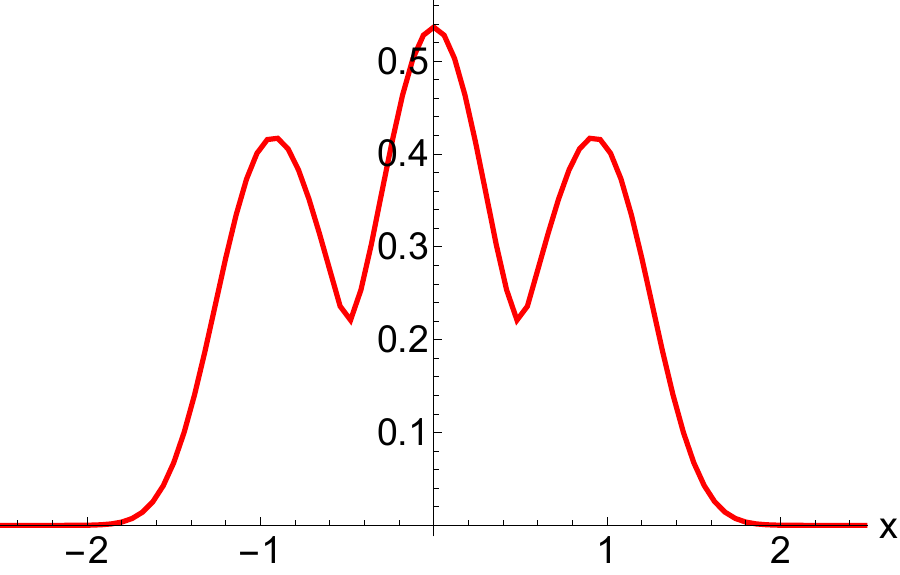}
        \caption{$\lambda_{c}^{n=1}=1.4209$}
    \end{subfigure}\hfill
    \begin{subfigure}[b]{0.23\textwidth}
        \includegraphics[width=\textwidth]{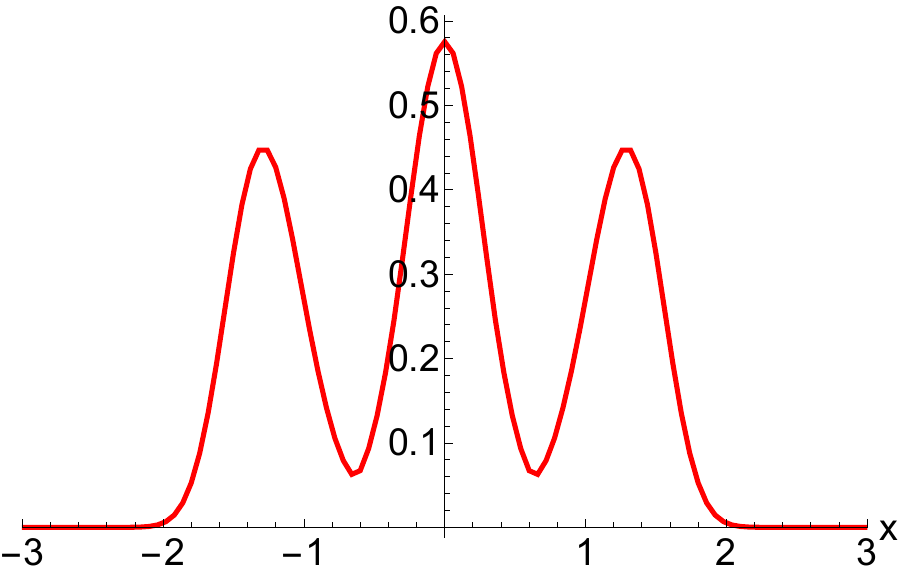}
        \caption{$\lambda=4$}
    \end{subfigure}

    \vspace{0.5cm} 

    \begin{subfigure}[b]{0.23\textwidth}
        \includegraphics[width=\textwidth]{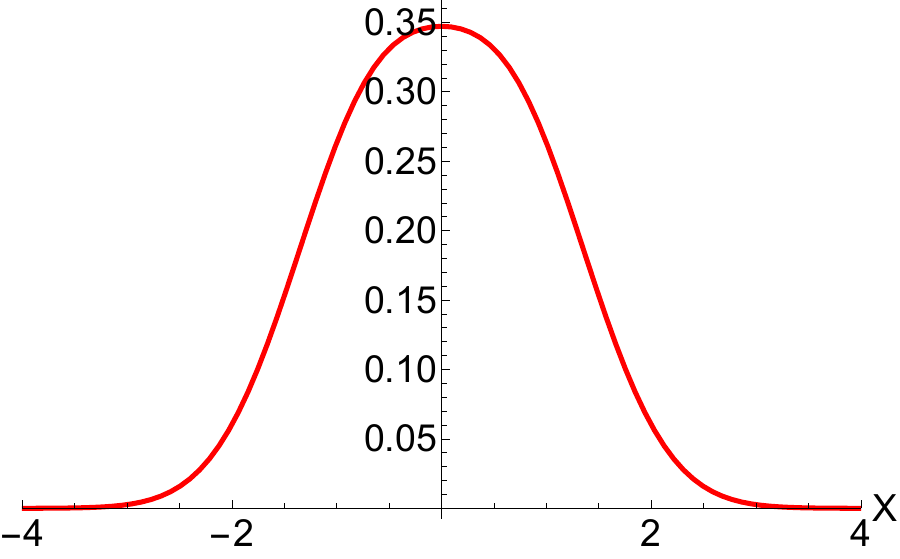}
        \caption{$\lambda=-\frac{3}{4}$}
    \end{subfigure}\hfill
    \begin{subfigure}[b]{0.23\textwidth}
        \includegraphics[width=\textwidth]{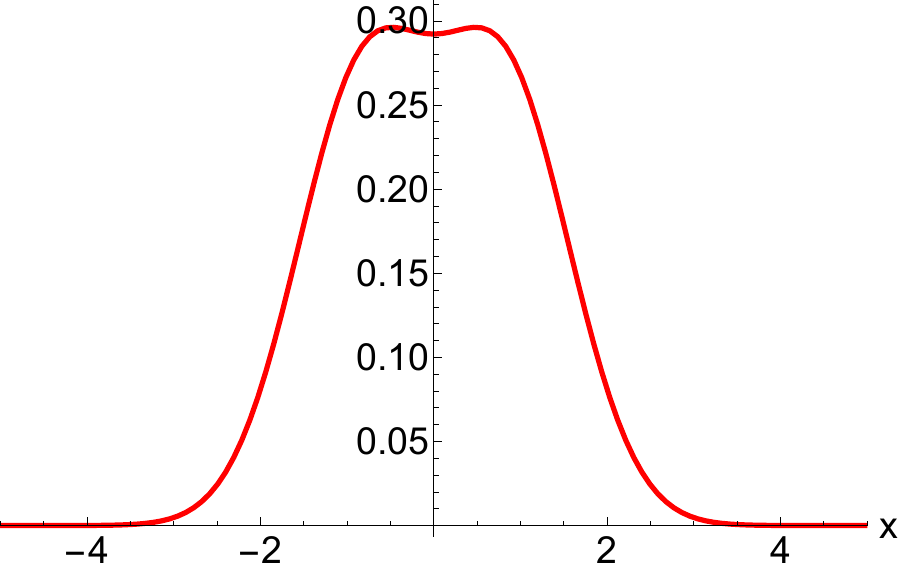}
        \caption{$\lambda=\frac{3}{4}$}
    \end{subfigure}\hfill
    \begin{subfigure}[b]{0.23\textwidth}
        \includegraphics[width=\textwidth]{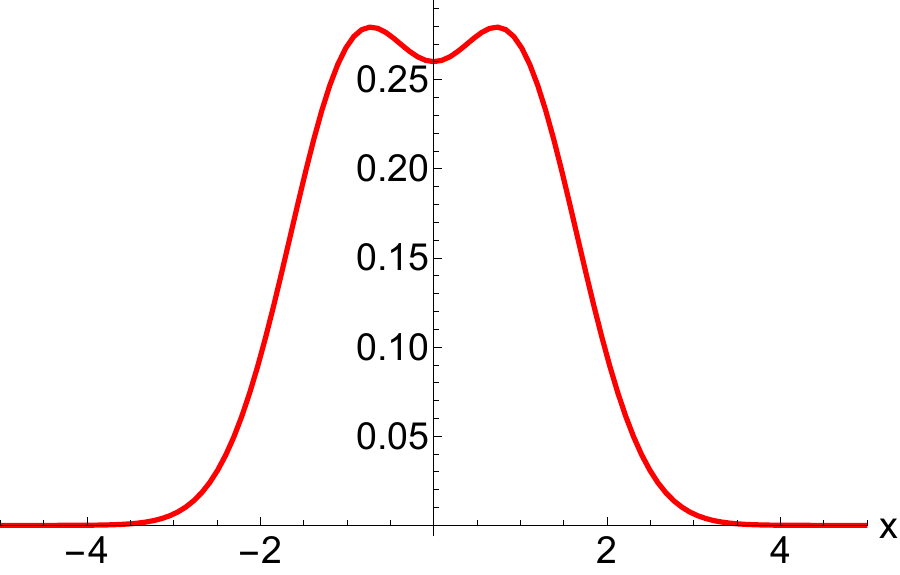}
        \caption{$\lambda_{c}^{n=1}=1.4209$}
    \end{subfigure}\hfill
    \begin{subfigure}[b]{0.23\textwidth}
        \includegraphics[width=\textwidth]{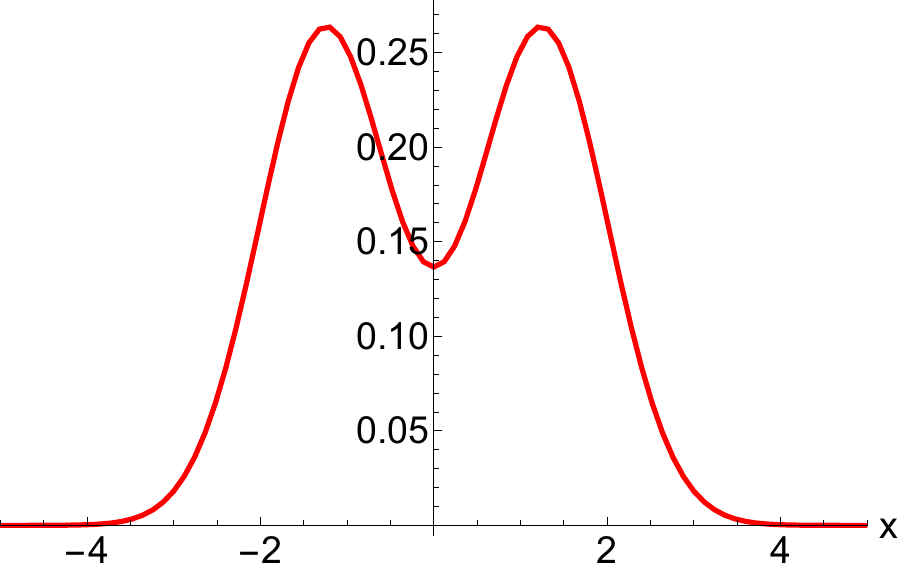}
        \caption{$\lambda=4$}
    \end{subfigure}

    \caption{First excited state $n=1$. Position-space marginals $Q_x$ of:  
(a)–(d) the Wigner function $W$,  
(e)–(h) its modulus $|W|$, and  
(i)–(l) the Husimi distribution $H$,  
for various values of $\lambda$.
}
    \label{marginalesX1}
\end{figure}

\begin{figure}[H]
    \centering
    \begin{subfigure}[b]{0.48\textwidth}
        \centering
        \includegraphics[width=\textwidth]{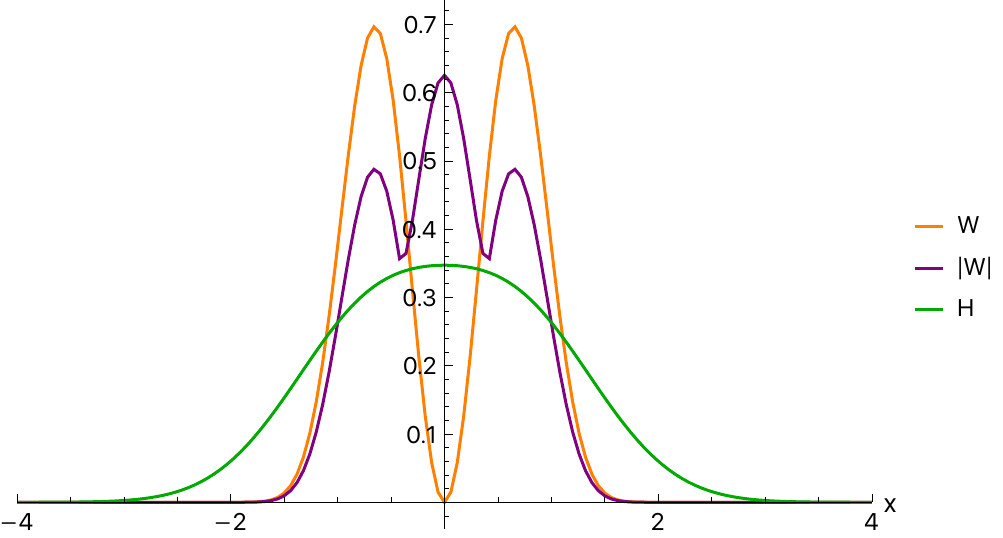}
        \caption{}
        \label{Den1m34x}
    \end{subfigure}
    \begin{subfigure}[b]{0.48\textwidth}
        \centering
        \includegraphics[width=\textwidth]{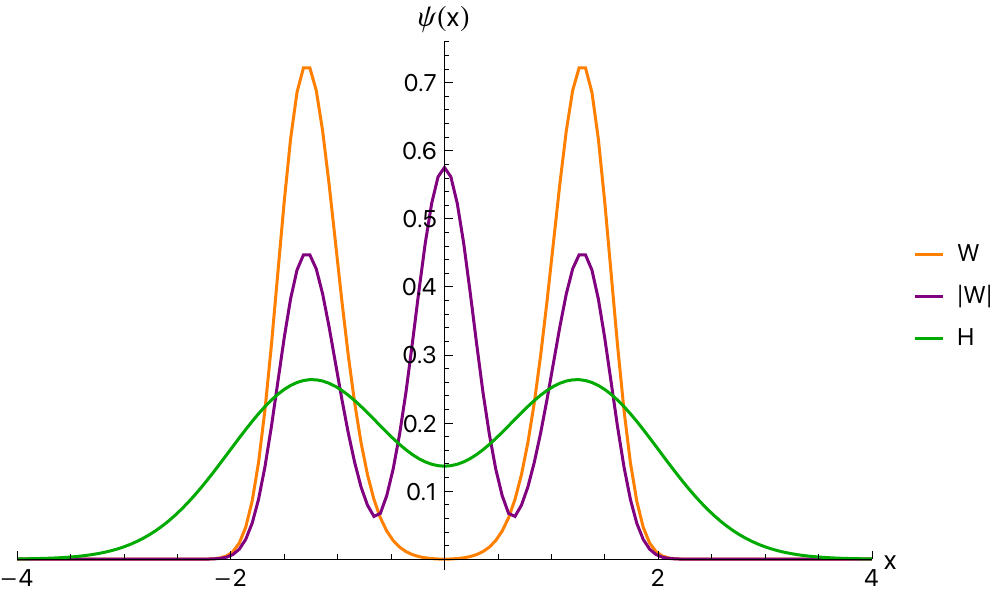}
        \caption{}
        \label{Den14x}
    \end{subfigure}
    \caption{Position-space marginals $Q_x$ of $W$, $|W|$, and $H$ for the first excited state ($n=1$), shown on a common scale for (a) $\lambda = -\tfrac{3}{4}$ and (b) $\lambda = 4$.}
    \label{densidadesjuntasX1}
\end{figure}

Figures~\ref{marginalesX0}-\ref{densidadesjuntasX1} illustrate
$Q_{x}(x)$ for the ground and first excited states. In Figs. \ref{marginalesX0} and \ref{marginalesX1} we adopt different scales on the axes to better highlight the overall form of the marginals, while in Figs. \ref{densidadesjuntasX0} and \ref{densidadesjuntasX1} we use a common scale to bring out additional structural details.
Several general
trends emerge across $\lambda$:

\begin{itemize}
    \item \textbf{Wigner marginal $W_{x}(x) = |\psi(x)|^{2}$.}  
    This is the only distribution that retains the nodal and parity structure of the
    true quantum state. For the ground state, $W_{x}(x)$ transitions from a 
    single-peaked profile to a bimodal distribution as $\lambda$ increases and the 
    double well becomes pronounced. For the first excited state, the central node at 
    $x=0$ is preserved for all values of the coupling.

    \item \textbf{Modulus marginal $|W|_{x}(x)$.}  
    Removing the sign structure from $W$ causes the interference-induced dips in
    $W_{x}$ to fill in. For $n=0$, the two peaks appear earlier as $\lambda$
    increases, because $|W|$ enhances separated structures by eliminating destructive
    interference. For $n=1$, the central depression is significantly reduced but not
    eliminated, since the antisymmetric shape of the state still influences the
    underlying geometry of $W$.

    \item \textbf{Husimi marginal $H_{x}(x)$.}  
    This represents the most smoothed version of the position density. The transition
    from a single peak to a two-peak structure occurs at larger $\lambda$ than in
    $W_{x}$ or $|W|_{x}$, reflecting the smoothing imposed by Gaussian convolution.
    Even at large $\lambda$, the peaks of $H_{x}$ remain significantly broader than
    those of $W_{x}$, showing that coarse graining suppresses sharp features related to
    tunneling-induced localization.
\end{itemize}

These differences highlight the role of negativity (present in $W$) and smoothing (present
in $H$) in shaping the observable distribution along the $x$ coordinate. The modulus
distribution $|W|_{x}$ provides an intermediate behaviour: it eliminates interference
without erasing the structural splitting associated with the classical minima of the
double-well potential.

\subsection{Momentum Marginals}

\begin{figure}[H]
    \centering

    \begin{subfigure}[b]{0.23\textwidth}
        \includegraphics[width=\textwidth]{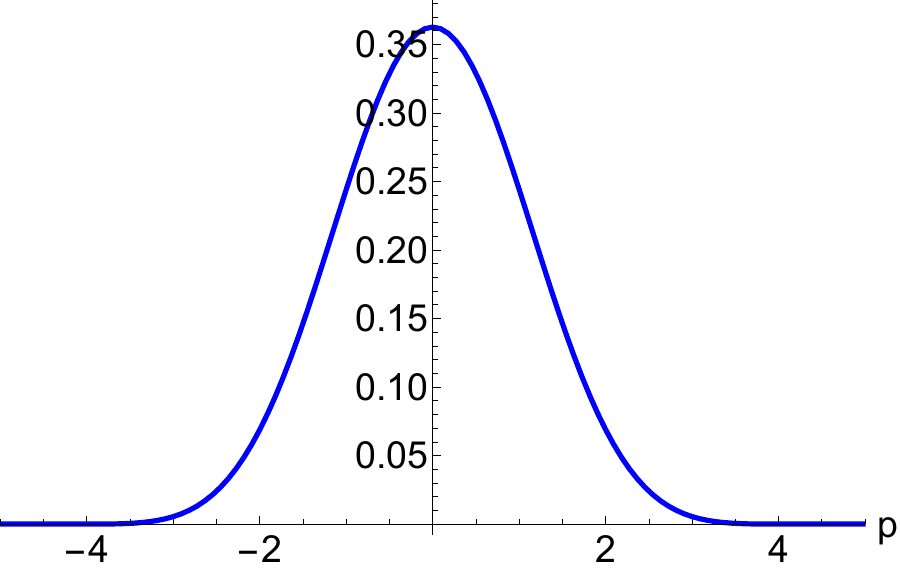}
        \caption{$\lambda=-\frac{3}{4}$}
    \end{subfigure}\hfill
    \begin{subfigure}[b]{0.23\textwidth}
        \includegraphics[width=\textwidth]{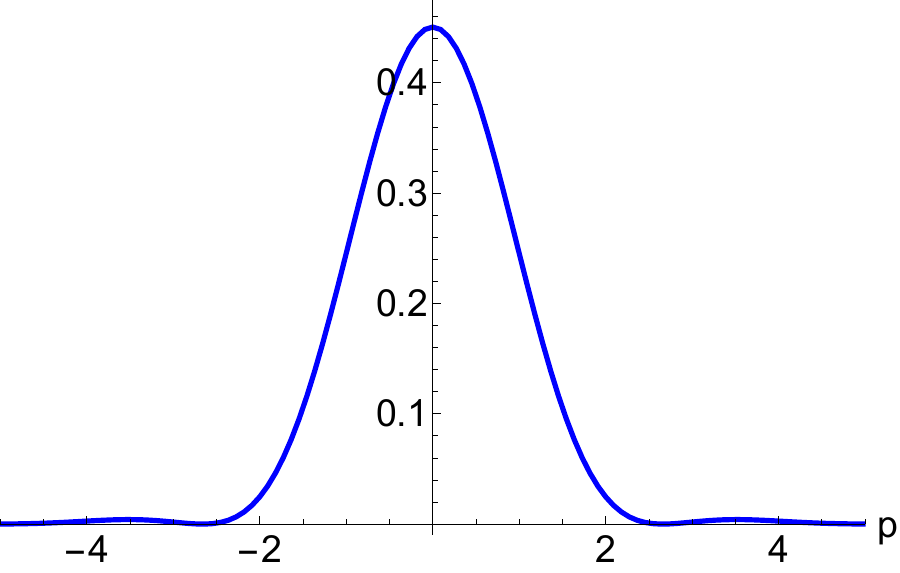}
        \caption{$\lambda_{c}^{n=0}=0.7329$}
    \end{subfigure}\hfill
    \begin{subfigure}[b]{0.23\textwidth}
        \includegraphics[width=\textwidth]{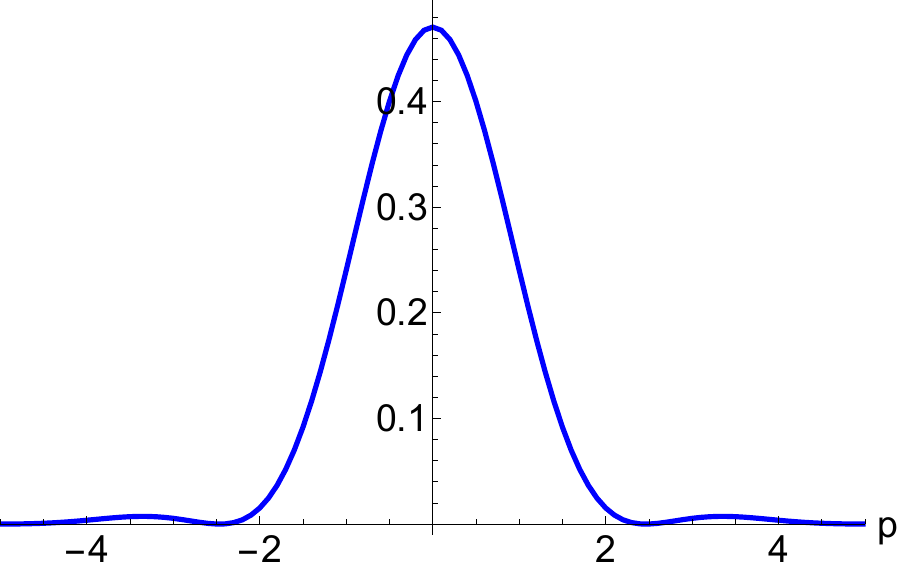}
        \caption{$\lambda=1$}
    \end{subfigure}\hfill
    \begin{subfigure}[b]{0.23\textwidth}
        \includegraphics[width=\textwidth]{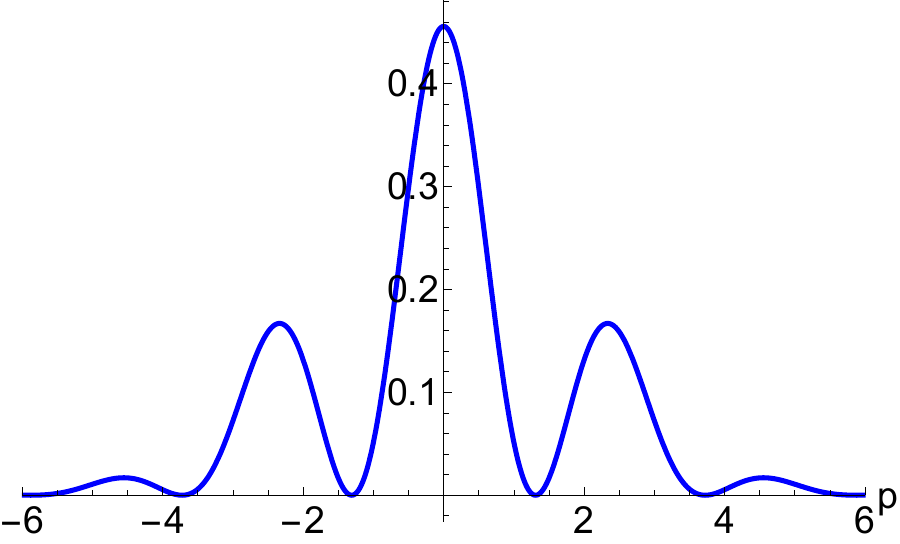}
        \caption{$\lambda=4$}
    \end{subfigure}

    \vspace{0.5cm} 

    \begin{subfigure}[b]{0.23\textwidth}
        \includegraphics[width=\textwidth]{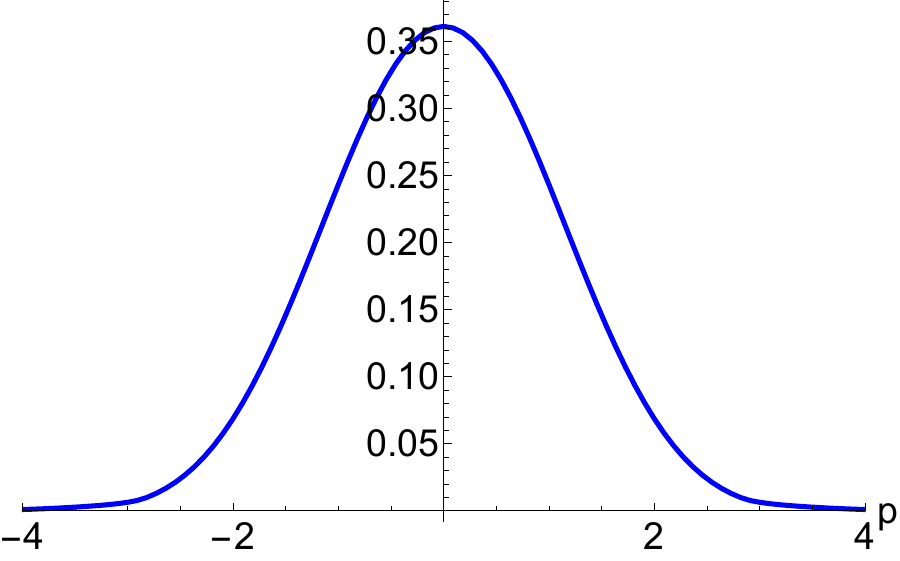}
        \caption{$\lambda=-\frac{3}{4}$}
    \end{subfigure}\hfill
    \begin{subfigure}[b]{0.23\textwidth}
        \includegraphics[width=\textwidth]{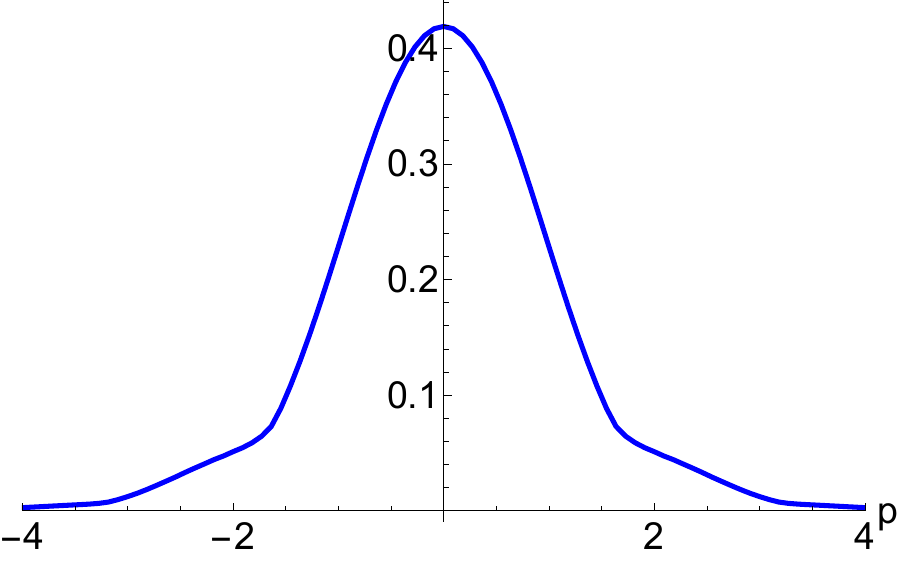}
        \caption{$\lambda_{c}^{n=0}=0.7329$}
    \end{subfigure}\hfill
    \begin{subfigure}[b]{0.23\textwidth}
        \includegraphics[width=\textwidth]{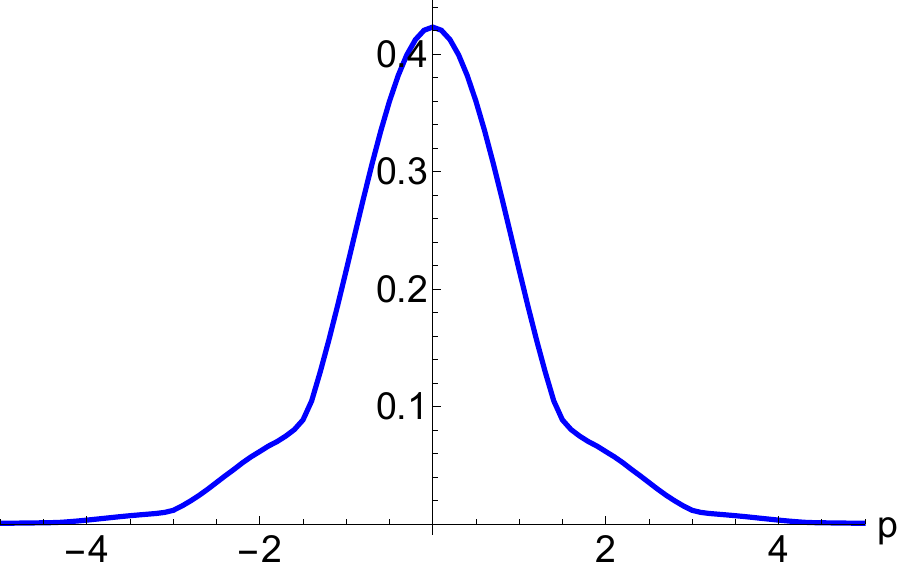}
        \caption{$\lambda=1$}
    \end{subfigure}\hfill
    \begin{subfigure}[b]{0.23\textwidth}
        \includegraphics[width=\textwidth]{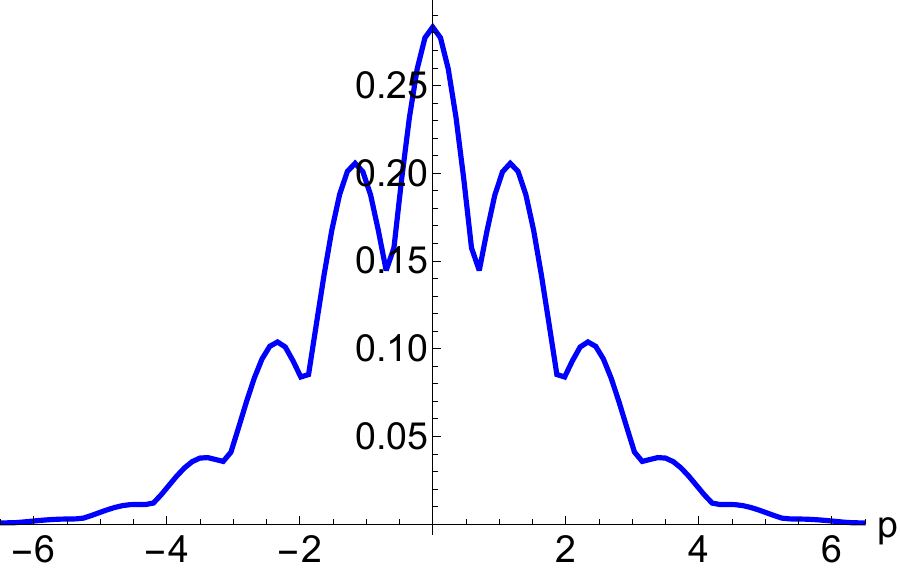}
        \caption{$\lambda=4$}
    \end{subfigure}

    \vspace{0.5cm} 

    \begin{subfigure}[b]{0.23\textwidth}
        \includegraphics[width=\textwidth]{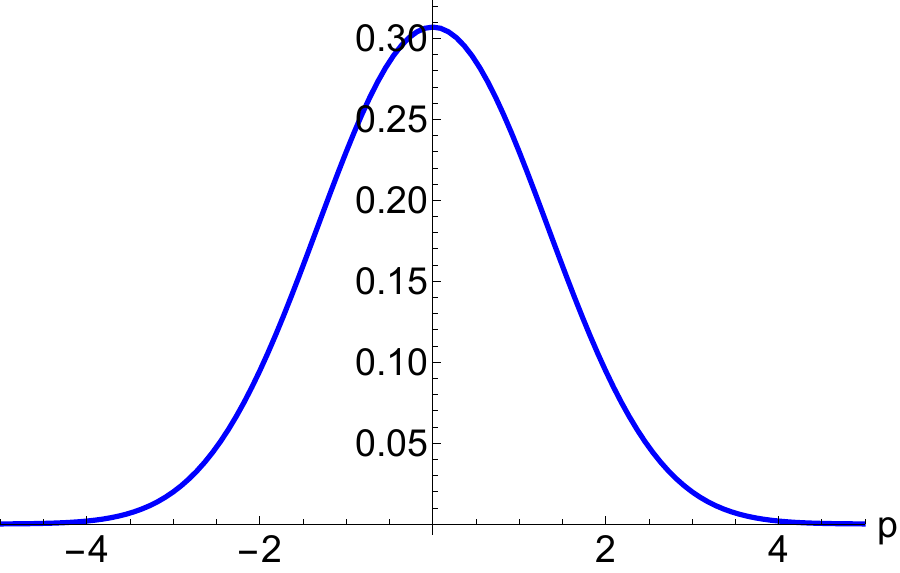}
        \caption{$\lambda=-\frac{3}{4}$}
    \end{subfigure}\hfill
    \begin{subfigure}[b]{0.23\textwidth}
        \includegraphics[width=\textwidth]{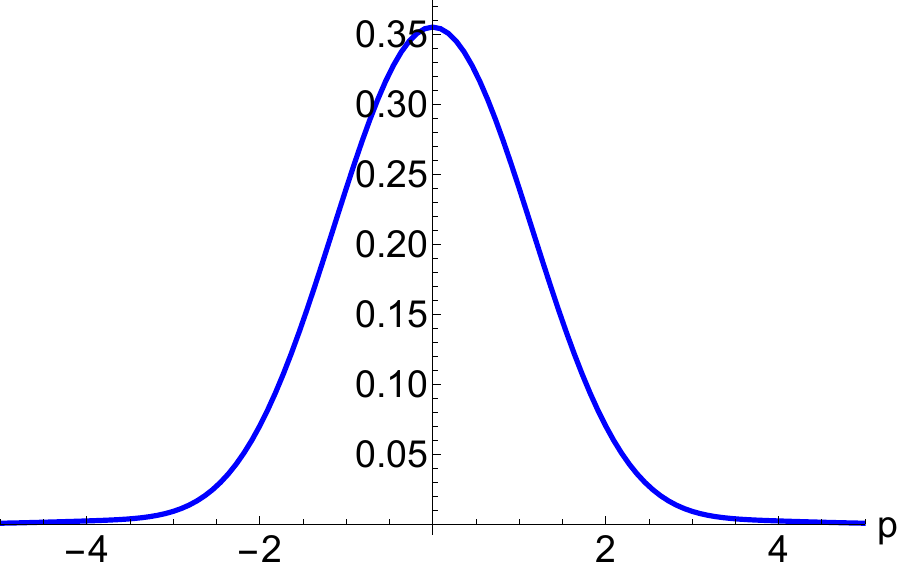}
        \caption{$\lambda_{c}^{n=0}=0.7329$}
    \end{subfigure}\hfill
    \begin{subfigure}[b]{0.23\textwidth}
        \includegraphics[width=\textwidth]{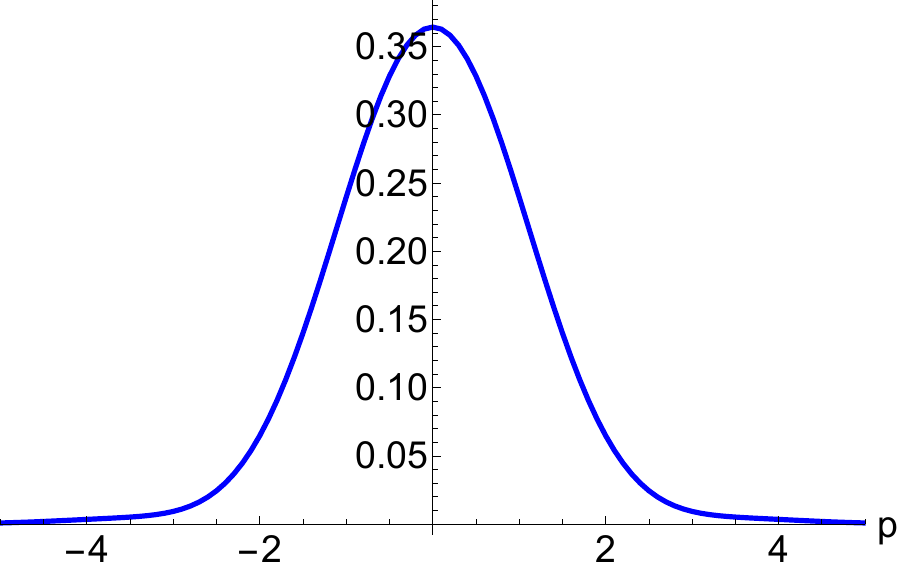}
        \caption{$\lambda=1$}
    \end{subfigure}\hfill
    \begin{subfigure}[b]{0.23\textwidth}
        \includegraphics[width=\textwidth]{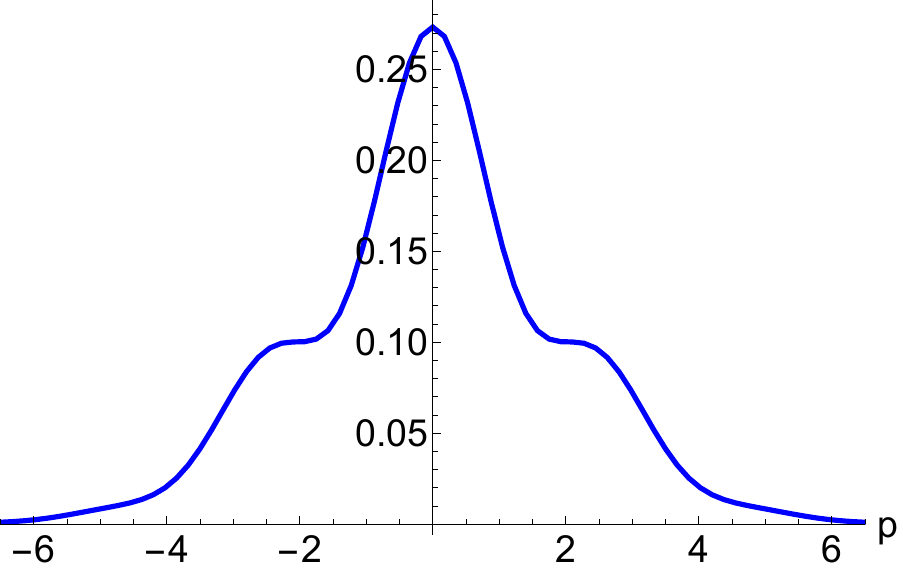}
        \caption{$\lambda=4$}
    \end{subfigure}

    \caption{\small Ground state $n=0$. Momentum-space marginals $Q_p$ of:  
(a)–(d) the Wigner function $W$,  
(e)–(h) its modulus $|W|$, and  
(i)–(l) the Husimi distribution $H$, shown for different values of $\lambda$.
}
    \label{marginalesP0}
\end{figure}

\begin{figure}[H]
    \centering
    \begin{subfigure}[b]{0.48\textwidth}
        \centering
        \includegraphics[width=\textwidth]{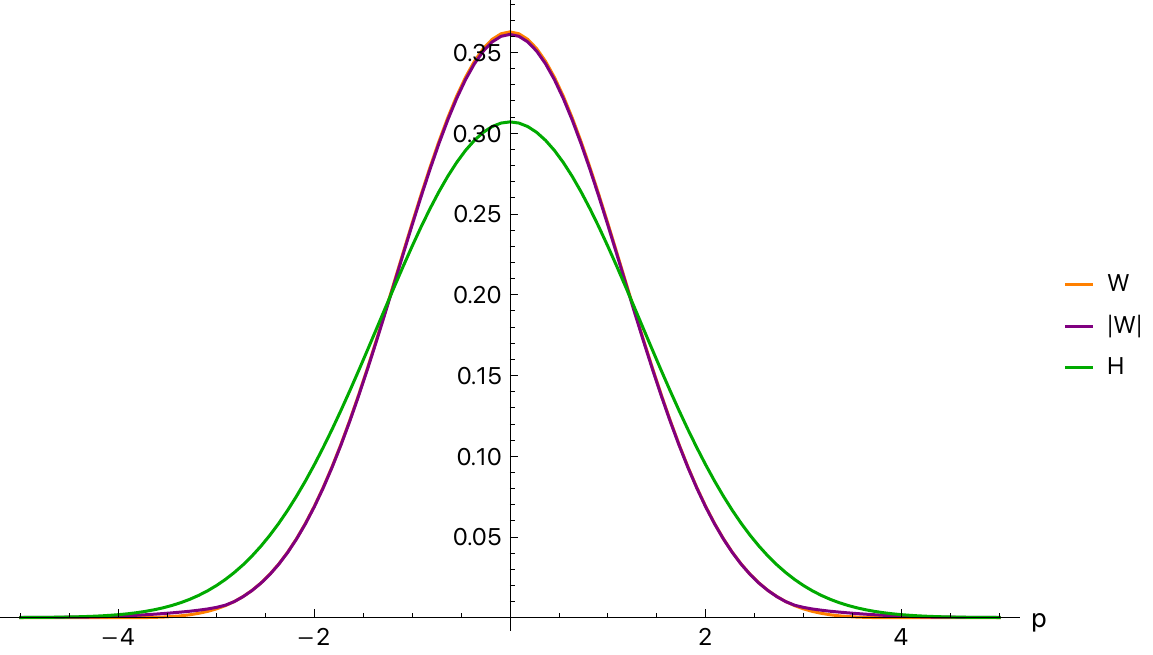}
        \caption{}
        \label{Den0m34p}
    \end{subfigure}
    \begin{subfigure}[b]{0.48\textwidth}
        \centering
        \includegraphics[width=\textwidth]{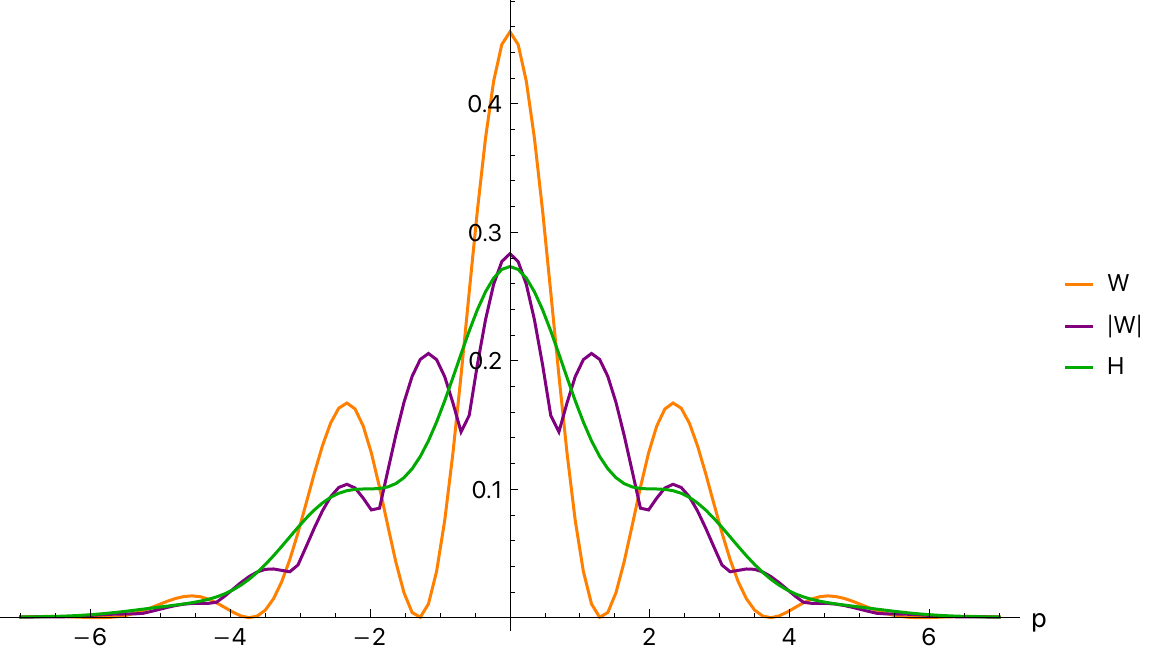}
        \caption{}
        \label{Den04p}
    \end{subfigure}
    \caption{Momentum-space marginals $Q_p$ of $W$, $|W|$, and $H$ for the ground state ($n=0$), shown on a common scale for (a) $\lambda = -\tfrac{3}{4}$ and (b) $\lambda = 4$.
}
    \label{densidadesjuntasP0}
\end{figure}

\begin{figure}[H]
    \centering

    \begin{subfigure}[b]{0.23\textwidth}
        \includegraphics[width=\textwidth]{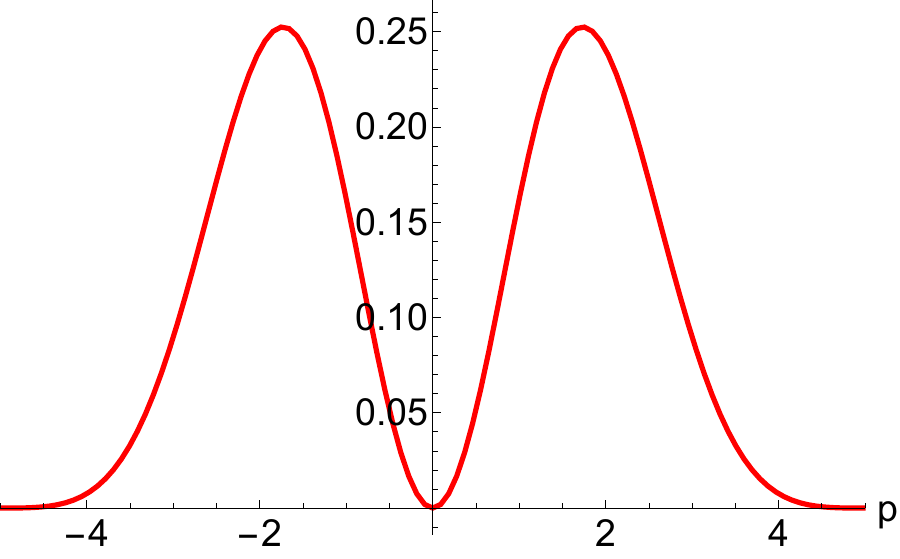}
        \caption{$\lambda=-\frac{3}{4}$}
    \end{subfigure}\hfill
    \begin{subfigure}[b]{0.23\textwidth}
        \includegraphics[width=\textwidth]{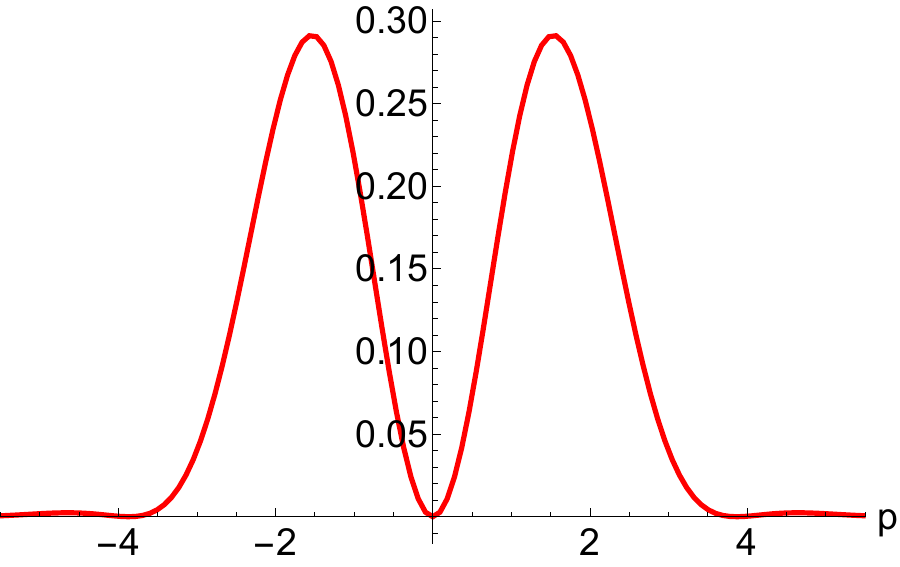}
        \caption{$\lambda=\frac{3}{4}$}
    \end{subfigure}\hfill
    \begin{subfigure}[b]{0.23\textwidth}
        \includegraphics[width=\textwidth]{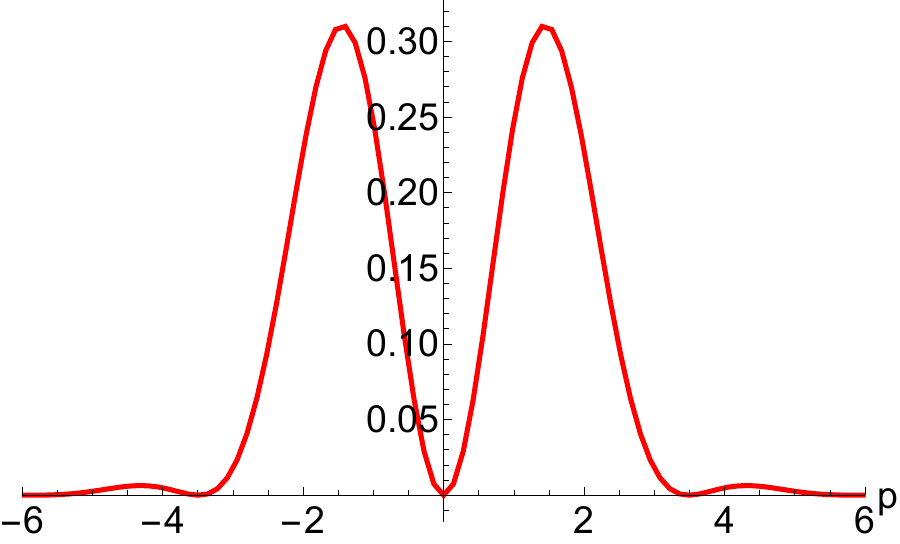}
        \caption{$\lambda_{c}^{n=1}=1.4209$}
    \end{subfigure}\hfill
    \begin{subfigure}[b]{0.23\textwidth}
        \includegraphics[width=\textwidth]{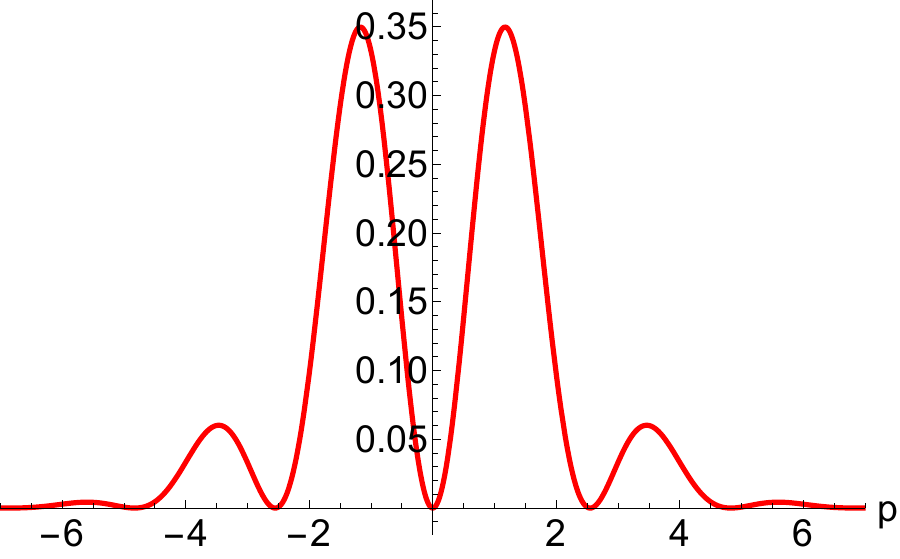}
        \caption{$\lambda=4$}
    \end{subfigure}

    \vspace{0.5cm} 

    \begin{subfigure}[b]{0.23\textwidth}
        \includegraphics[width=\textwidth]{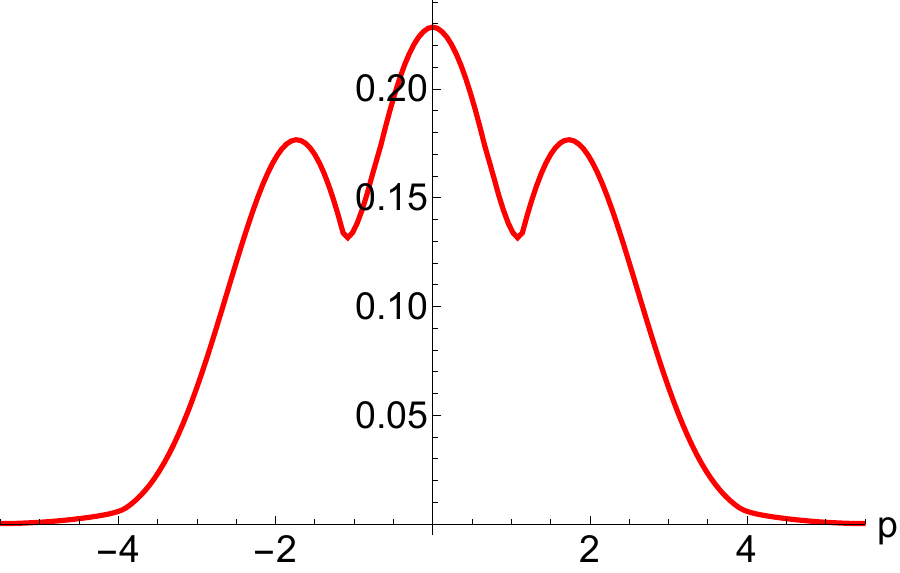}
        \caption{$\lambda=-\frac{3}{4}$}
    \end{subfigure}\hfill
    \begin{subfigure}[b]{0.23\textwidth}
        \includegraphics[width=\textwidth]{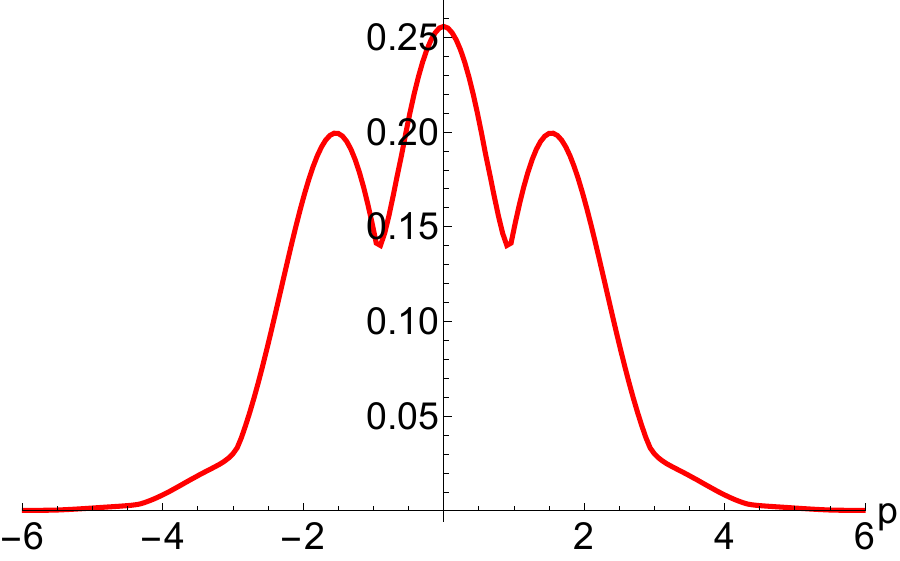}
        \caption{$\lambda=\frac{3}{4}$}
    \end{subfigure}\hfill
    \begin{subfigure}[b]{0.23\textwidth}
        \includegraphics[width=\textwidth]{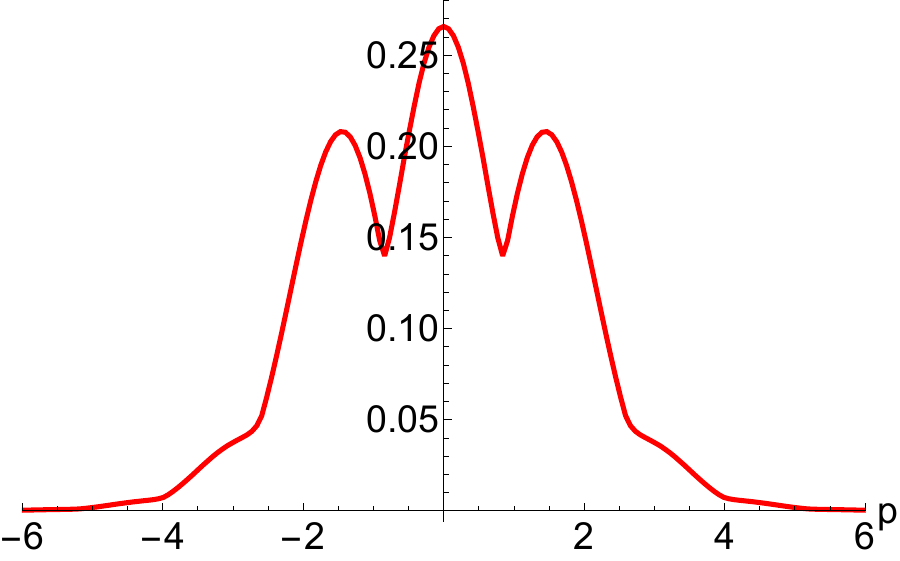}
        \caption{$\lambda_{c}^{n=1}=1.4209$}
    \end{subfigure}\hfill
    \begin{subfigure}[b]{0.23\textwidth}
        \includegraphics[width=\textwidth]{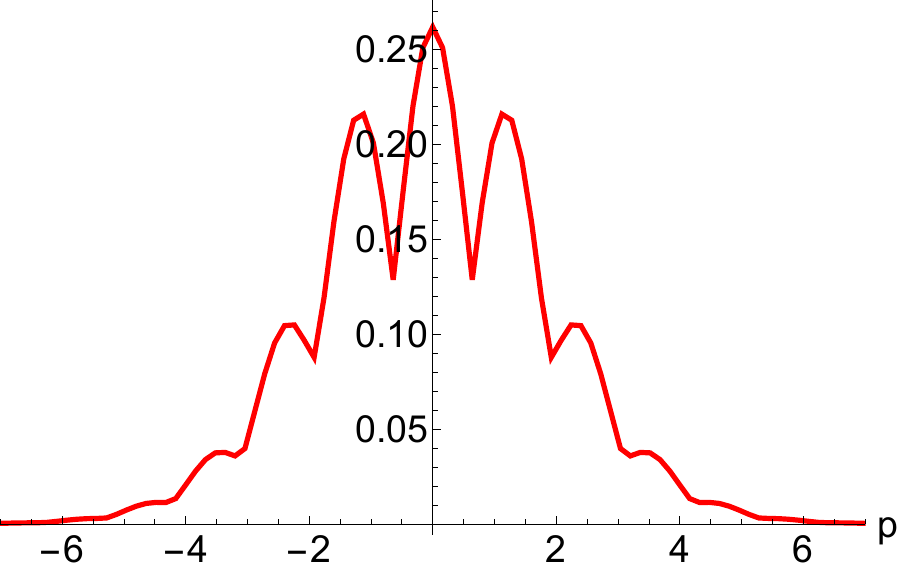}
        \caption{$\lambda=4$}
    \end{subfigure}

    \vspace{0.5cm} 

    \begin{subfigure}[b]{0.23\textwidth}
        \includegraphics[width=\textwidth]{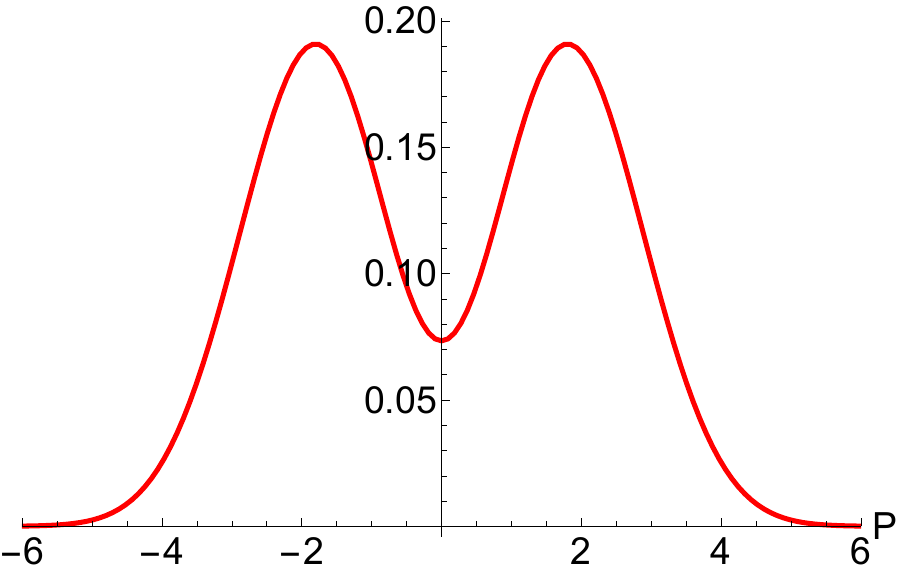}
        \caption{$\lambda=-\frac{3}{4}$}
    \end{subfigure}\hfill
    \begin{subfigure}[b]{0.23\textwidth}
        \includegraphics[width=\textwidth]{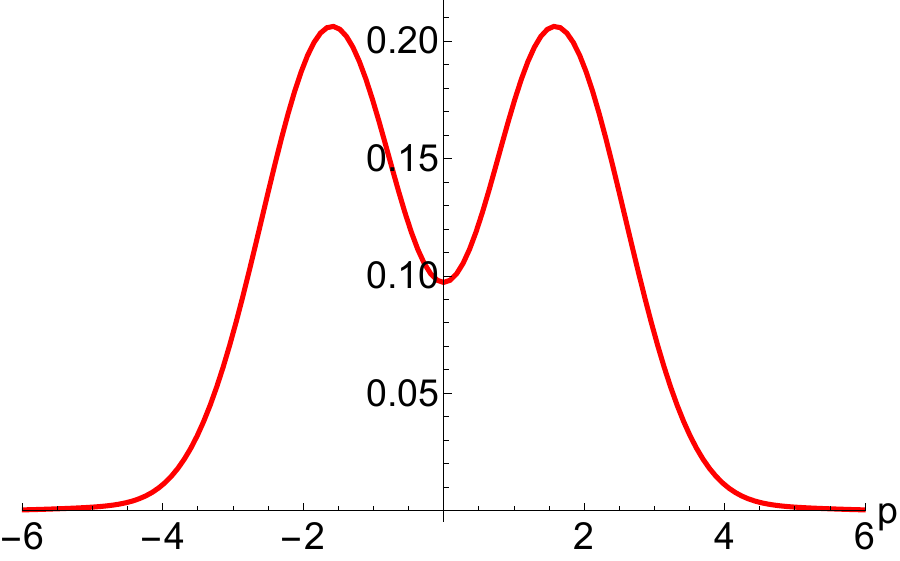}
        \caption{$\lambda=\frac{3}{4}$}
    \end{subfigure}\hfill
    \begin{subfigure}[b]{0.23\textwidth}
        \includegraphics[width=\textwidth]{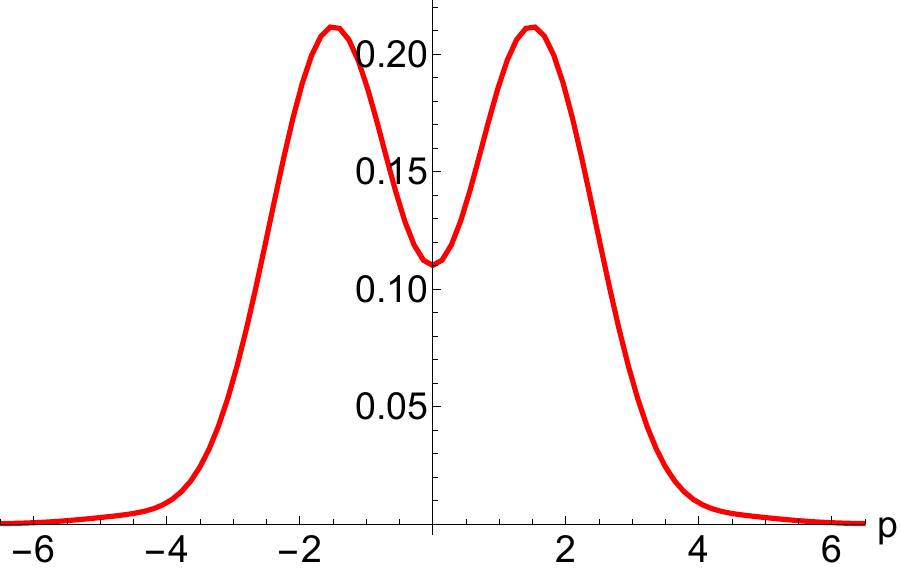}
        \caption{$\lambda_{c}^{n=1}=1.4209$}
    \end{subfigure}\hfill
    \begin{subfigure}[b]{0.23\textwidth}
        \includegraphics[width=\textwidth]{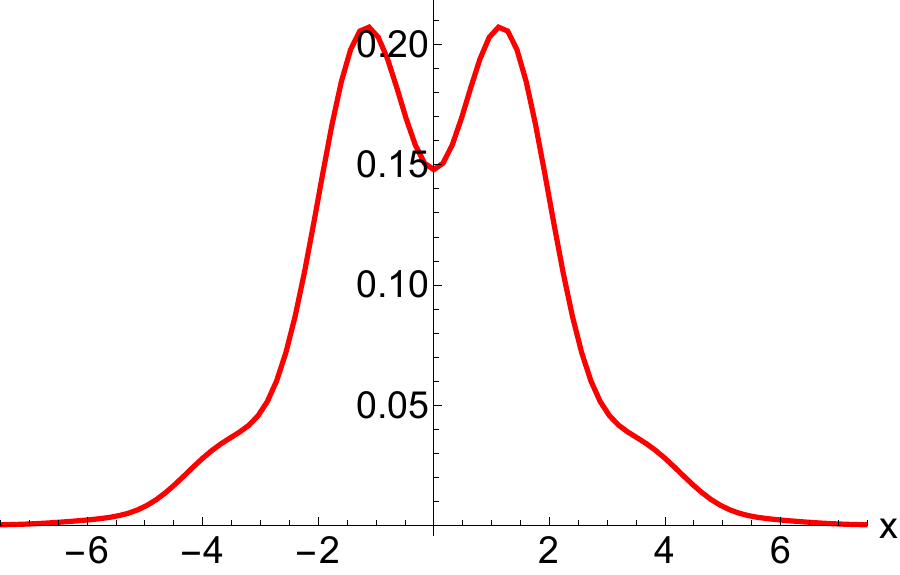}
        \caption{$\lambda=4$}
    \end{subfigure}

    \caption{First excited state $n=1$. Momentum-space marginals $Q_x$ of:  
(a)–(d) the Wigner function $W$,  
(e)–(h) its modulus $|W|$, and  
(i)–(l) the Husimi distribution $H$, shown for different values of $\lambda$.
}
    \label{marginalesP1}
\end{figure}

\begin{figure}[H]
    \centering
    \begin{subfigure}[b]{0.48\textwidth}
        \centering
        \includegraphics[width=\textwidth]{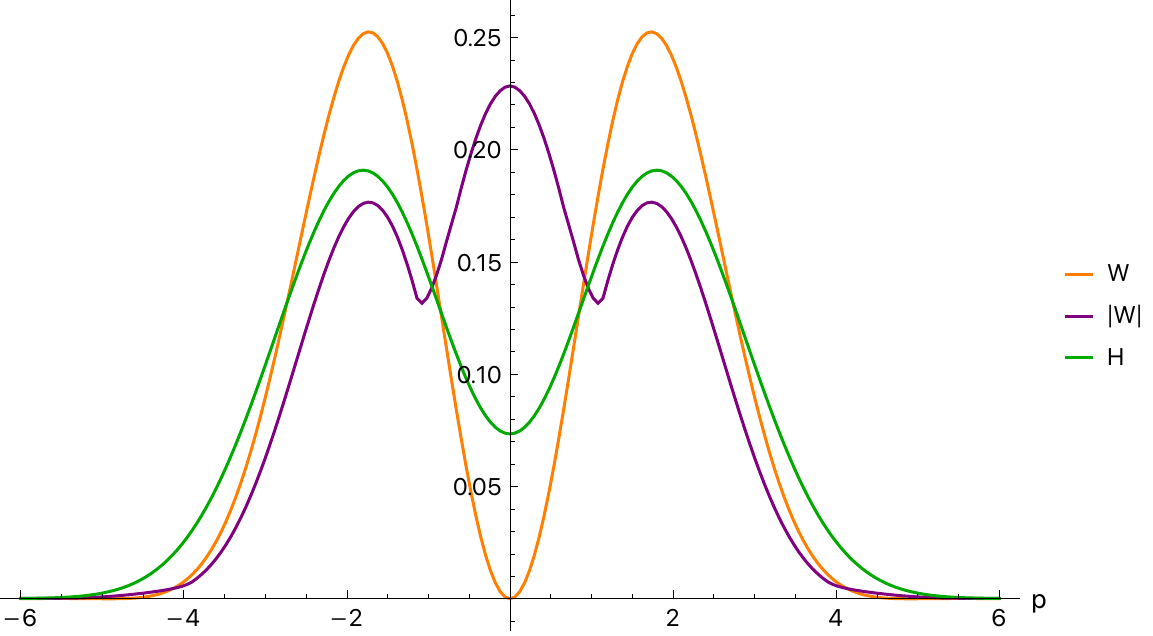}
        \caption{}
        \label{Den1m34p}
    \end{subfigure}
    \begin{subfigure}[b]{0.48\textwidth}
        \centering
        \includegraphics[width=\textwidth]{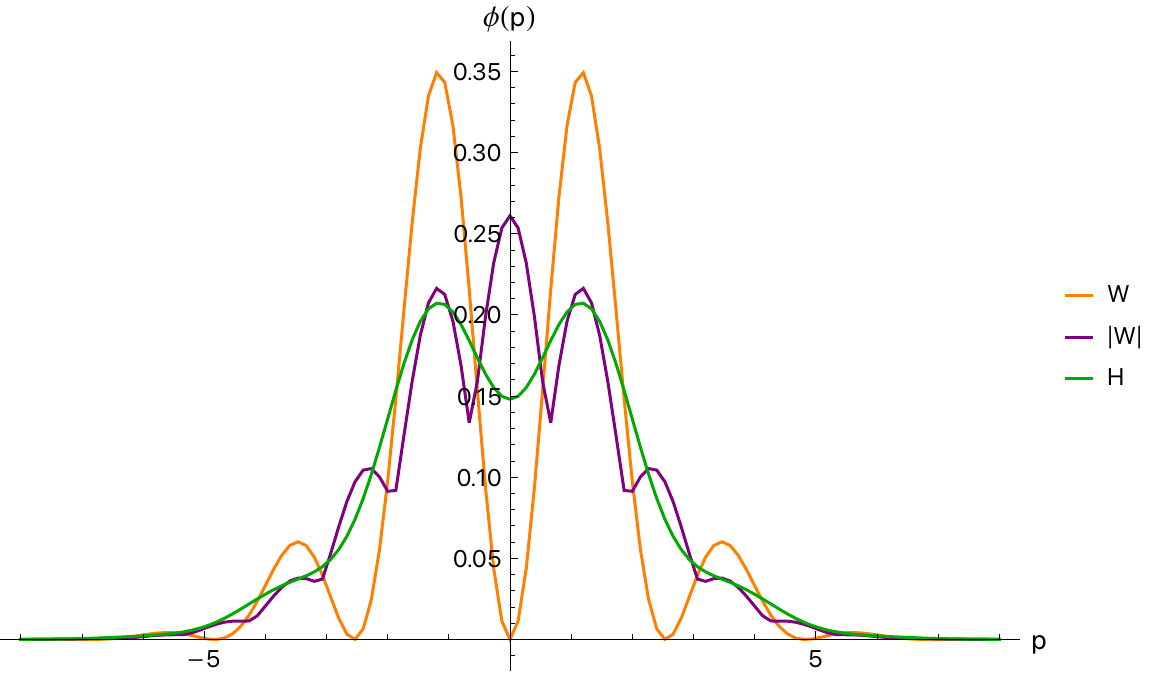}
        \caption{}
        \label{Den14p}
    \end{subfigure}
    \caption{Momentum-space marginals $Q_p$ of $W$, $|W|$, and $H$ for the first excited state ($n=1$), shown on a common scale for (a) $\lambda = -\tfrac{3}{4}$ and (b) $\lambda = 4$.
}
    \label{densidadesjuntasP1}
\end{figure}

The momentum marginals ${Q}_{p}(p)$, shown in Figures \ref{marginalesP0}-\ref{densidadesjuntasP1}, exhibit trends complementary to those observed in position space. In Figs. \ref{marginalesP0} and \ref{marginalesP1} we adopt different scales on the axes to better highlight the overall form of the marginals, while in Figs. \ref{densidadesjuntasP0} and \ref{densidadesjuntasP1} we use a common scale to bring out additional structural details. Some remarks are in order:

\begin{itemize}
    \item \textbf{Wigner marginal $W_{p}(p) = |\phi(p)|^{2}$.}  
    The momentum-space distributions reflect the Fourier-transform structure of
    $\psi(x)$, and therefore provide a sensitive probe of interference between separated
    spatial lobes. As $\lambda$ increases and the state becomes more localized in 
    position, $W_{p}(p)$ broadens and develops oscillations associated with coherent
    superpositions of left- and right-localized components.

    \item \textbf{Modulus marginal $|W|_{p}(p)$.}  
    The removal of oscillatory signs reduces the amplitude of interference fringes in 
    momentum space, but does not eliminate them entirely. For $n=1$, the central dip 
    at $p=0$ is partially filled, reflecting the fact that the antisymmetric character of 
    the state is less strongly encoded in $|W|$ than in $W$.

    \item \textbf{Husimi marginal $H_{p}(p)$.}  
    This marginal is substantially smoother than the other two and typically exhibits a 
    single broad peak for the ground state and a shallow central depression for the 
    first excited state. The fine oscillatory structure present in $W_{p}(p)$ is entirely 
    washed out, consistent with the coarse-grained nature of $H$.
\end{itemize}

As in position space, $H_{p}(p)$ gives the most classical-looking profile, while 
$W_{p}(p)$ retains the full oscillatory structure. The modulus distribution occupies an 
intermediate regime in which peak structure and overall width resemble those of $W$, 
but interference-induced oscillations are weakened.

\clearpage

\subsection{Comparative interpretation}

The systematic comparison of the marginals yields several general observations relevant 
to the later information-theoretic analysis:

\begin{itemize}

\item \textbf{Position marginals before tunneling (\(\lambda \ll \lambda_c\)):}  
For \(n=0\), all three marginals \(W_x\), \(|W|_x\), and \(H_x\) are nearly identical in structure (Fig.~\ref{marginalesX0}), each showing a single central peak.  
For \(n=1\), the three already differ (Fig.~\ref{marginalesX1}): \(W_x\) preserves the nodal structure at \(x=0\), \(|W|_x\) partially fills the central dip but keeps a two-peak shape, and \(H_x\) produces the broadest, most smoothed profile.  
\emph{For \(n=1\), the qualitative structure of \(W_x\) (one node, two lobes) remains unchanged with \(\lambda\); only the width, peak separation, and interference amplitude evolve, reflecting non-structural—rather than topological—modifications.}

\item \textbf{Position marginals after tunneling (\(\lambda > \lambda_c\)):}  
Both states develop clear bimodality, and the three representations diverge (Figs.~\ref{marginalesX0}-\ref{marginalesX1}): \(W_x\) shows the sharpest peaks and most pronounced splitting, \(|W|_x\) enhances the two-peak structure by removing destructive interference, and \(H_x\) delays and smooths the appearance of bimodality, giving the broadest double-humped distribution.

\item \textbf{Momentum marginals before tunneling:}  
For \(n=0\), \(W_p\), \(|W|_p\), and \(H_p\) differ only slightly (Fig.~\ref{marginalesP0}): \(W_p\) and \( |W|_p\) have similar widths, while \(H_p\) is slightly broader.  
For \(n=1\), \(W_p\) exhibits a central dip reflecting the spatial node (Fig.~\ref{marginalesP1}), \(|W|_p\) partially fills this dip, and \(H_p\) yields a single smooth peak with minimal structure.  
\emph{As in position space, the shape of \(W_p\) for \(n=1\) does not change qualitatively with \(\lambda\); the node in \(W_x\) simply translates into amplitude and width changes in \(W_p\), not a modification of its underlying structure.} 

\item \textbf{Momentum marginals after tunneling:}  
For both states, \(W_p\) shows the strongest oscillatory structure due to left–right interference, \(|W|_p\) retains a reduced remnant of these oscillations as strictly positive modulations, and \(H_p\) smooths them out almost completely (Figs.~\ref{marginalesP0}-\ref{marginalesP1}).  
The width of all marginals increases with \(\lambda\), but the smoothing hierarchy is preserved.

\item \textbf{Cross-state comparison:}  
After tunneling, \(|W|_x\) and \(|W|_p\) for \(n=0\) and \(n=1\) become more similar than the corresponding \(W\) or \(H\) marginals.  
Removing the sign structure erases the parity-induced contrast that remains visible in \(W_x\), \(W_p\), and, to a weaker extent, in the Husimi marginals.

\end{itemize}

These qualitative differences have direct consequences for the entropic and
mutual-information measures studied in the next sections, where the marginals act as the
primary inputs for quantifying the spreading, coherence, and structural complexity of the
quantum state.


\section{Entropic measures}
\label{SEC6}

Entropy-based quantities provide compact global indicators of the structural changes
that the quasiprobability distributions undergo as the coupling parameter $\lambda$
modifies the shape of the sextic potential. Since $W$, $|W|$, and $H$ encode different
degrees of smoothing and interference suppression, their entropies quantify, in a
comparative and systematic way, how each representation captures delocalization,
peak separation, and the emergence of double-well structure. In this section we examine
three types of entropy: the phase–space Shannon entropy computed directly from each
quasiprobability, the position- and momentum-space entropies extracted from their
marginals, and the combined total entropy $S_{t}=S_{x}+S_{p}$. In fact, $S_t$ may be interpreted as the entropy of a separable phase-space distribution $Q(x,p)=Q_x(x)Q_p(p)$. When $Q(x,p)=W(x,p)$, this quantity is known as the Leipnik entropy \cite{Leipnik1960}.

The phase-space entropies are defined as
\begin{equation}
    S[Q] = - \int_{-\infty}^{\infty} \int_{-\infty}^{\infty} Q(x,p)\,\ln Q(x,p)\, dx\, dp,
\end{equation}
while the entropies of their marginals in position and in momentum space are
\begin{equation}
S_x[Q] = - \int_{-\infty}^{\infty} Q_x(x)\,\ln Q_x(x)\, dx,
\qquad
S_p[Q] = - \int_{-\infty}^{\infty} Q_p(p)\,\ln Q_p(p)\, dp.
\end{equation}

\subsection{Shannon entropy}

We begin by recalling that the quantity $S[W]$ does not constitute a strict Shannon 
entropy, since the Wigner function is not a bona fide probability density and may assume 
negative values. Rather, following the information–theoretic viewpoint adopted in the earlier 
work \cite{Laguna2010}, $S[W]$ is best interpreted as a 
global delocalization functional defined directly on phase space. Within this framework, 
the real part $\mathrm{Re}\,S_W$ provides a meaningful measure of the spatial and momentum 
spreading encoded in $W(x,p)$, while the imaginary component $\mathrm{Im}\,S_W$—which 
remains small across all values of $\lambda$ considered here—acts as a sensitive indicator 
of sign–changing regions associated with interference and Wigner negativity. Moreover, the imaginary component is proportional to the volume of the negative regions in phase-space\cite{Salazar2023}. Note that this component increases with $\lambda$, illustrating that the volume of the negative regions also grows as $\lambda$ increases.

Thus, even 
though $S[W]$ lacks the formal probabilistic status of $S[H]$ or of the one-dimensional 
Shannon entropies, its decomposition into real and imaginary parts yields complementary 
insights into phase–space structure, combining delocalization information with a diagnostic 
of quantum interference.

\begin{figure}[H]
    \centering
    \begin{subfigure}[b]{0.48\textwidth}
        \centering
        \includegraphics[width=\textwidth]{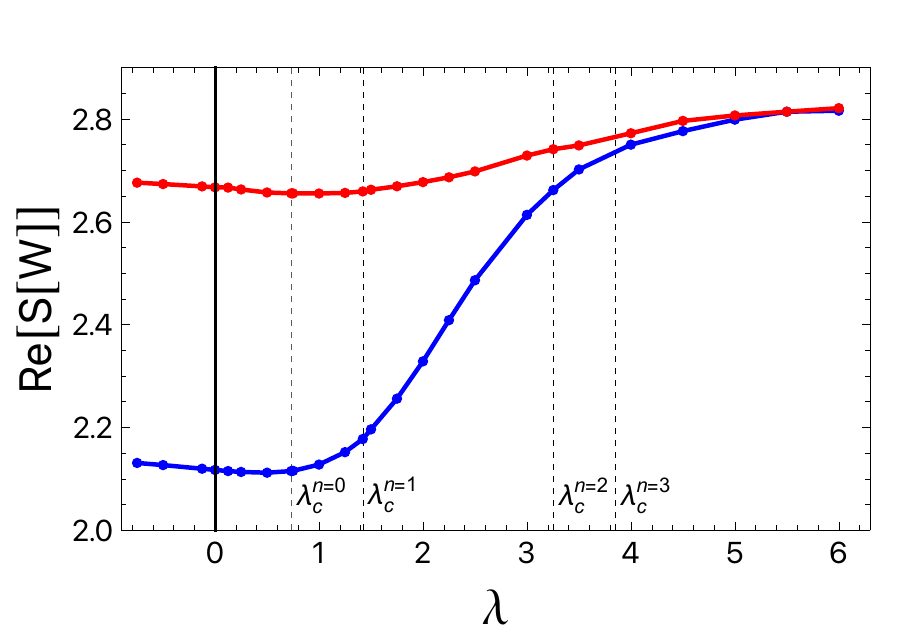}
        \caption{}
        \label{ReSW}
    \end{subfigure}
    \begin{subfigure}[b]{0.48\textwidth}
        \centering
        \includegraphics[width=\textwidth]{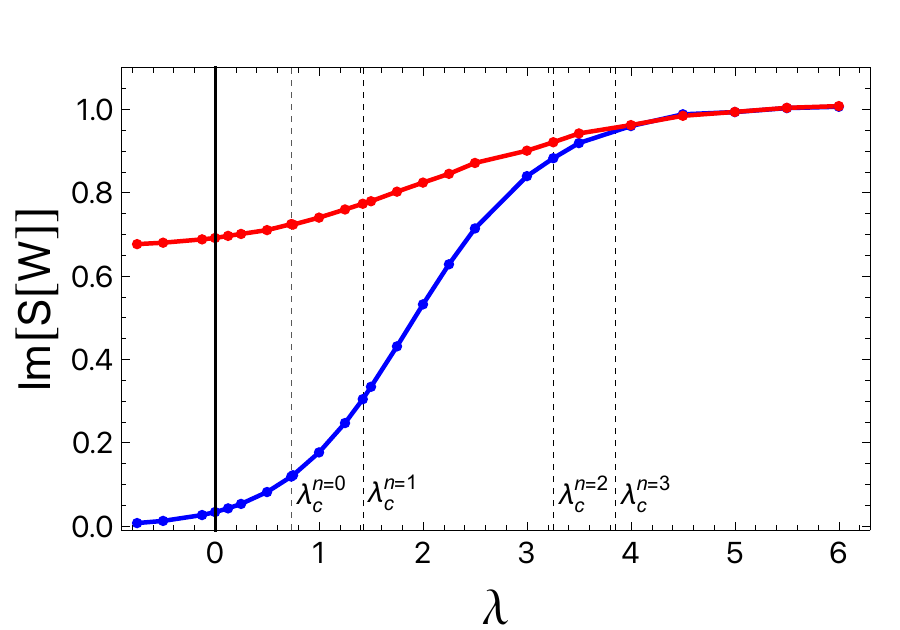}
        \caption{}
        \label{ImSW}
    \end{subfigure}
    \caption{Shannon entropy of the Wigner distribution for the ground $n=0$ (blue curve) and first excited $n=1$ (red curve) states as a function of $\lambda$.  
Panels (a)–(b) show $\mathrm{Re}\,S_W$ and $\mathrm{Im}\,S_W$, illustrating increased phase-space delocalization with $\lambda$.
}
    \label{shannonW}
\end{figure}

\begin{figure}[H]
    \centering
    \begin{subfigure}[b]{0.32\textwidth}
        \centering    \includegraphics[width=\textwidth]{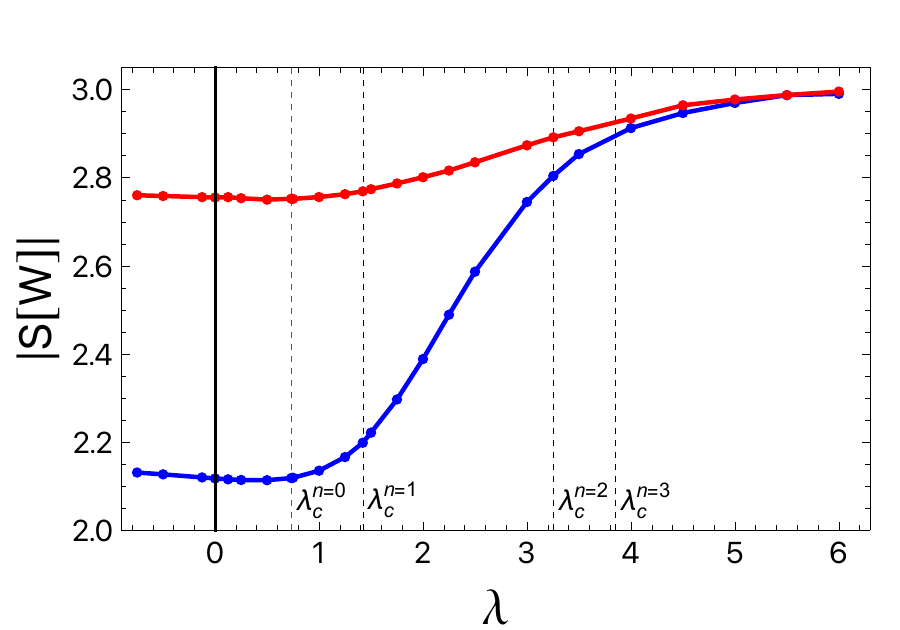}
        \caption{}
        \label{AbsSw}
    \end{subfigure}
    \begin{subfigure}[b]{0.32\textwidth}
        \centering       \includegraphics[width=\textwidth]{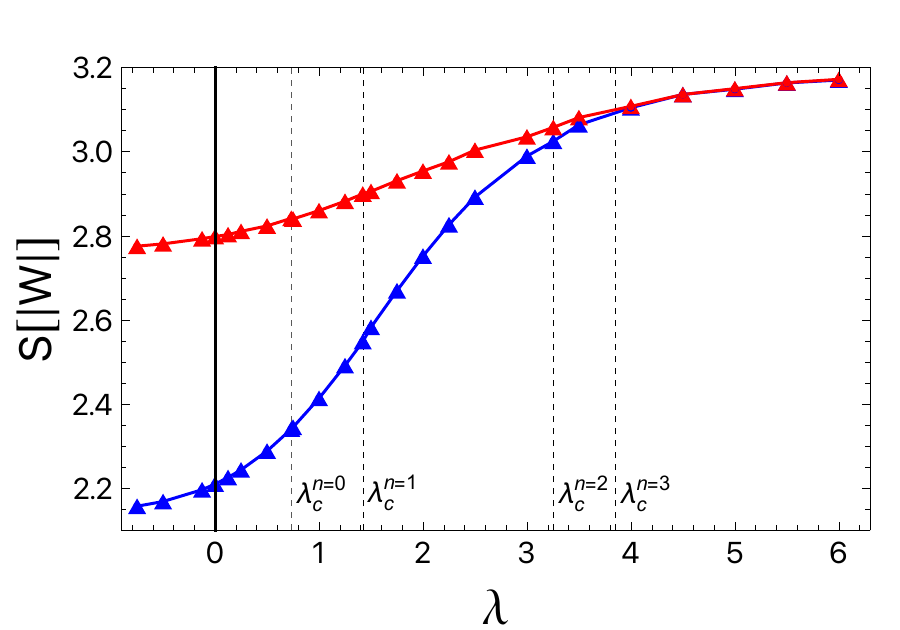}
        \caption{}
        \label{ReSW}
    \end{subfigure}
    \begin{subfigure}[b]{0.32\textwidth}
        \centering
        \includegraphics[width=\textwidth]{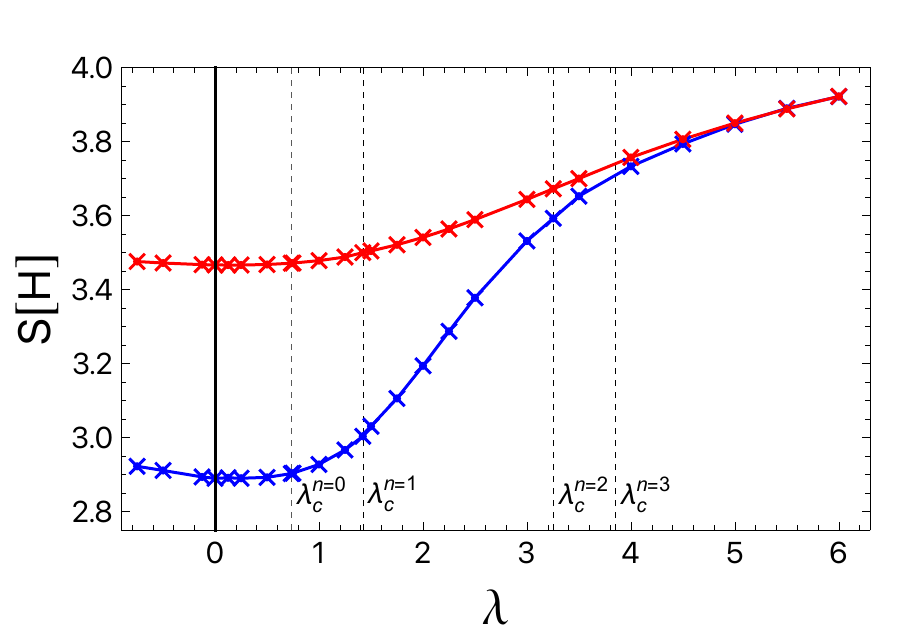}
        \caption{}
        \label{ImSW}
    \end{subfigure}
    \caption{Shannon entropy of the quasiprobability distributions vs.\ $\lambda$ for the ground $n=0$ (blue curve) and first excited $n=1$ (red curve) states. Panels (a)–(c) display $|S[W]|$, $S[|W|]$ and $S[H]$, respectively, all showing increasing phase-space delocalization with $\lambda$.}
    \label{shannonAbsWH}
\end{figure}

Figures~\ref{shannonW} and \ref{shannonAbsWH} summarize the dependence of the
phase–space Shannon entropy on $\lambda$ for the ground and first excited states.
Because the Wigner function can take negative values, its Shannon entropy may become
complex valued; accordingly, Fig.~\ref{shannonW} displays its real and imaginary parts.
The real part grows with $\lambda$, reflecting the increased phase–space complexity
associated with lobe separation and interference. The imaginary part remains small but
nonzero, indicating the persistence of oscillatory regions where $W(x,p)$ changes sign. It is striking that there are broad minima in ${\rm Re}\,(S[\,W\,])$, $\vert S[ W ] \vert$ and $S[H]$ in the ground state, but absent in  $ S[\vert W \vert ]$. These roughly coincide in the regions around the $\lambda$ critical point.

Figure~\ref{shannonAbsWH} compares the entropy of the modulus $|W|$ with that of
the Husimi distribution. Removing negativity consistently yields larger entropy in
$|W|$, reflecting the enhancement of geometric peak separation once destructive
interference has been eliminated. The Husimi entropy is the largest across all regimes:
Gaussian smearing broadens the distribution in both variables, maximizing uncertainty.
These trends support a clear hierarchy across all values of $\lambda$,
\[
{\rm Re}\,(S[\,W\,]) \ <\  \mid S[\,W\,]\mid \ < \  S[\,|W|\,] \ < \ S[\,H\,]\,,
\]
showing that each representation introduces a progressively stronger degree of smoothing. Also, there is an overall increasing tendency of the entropies with $\lambda$ which can be interpreted as phase-space delocalization.

\subsection{Shannon entropies of marginals distributions}

The position- and momentum-space entropies listed in
Tables~\ref{tab:SxSpn0} and \ref{tab:SxSpn1} provide a more detailed view of how
localization evolves in each representation. For the Wigner function, $S_{x}[W]$
initially decreases as the potential narrows but increases again as the state becomes
bimodal. The corresponding momentum entropy $S_{p}[W]$ follows the expected
Fourier-dual trend: it increases whenever spatial localization becomes stronger.

For $|W|$, both $S_{x}$ and $S_{p}$ tend to be larger than the Wigner values.
Eliminating interference fills in destructive dips and enhances the prominence of 
structural splitting, especially in the double-well regime.  
The Husimi entropies $S_{x}[H]$ and $S_{p}[H]$ are systematically the largest,
reflecting the broadest and most smoothed marginals.


\begin{table}[ht]
  \centering
  \renewcommand{\arraystretch}{0.3}
  \resizebox{\textwidth}{!}{
  \begin{tabular}{c|cccc}
    \hline
    \rowcolor{gray!10} \multicolumn{5}{c}{\textbf{$S_{x}$}} \\
    \hline
    \textbf{Distributions} & \textbf{$\lambda=-\frac{3}{4}$} & \textbf{$\lambda_{c}^{n=0}=0.7329$} & \textbf{$\lambda=1$} & \textbf{$\lambda=4$} \\
    \hline
    $W$  & 0.669446 & 0.938019 &0.993322  & 0.852065 \\
    $|W|$  & 0.668119 & 0.917103 & 0.966693 & 1.15412 \\
    $H$  & 1.25773 & 1.36759 & 1.4004 & 1.70738 \\
    \hline
    \rowcolor{gray!10} \multicolumn{5}{c}{\textbf{$S_{p}$}} \\
    \hline
    \textbf{Distributions} & \textbf{$\lambda=-\frac{3}{4}$} & \textbf{$\lambda_{c}^{n=0}=0.7329$} & \textbf{$\lambda=1$} & \textbf{$\lambda=4$} \\
    \hline
     $W$  & 1.48341 & 1.28365 & 1.2624 & 1.69748 \\
    $|W|$  & 1.50555 & 1.4778 & 1.50888 & 2.00614 \\
    $H$  & 1.66566 & 1.54541 & 1.54159 & 2.05315 \\
    \hline
  \end{tabular}
  }
  \caption{Shannon Entropy in position and momentum space for some $\lambda$, $n=0$}
  \label{tab:SxSpn0}
\end{table}

\begin{table}[ht]
  \centering
  \renewcommand{\arraystretch}{0.3}
  \resizebox{\textwidth}{!}{
  \begin{tabular}{c|cccc}
    \hline
    \rowcolor{gray!10} \multicolumn{5}{c}{\textbf{$S_{x}$}} \\
    \hline
    \textbf{Distributions} & \textbf{$\lambda=-\frac{3}{4}$} & \textbf{$\lambda=\frac{3}{4}$} & \textbf{$\lambda_{c}^{n=1}=1.4209$} & \textbf{$\lambda=4$} \\
    \hline
    $W$  & 0.799305 & 0.894196 & 0.919881 & 0.828591 \\
    $|W|$  & 0.907124 & 1.03248 & 1.08487 & 1.1505 \\
    $H$  & 1.44131 & 1.52491 & 1.56973 & 1.70813 \\
    \hline
    \rowcolor{gray!10} \multicolumn{5}{c}{\textbf{$S_{p}$}} \\
    \hline
    \textbf{Distributions} & \textbf{$\lambda=-\frac{3}{4}$} & \textbf{$\lambda=\frac{3}{4}$} & \textbf{$\lambda_{c}^{n=1}=1.4209$} & \textbf{$\lambda=4$} \\
    \hline
     $W$  & 1.81374 & 1.67882 & 1.63925 & 1.71753 \\
    $|W|$  &1.93029  & 1.87034 & 1.87503 & 2.01796 \\
    $H$  & 2.06706 & 1.98019 & 1.96378 & 2.07535 \\
    \hline
  \end{tabular}
  }
  \caption{Shannon Entropy in position and momentum space for some $\lambda$, $n=1$}
  \label{tab:SxSpn1}
\end{table}

The Shannon entropies in Tables~\ref{tab:SxSpn0} and \ref{tab:SxSpn1} quantify the smoothing hierarchy observed
across the three representations. For all $\lambda$ and for both $n=0$ and $n=1$,
the ordering $S_x[W] < S_x[|W|] < S_x[H]$ (and likewise for $S_p$) holds
consistently: for example, at $\lambda=4$ in the ground state we find
$S_x[W]=0.852$, $S_x[|W|]=1.154$, and $S_x[H]=1.707$, with momentum entropies
$S_p[W]=1.697$, $S_p[|W|]=2.006$, and $S_p[H]=2.053$. Near the critical couplings
$\lambda_c^{(0)}\approx0.73$ and $\lambda_c^{(1)}\approx1.42$, the position entropies
reach clear minima, reflecting maximal spatial localization before the emergence of
double-well splitting. The systematic increase of $S_p$ with $\lambda$ signals the
stronger momentum-space oscillations associated with tunneling, while the consistently
larger Husimi entropies highlight the extent to which Gaussian coarse--graining
broadens both position and momentum marginals. Collectively, these trends provide a
quantitative backbone for the structural differences observed among $W_n$, $|W_n|$,
and $H_n$.

Interestingly, the Fig. \ref{Shannon} shows that the maximum in $S_x[W]$ is also present in $S_x[|W|]$, but not in $S_x[H]$. This illustrates how the structure present in the Wigner-based distributions is largely washed out in the Husimi distribution. On the other hand, the minimum in $S_p[Q]$ appears in all three cases. 

\subsection{Total Shannon entropy}

\begin{figure}[H]
    \centering

    \begin{subfigure}[b]{0.32\textwidth}
        \includegraphics[width=\textwidth]{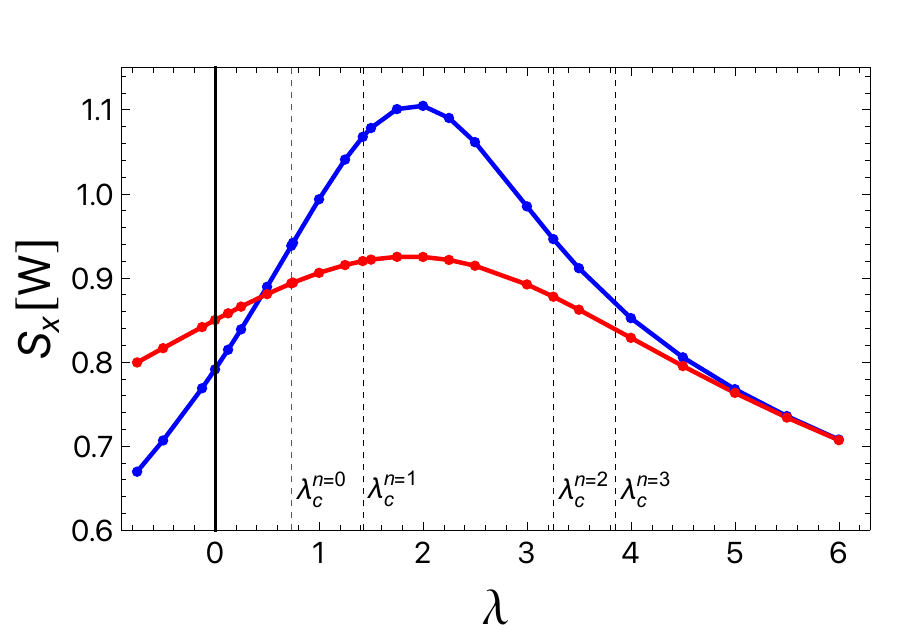}
        \caption{$S_{x}[W]$}
    \end{subfigure}\hfill
    \begin{subfigure}[b]{0.32\textwidth}
        \includegraphics[width=\textwidth]{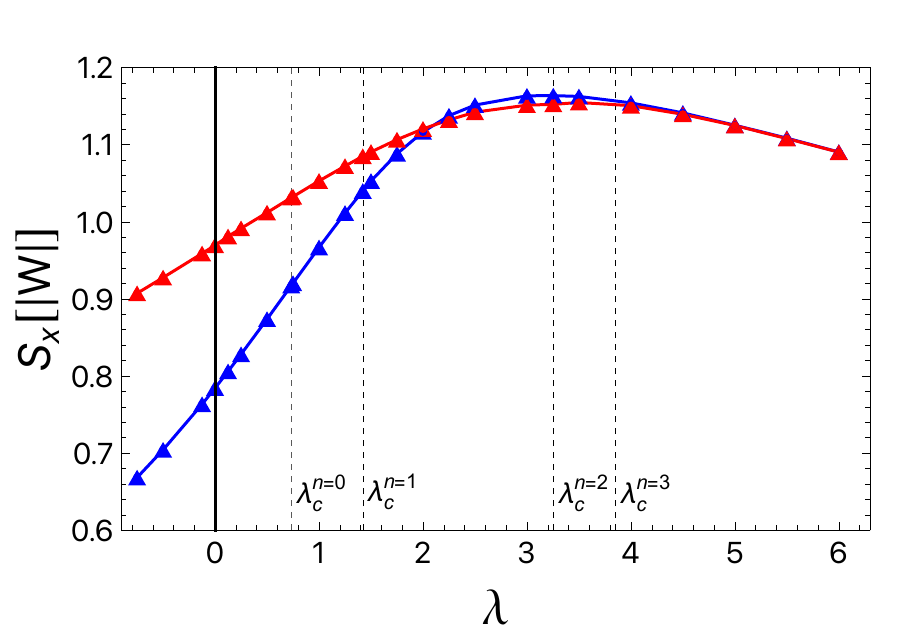}
        \caption{$S_{x}[|W|]$}
    \end{subfigure}\hfill
    \begin{subfigure}[b]{0.32\textwidth}
        \includegraphics[width=\textwidth]{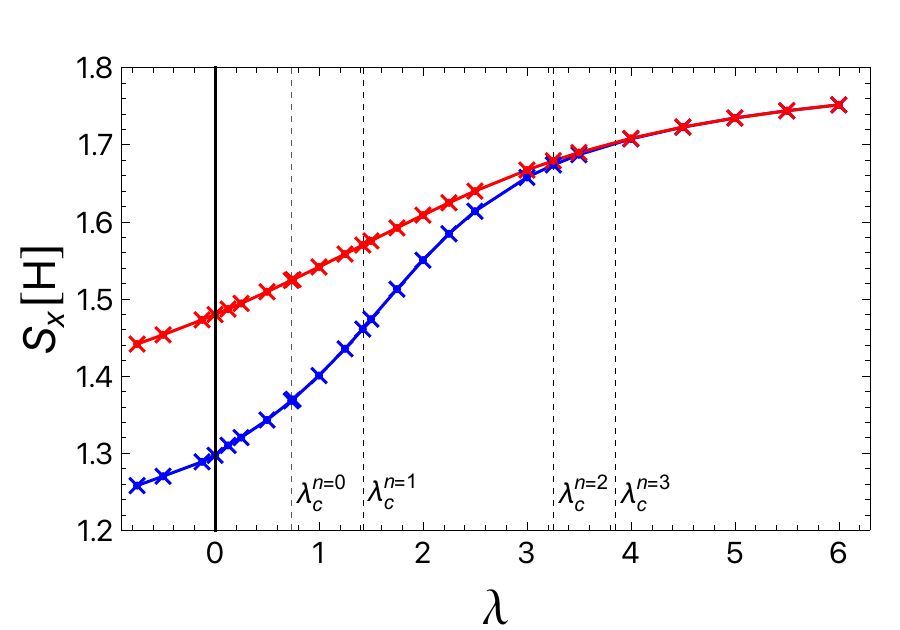}
        \caption{$S_{x}[H]$}
    \end{subfigure}

    \vspace{0.5cm}

    \begin{subfigure}[b]{0.32\textwidth}
        \includegraphics[width=\textwidth]{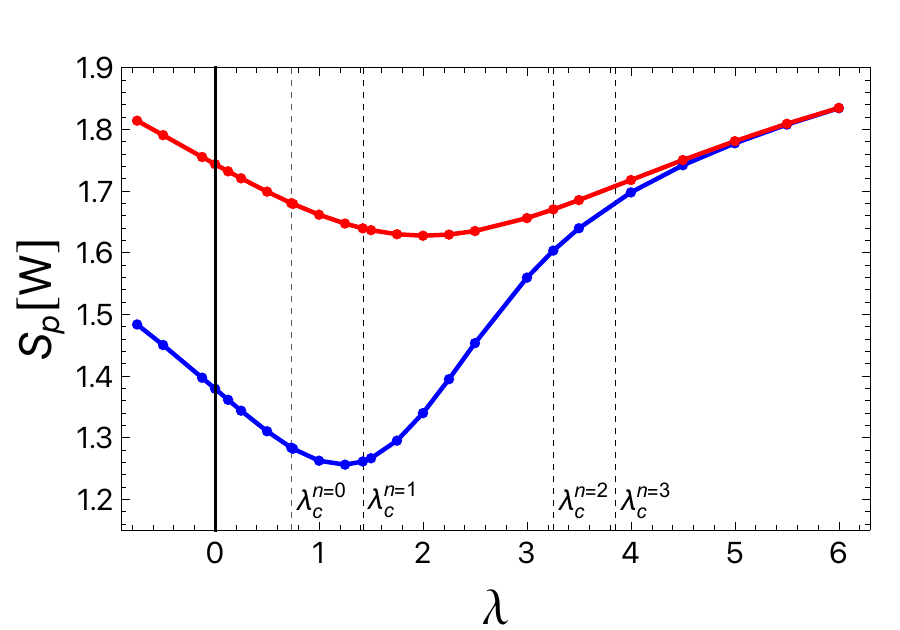}
        \caption{$S_{p}[W]$}
    \end{subfigure}\hfill
    \begin{subfigure}[b]{0.32\textwidth}
        \includegraphics[width=\textwidth]{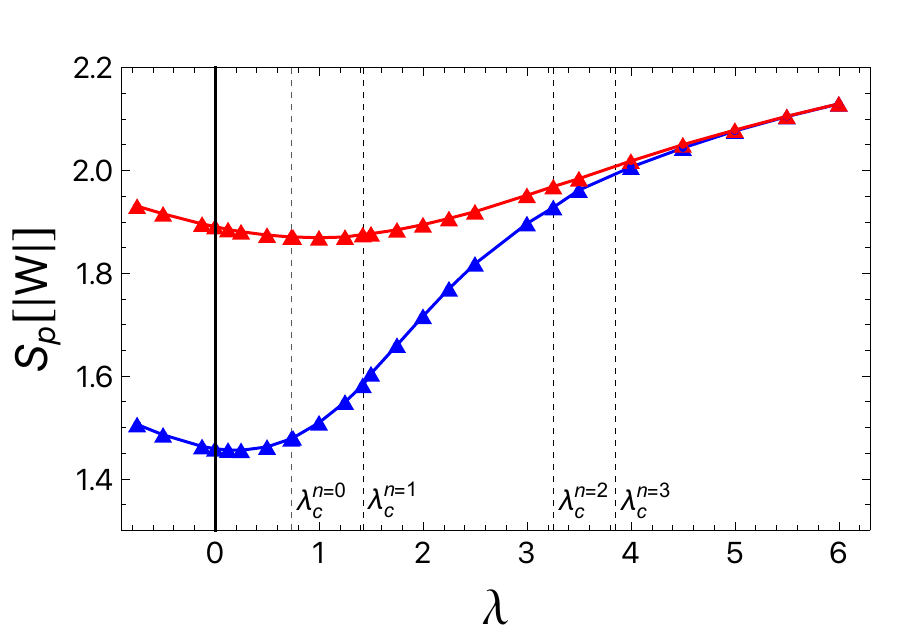}
        \caption{$S_{p}[|W|]$}
    \end{subfigure}\hfill
    \begin{subfigure}[b]{0.32\textwidth}
        \includegraphics[width=\textwidth]{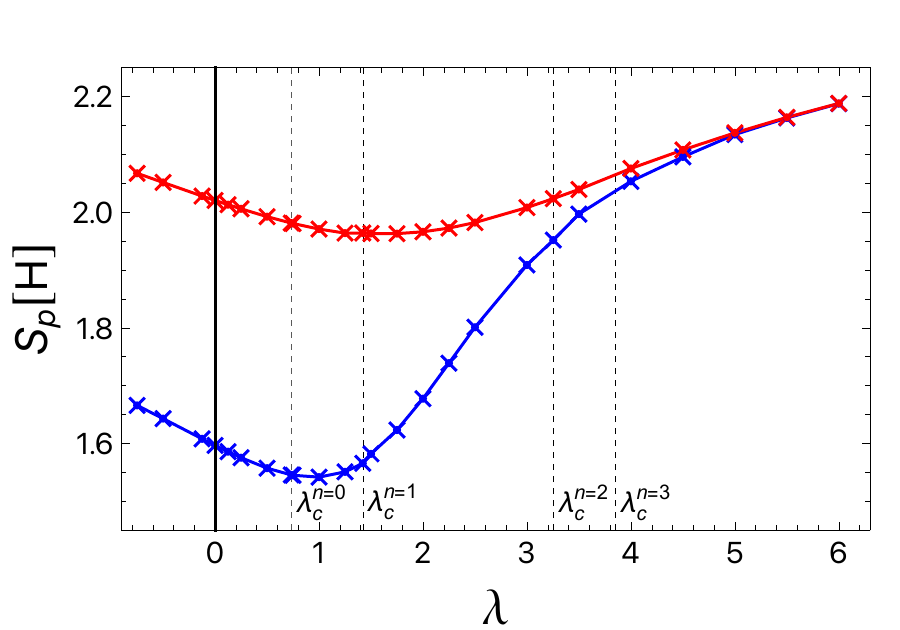}
        \caption{$S_{p}[H]$}
    \end{subfigure}

    \vspace{0.5cm}

    \begin{subfigure}[b]{0.32\textwidth}
        \includegraphics[width=\textwidth]{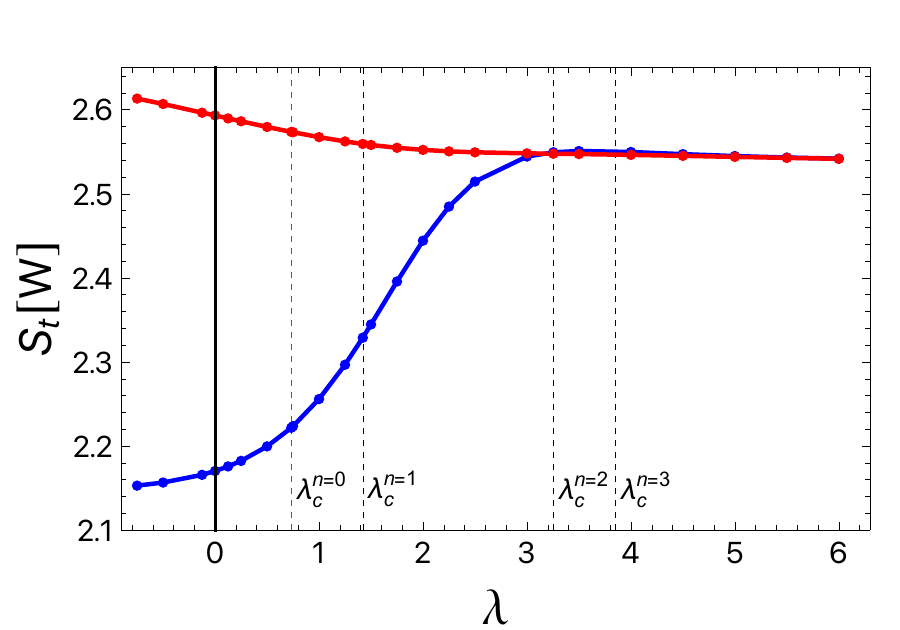}
        \caption{$S_{t}[W]$}
    \end{subfigure}\hfill
    \begin{subfigure}[b]{0.32\textwidth}
        \includegraphics[width=\textwidth]{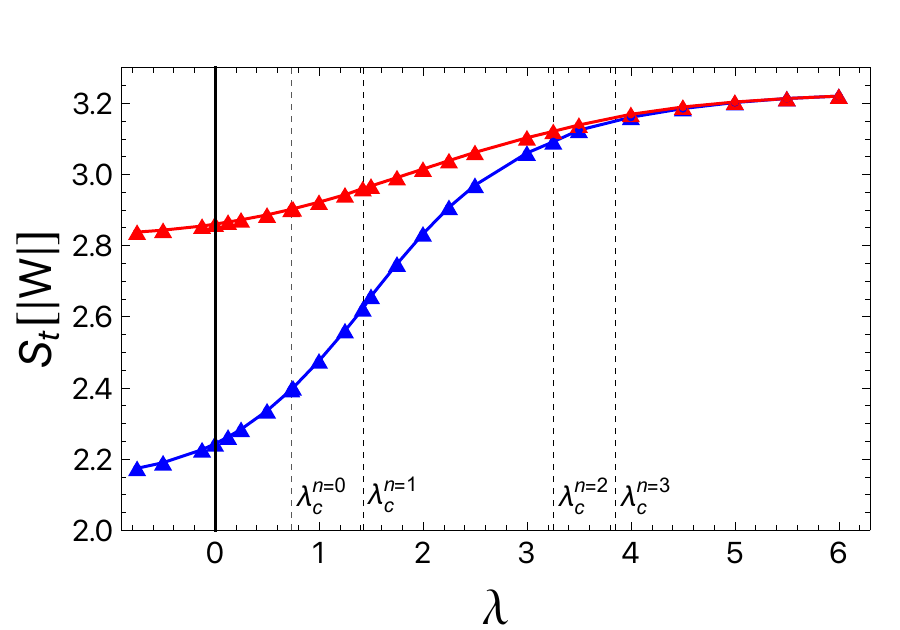}
        \caption{$S_{t}[|W|]$}
    \end{subfigure}\hfill
    \begin{subfigure}[b]{0.32\textwidth}
        \includegraphics[width=\textwidth]{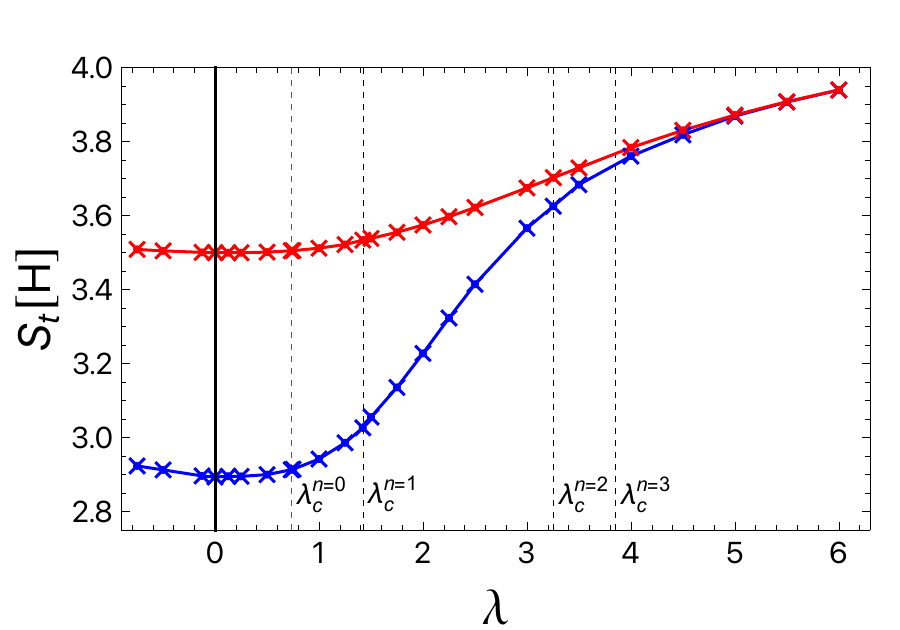}
        \caption{$S_{t}[H]$}
    \end{subfigure}

    \caption{\small Position, momentum, and total entropies $S_x$, $S_p$, and 
$S_t$ versus $\lambda$ for $n=0$ (blue curve) and $n=1$ (red curve), showing the consistent hierarchy 
$S[W_n] < S[|W_n|] < S[H_n]$ and the growth of global spreading as the 
potential evolves toward a double-well shape.
}
    \label{Shannon}
\end{figure}

The total entropy
\begin{equation}
S_{t} \ = \  S_{x} \  + \ S_{p} \ ,
\end{equation}
provides a single measure of global spreading in the $x$–$p$ plane. The behavior shown
in Fig.~\ref{Shannon} reflects the same hierarchical smoothing already observed in the
individual marginals. For both quantum states, $S_{t}$ increases with $\lambda$,
indicating that the transition to a double-well configuration introduces additional 
complexity in both conjugate variables. Once again, the Wigner function marginals yield the 
smallest entropy sum, $|W|$ gives intermediate values, and the Husimi function produces 
the broadest distributions and largest uncertainties. Finally, $S_t[H]$ exhibits a minimum that is not present in the other two (quasi)distributions.

\clearpage

Mutual information quantifies the statistical dependence between position and
momentum in a given phase–space representation. Since $W$, $|W|$, and $H$
encode different amounts of interference, smoothing, and geometric structure, the
corresponding values of $I_Q$ reflect how strongly each quasiprobability couples
the variables $x$ and $p$. It is defined as
\begin{equation}
I_Q = \int_{-\infty}^{\infty} \int_{-\infty}^{\infty} Q(x,p)\,\ln\!\left[\frac{Q(x,p)}{Q_x(x)\,Q_p(p)}\right] dx\,dp
    = S_x[Q] + S_p[Q] - S[Q],
\end{equation}
so that larger values indicate a greater departure from a factorized, product form of the phase-space distribution.

\begin{figure}[H]
    \centering
    \begin{subfigure}[b]{0.32\textwidth}
        \centering
        \includegraphics[width=\textwidth]{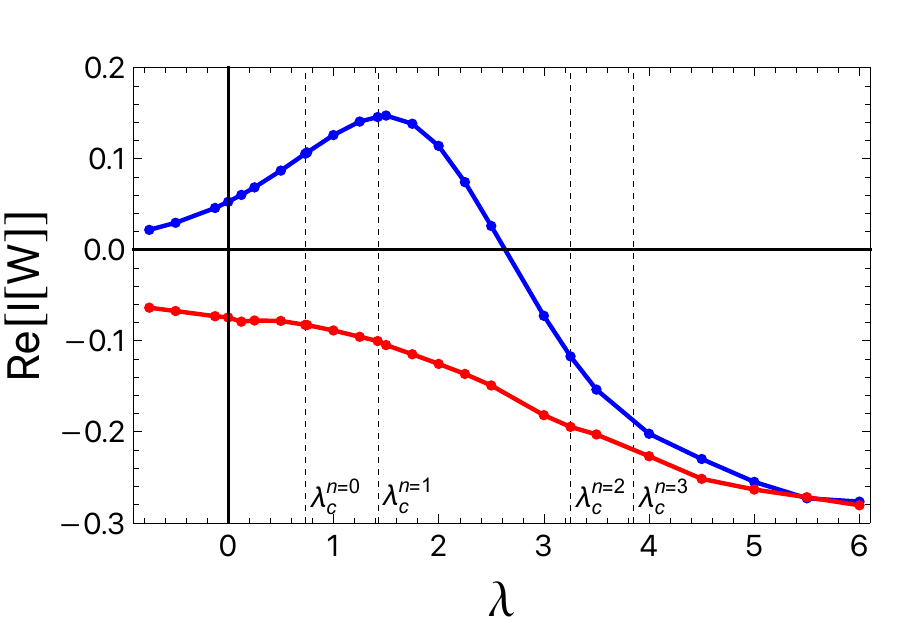}
        \caption{}
        \label{IMReW}
    \end{subfigure}
    \begin{subfigure}[b]{0.32\textwidth}
        \centering
        \includegraphics[width=\textwidth]{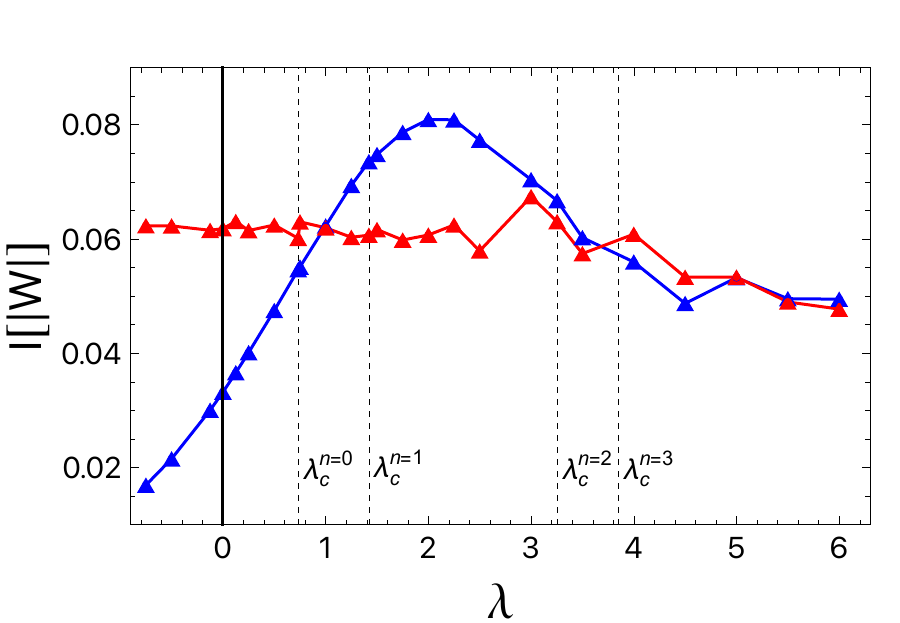}
        \caption{}
        \label{IMAbsW}
    \end{subfigure}
    \begin{subfigure}[b]{0.32\textwidth}
        \centering
        \includegraphics[width=\textwidth]{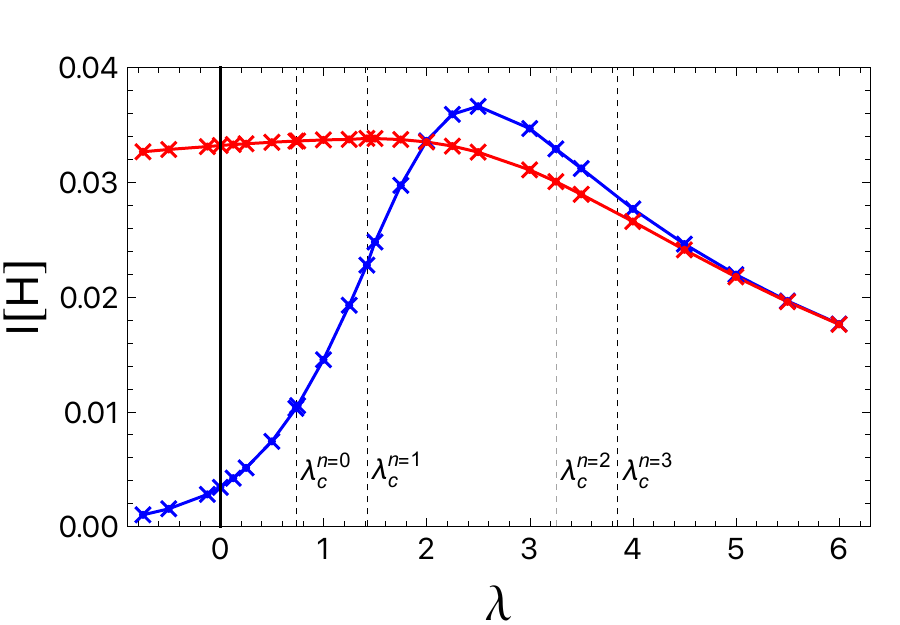}
        \caption{}
        \label{IMH}
    \end{subfigure}
    \caption{Mutual information $I$ of the (quasi)probability distributions:  (a) $W$, (b) $|W|$ and (c) $H$ as a function of
    the coupling parameter $\lambda$ for the ground $n = 0$ (blue curve) and first excited $n = 1$ (red curve) states.}
    \label{MutualInformation}
\end{figure}

Figure~\ref{MutualInformation} shows $I_Q(\lambda)$ for the three
representations and the two lowest states. Only the real part of $I_W$ is
displayed, as its imaginary component is the negative of ${\rm Im}[S[W]]$
presented in Fig.~\ref{shannonW}. For the Wigner function, $I_W$ increases
monotonically as the system evolves from a single well to a double well,
reflecting the strengthening of $x$--$p$ correlations induced by spatial
separation and the associated interference fringes, which sharpen near and
beyond the critical couplings $\lambda_c^{(n)}$.

For $|W|$, the removal of negativity suppresses the finest oscillations while
preserving the underlying bimodal geometry. The resulting $I_{|W|}$ therefore
lies between the Wigner and Husimi values: reduced relative to $W$ yet still
substantially larger than $I_H$. This indicates that geometric localization
alone produces appreciable $x$--$p$ dependence even in the absence of
interference.

The Husimi function, shaped by Gaussian coarse graining, yields the smallest
mutual information. Here the joint entropy $S$ grows faster than the marginals
$S_x$ and $S_p$, leaving only broad semiclassical correlations. The hierarchy
\[
{\rm Re}[I_W] \;>\; I_{|W|} \;>\; I_H
\]
thus captures how interference, geometric splitting, and smoothing generate
strong, intermediate, and weak phase–space correlations. All three
$I_Q(\lambda)$ curves exhibit maxima in the ground state, though not at the
same coupling values.

This ordering also has a structural interpretation. Mutual information measures
the departure from a separable portrait $Q(x,p)=Q_x(x)Q_p(p)$ and therefore
characterizes an entanglement–like nonseparability between $x$ and $p$.
Interference fringes and multimodal structures in $W(x,p)$ introduce multiple
relevant spatial scales, producing the largest effective correlation dimension.
The modulus $|W|$ retains only the geometric splitting and therefore yields an
intermediate dimension, while the Husimi function is dominated by a single
smooth, semiclassical scale. This behavior parallels the entropy hierarchy
\[
S[W] < S[|W|] < S[H],
\]
and confirms that the transition to the double–well regime enhances both the
nonseparability of the phase–space variables and the structural complexity of
the quasiprobability representations.

\subsection{Cumulative Residual Jeffreys (CRJ) Divergence}

The Cumulative Residual Jeffreys (CRJ) divergence provides a symmetric and 
numerically stable measure of dissimilarity between two probability densities 
$\rho_m$ and $\rho_n$~\cite{HSR2019}.  
For any density $\rho_n$, we define the associated survival (cumulative–residual) 
function
\(
S_n(x)=\int_x^\infty \rho_n(t)\, dt ,
\)
and analogously for $S_m(x)$.  
The CRJ divergence is then given by
\begin{equation}
    D_{\mathrm{CRJ}}(\rho_m \,\|\, \rho_n)
    =
    \int_{-\infty}^{\infty}
        S_n(x)\,\ln\!\left(\frac{S_n(x)}{S_m(x)}\right)\, dx
    +
    \int_{-\infty}^{\infty}
        S_m(x)\,\ln\!\left(\frac{S_m(x)}{S_n(x)}\right)\, dx ,
\end{equation}
which is manifestly symmetric under 
$\rho_m \leftrightarrow \rho_n$.  
Unlike the KL divergence, $D_{\mathrm{CRJ}}$ remains finite even when the densities 
possess different supports or nodal structures, since it compares their 
\emph{tails} rather than their pointwise values.

In the present context, we use $D_{\mathrm{CRJ}}$ to quantify how strongly the 
marginals of the modulus distribution $|W|$ and the Husimi distribution $H$ deviate 
from those of the Wigner function $W$, in both position and momentum space.  
Because these three quasiprobabilities differ in their treatment of negativity, 
interference, and Gaussian smoothing, the CRJ divergence provides a precise and 
robust diagnostic of how these structural differences manifest in their one-dimensional 
marginals.

Tables~\ref{tab:CRJn0} and~\ref{tab:CRJn1} summarize the divergences for the ground 
and first excited states across representative values of~$\lambda$.  
Several clear trends emerge:

\begin{itemize}
    \item \textbf{Near-coincidence of $|W_n|$ and $W_n$ at low coupling.}  
    For small~$\lambda$, $D_{\mathrm{CRJ}}(|W_n|,W_n)$ is extremely small 
    (e.g., $9.7\times10^{-7}$ in $x$ for $n=0$ at $\lambda=-\tfrac{3}{4}$), showing 
    that sign removal barely affects the Wigner marginals.

    \item \textbf{Steady growth of $D_{\mathrm{CRJ}}(|W_n|,W_n)$ with well formation.}  
    As $\lambda$ increases and the double well develops, the divergences rise 
    (e.g., $\approx1.3\times10^{-1}$ in $x$ at $\lambda=4$), reflecting enhanced 
    interference and lobe separation.

    \item \textbf{CRJ as a tunneling indicator.}  
    The monotonic increase of $D_{\mathrm{CRJ}}$ tracks the emergence of left--right 
    coherence and the suppression of interference under smoothing.

    \item \textbf{Largest deviations for Husimi marginals.}  
    Husimi smoothing produces the strongest departures from $W$.  
    For $n=0$, $D_{\mathrm{CRJ}}(H_n,W_n)$ grows from $0.351$ to $0.595$ as 
    $\lambda$ increases from $-\tfrac{3}{4}$ to $4$, roughly an order of magnitude 
    larger than the corresponding $|W_n|$–$W_n$ values.

    \item \textbf{Consistently smaller divergences in momentum space.}  
    Momentum-space divergences remain low (typically $\lesssim0.07$), reflecting the 
    weaker oscillatory structure of the momentum marginals and the reduced impact of 
    Husimi smoothing in~$p$.

    \item \textbf{Greater sensitivity of the first excited state.}  
    The nodal structure of $n=1$ yields systematically larger divergences, e.g., 
    $D_{\mathrm{CRJ}}(|W_1|,W_1)=0.133$ and $D_{\mathrm{CRJ}}(H_1,W_1)=0.580$ at 
    $\lambda=4$, exceeding the corresponding ground-state values.
\end{itemize}

\begin{table}[ht]
  \centering
  \renewcommand{\arraystretch}{0.3}
  \resizebox{\textwidth}{!}{
  \begin{tabular}{c|cccccc}
    \hline
    \rowcolor{gray!10} \multicolumn{7}{c}{\textbf{Position space}} \\
    \hline
    \textbf{$D_{CRJ}(\rho_{m}||\rho_{n})$} & \textbf{$\lambda=-\frac{3}{4}$} & \textbf{$\lambda_{c}^{n=0}=0.7329$} & \textbf{$\lambda=1$} & \textbf{$\lambda=2.5$} & \textbf{$\lambda=4$} & \textbf{$\lambda=5$} \\
    \hline
    $(|W|,W)$  & $9.72868 \times 10^{-7}$ & 0.000839818 & 0.00214324 & 0.0579798 & 0.131541 & 0.157213 \\
    $(H,W)$  & 0.350672 & 0.287128 & 0.28866 & 0.431972 & 0.595489 & 0.638869 \\
    $(H,|W|)$  & 0.346475 & 0.301167 & 0.308648 & 0.488934 & 0.672753 & 0.726098 \\
    \hline
    \rowcolor{gray!10} \multicolumn{7}{c}{\textbf{Momentum space}} \\
    \hline
    \textbf{$D_{CRJ}(\rho_{m}||\rho_{n})$} & \textbf{$\lambda=-\frac{3}{4}$} & \textbf{$\lambda_{c}^{n=0}=0.7329$} & \textbf{$\lambda=1$} & \textbf{$\lambda=2.5$} & \textbf{$\lambda=4$} & \textbf{$\lambda=5$} \\
    \hline
    $(|W|,W)$  & 0.00289026 & 0.0434621 & 0.0589254 & 0.0757902 & 0.0431918 & 0.0337971 \\
    $(H,W)$  & 0.0476579 & 0.0658988 & 0.0667597 & 0.0491116 & 0.0369542 & 0.0325187 \\
    $(H,|W|)$  & 0.0333811 & 0.0880575 & 0.00589606 & 0.00879494 & 0.00853151 & 0.00753547 \\
    \hline
  \end{tabular}
  }
  \caption{CRJ divergences for the ground state ($n=0$) in position and momentum space for selected values of $\lambda$.}
  \label{tab:CRJn0}
\end{table}

\begin{table}[ht]
  \centering
  \renewcommand{\arraystretch}{0.3}
  \resizebox{\textwidth}{!}{
  \begin{tabular}{c|cccccc}
    \hline
    \rowcolor{gray!10} \multicolumn{7}{c}{\textbf{Position space}} \\
    \hline
    \textbf{$D_{CRJ}(\rho_{m}||\rho_{n})$} & \textbf{$\lambda=-\frac{3}{4}$} & \textbf{$\lambda=\frac{3}{4}$} & \textbf{$\lambda_{c}^{n=1}=1.4209$}  & \textbf{$\lambda=2.5$} & \textbf{$\lambda=4$} & \textbf{$\lambda=5$} \\
    \hline
    $(|W|,W)$  & 0.0390777 & 0.0518047 & 0.063286 & 0.0910459 & 0.13334 & 0.157484 \\
    $(H,W)$  & 0.394839 & 0.40362 & 0.425713 & 0.492836 & 0.579705 & 0.637086 \\
    $(H,|W|)$  & 0.432782 & 0.462339 & 0.485834 & 0.558428 & 0.657533 & 0.724503 \\
    \hline
    \rowcolor{gray!10} \multicolumn{7}{c}{\textbf{Momentum space}} \\
    \hline
    \textbf{$D_{CRJ}(\rho_{m}||\rho_{n})$} & \textbf{$\lambda=-\frac{3}{4}$} & \textbf{$\lambda=\frac{3}{4}$} & \textbf{$\lambda_{c}^{n=1}=1.4209$} &\textbf{$\lambda=2.5$} & \textbf{$\lambda=4$} & \textbf{$\lambda=5$} \\
    \hline
    $(|W|,W)$  & 0.101457 & 0.0816687 & 0.0735118 & 0.059317 & 0.0417442 & 0.0333961 \\
    $(H,W)$  & 0.0459038 & 0.0542097 & 0.0541598 & 0.0471514 & 0.0372011 & 0.0319676 \\
    $(H,|W|)$  & 0.122977 & 0.078894 & 0.0488316 & 0.0225922 & 0.00993492 & 0.00765889 \\
    \hline
  \end{tabular}
  }
  \caption{CRJ divergences for the first excited state ($n=1$) in position and momentum space for selected values of $\lambda$.}
  \label{tab:CRJn1}
\end{table}

\section{Key Findings}
\label{SEC7}

This study presented a systematic comparison of three phase-space representations of a quantum state—the Wigner function $W(x,p)$, its modulus $|W(x,p)|$, and the Husimi distribution $H(x,p)$—using the quasi-exactly solvable (QES) sextic potential as a model system. Our analysis, based on variational wavefunctions for the ground and first excited states, led to the following three main discoveries:

\begin{enumerate}
    \item \textbf{Intermediate Role of the Modulus-Wigner Function:} The modulus of the Wigner function, $|W(x,p)|$, provides a meaningful intermediate phase-space representation. It preserves the geometric structure and localization features of the quantum state while eliminating the sign-changing interference fringes present in the original Wigner function. This representation bridges the detailed, oscillatory Wigner function and the smoothed, positive-definite Husimi distribution.

    \item \textbf{Hierarchical Ordering of Coherence and Delocalization:} Shannon entropy and mutual information analyses reveal a consistent hierarchy across the three phase-space distributions. The Wigner function shows the strongest coherence and sharpest localization (lowest entropy, highest mutual information), followed by $|W(x,p)|$, with the Husimi function displaying the most delocalization and weakest correlations. This ordering—$W > |W| > H$—holds across the transition from single- to double-well regimes.

    \item \textbf{CRJ Divergence as a Diagnostic Tool:} The Cumulative Residual Jeffreys (CRJ) divergence was introduced as a robust and symmetric metric to quantify differences between phase-space representations. It effectively tracks tunneling-induced changes, with $D_{\text{CRJ}}(H, W)$ increasing significantly as the double-well structure emerges. In contrast, $|W|$ remains closer to $W$, confirming its utility as a structure-preserving yet interference-filtering alternative.
\end{enumerate}

The distinct ability of the three Shannon entropies to capture the structural transition is summarized in Table~\ref{tab:entropy_comparison}.

\begin{table}[h]
\caption{Sensitivity of Shannon entropies to the single-- to double--well transition and the onset of tunneling.}
\begin{tabular}{lccc}
Entropy & Well splitting & Tunneling / interference & Transition signature \\
\hline
$S[W]$     & Strong    & Strong    & Nonmonotonic, sharp \\
$S[|W|]$   & Moderate  & None      & Smooth, monotonic   \\
$S[H]$     & Weak      & None      & Broad, featureless  \\
\end{tabular}
\label{tab:entropy_comparison}
\end{table}

\clearpage

\section{Conclusions}
\label{sec:conclusions}

The comparative framework developed in this work sheds light on the structure, coherence, and delocalization properties encoded in different phase-space representations of quantum states. We showed that while the Wigner function offers a complete picture including quantum interference, the modulus-Wigner and Husimi distributions provide progressively smoothed alternatives, useful for emphasizing geometric structure or classical correspondence.

The hierarchy $W > |W| > H$ appears to be a general feature in systems where coherence and classicality compete, such as double-well and tunneling configurations. In particular, the $|W|$ function emerges as a valuable compromise, retaining fine structure without interference, and the CRJ divergence serves as a sensitive indicator of structural changes and information loss.

These insights have broader implications for phase-space methods in atomic, molecular, optical, and condensed-matter systems. Potential applications include quantum control, decoherence monitoring, and nonclassical state characterization. The tools demonstrated here—especially the use of entropy, mutual information, and CRJ divergence—are broadly applicable to analyzing localization, spreading, and the emergence of classicality in quantum systems.

Moreover, the present results provide a general framework for assessing the information-theoretic sensitivity of phase-space representations in systems with emerging tunneling and bimodality.

This framework can be extended to time-dependent dynamics to study how phase-space coherence and structure evolve under tunneling, interference, and decoherence. Applying the same tools to higher excited states or to other quasi-exactly solvable and anharmonic potentials could test the universality of the observed trends. Further comparisons with alternative quasiprobability distributions (e.g., Glauber–Sudarshan $P$ or Kirkwood–Rihaczek) may also deepen our understanding of quantum–classical transitions in diverse physical settings.

\section*{Funding declaration}

A. M. Escobar Ruiz would like to thank the support from Consejo Nacional de Humanidades, Ciencias y Tecnologías (CONAHCyT, now Secihti) of Mexico under Grant CF-2023-I-1496 and from UAM research grant DAI 2024-CPIR-0. A. N. Mendoza Tavera wishes to thank Secihti (México) for the financial support provided through a doctoral scholarship.

\section*{Data availability}
Data sharing is not applicable to this article as no new data were created or analyzed in this study.

\bibliography{Biblio}

\appendix

\section{Algebraic QES structure of the sextic potential}

In this Appendix, we summarize the Lie-algebraic foundation underlying the quasi-exact solvability of the sextic potential \cite{Turbiner1988}. For a discrete set of couplings $\lambda = N \in \mathbb{N}$, the Schr\"odinger operator can be gauge-rotated into an element of the universal enveloping algebra of $\mathfrak{sl}(2,\mathbb{R})$ acting on a finite-dimensional polynomial space, yielding $(N+1)$ exact even-parity eigenstates.

\subsection*{A.1 Rescaled Hamiltonian and gauge factor}

The Hamiltonian is
\begin{equation}
H = -\frac{1}{2}\frac{d^{2}}{dx^{2}}
+ \frac12\bigl(x^{6}+2x^{4}-2(2\lambda+1)x^{2}\bigr),
\end{equation}
and we rescale
\begin{equation}
\widetilde H = 2H
= -\frac{d^{2}}{dx^{2}}
+ x^{6} + 2x^{4} - 2(2\lambda+1)x^{2}.
\end{equation}

We perform the gauge transformation
\begin{equation}
\psi(x)=e^{-g(x)} P(x), \qquad
g(x)=\frac{x^{4}}{4}+\frac{x^{2}}{2},
\end{equation}
and define the gauge-rotated operator
\begin{equation}
h = e^{g}\,\widetilde H\,e^{-g}.
\end{equation}

A direct calculation gives
\begin{equation}
h = -\frac{d^{2}}{dx^{2}}
+ 2(x^{3}+x)\,\frac{d}{dx}
+ \bigl(1-4\lambda x^{2}\bigr).
\label{eq:hx}
\end{equation}
The original eigenvalue problem becomes
\begin{equation}
h \,P(x) = E\,P(x),
\end{equation}
with $P(x)$ a function to be determined.

\subsection*{A.2 Even sector and change of variable $z=x^{2}$}

For even-parity states we write
\begin{equation}
z=x^{2}, \qquad P(x)=\Phi(z),
\end{equation}
so that
\begin{equation}
\frac{d}{dx} = 2x\,\frac{d}{dz}, \qquad
\frac{d^{2}}{dx^{2}} = 2\frac{d}{dz} + 4z\,\frac{d^{2}}{dz^{2}}.
\end{equation}

Substituting these into~\eqref{eq:hx} yields an operator acting on $\Phi(z)$:
\begin{equation}
h =
-4z\,\frac{d^{2}}{dz^{2}}
+ (4z^{2}+4z-2)\,\frac{d}{dz}
+ \bigl(1-4\lambda z\bigr).
\label{eq:hz}
\end{equation}

\subsection*{A.3 $\mathfrak{sl}(2)$ generators and the QES condition}

Let
\begin{equation}
\mathcal{P}_{N}=\{1,z,z^{2},\dots,z^{N}\},
\end{equation}
and consider the differential realization of $\mathfrak{sl}(2,\mathbb{R})$ on $\mathcal{P}_{N}$:
\begin{equation}
J_-=\frac{d}{dz},\qquad
J_0=z\frac{d}{dz} - \frac{N}{2},\qquad
J_+=z^{2}\frac{d}{dz} - Nz,
\end{equation}
which satisfy
\begin{equation}
[J_0,J_\pm]=\pm J_\pm,\qquad
[J_+,J_-]=-2J_0.
\end{equation}
These generators map $\mathcal{P}_{N}$ into itself, so any polynomial expression in
$\{J_+,J_0,J_-\}$ preserves $\mathcal{P}_{N}$.

To see the quasi-exact solvability, we examine the action of~\eqref{eq:hz} on a highest-degree monomial $z^{N}$. The potentially degree-raising term in $h z^{N}$ is proportional to $z^{N+1}$ with coefficient
\begin{equation}
4N - 4\lambda,
\end{equation}
which vanishes precisely when
\begin{equation}
\lambda = N \in \mathbb{N}.
\end{equation}
Thus, for these discrete couplings, $h$ preserves $\mathcal{P}_{N}$.

Moreover, one can express $h$ as a quadratic element of the universal enveloping algebra $U(\mathfrak{sl}(2,\mathbb{R}))$. For $\lambda=N$ one finds explicitly
\begin{equation}
h = 4 J_+ - 4 J_0 J_- + (-2N-2)\,J_- + 4 J_0 + (2N+1)\,\mathbb{I},
\end{equation}
which acts invariantly on $\mathcal{P}_N$, i.e.
\begin{equation}
h:\mathcal{P}_{N} \longrightarrow \mathcal{P}_{N}.
\end{equation}

The eigenvalue problem
\begin{equation}
h\Phi(z)=E\Phi(z), \qquad \Phi(z)\in\mathcal{P}_{N},
\end{equation}
reduces to diagonalizing a finite $(N+1)\times(N+1)$ matrix. Therefore, for $\lambda=N$ the sextic oscillator possesses exactly $(N+1)$ even algebraic eigenstates, while all higher states lie outside $\mathcal{P}_{N}$ and are not algebraically solvable.

\section{On the convergence of variational calculations}
\label{A2}

\noindent
For illustrative purposes, we restrict our attention to the ground--state wavefunction.

\vspace{0.2cm}

\noindent
For $\lambda = 0$, the exactly solvable (QES) ground state ($n=0$) is
\[
\psi_0(x) = e^{-\frac{1}{4}x^{4} - \frac{1}{2}x^{2}}\ ,
\]
with the corresponding exact energy
\[
E_0 = \frac{1}{2}\ .
\]

\noindent
Our variational ansatz is constructed to reflect the analytic structure of the exact solution and is written as
\begin{equation}
\psi_{0}^{(\mathrm{var})}(x;\lambda)
= Q_{k}^{(0)}(x;\lambda)\, e^{-x^{4}/4},
\end{equation}
where the polynomial prefactor is defined by
\begin{equation}
Q_{k}^{(0)}(x;\lambda) = \sum_{i=0}^{k} c_i\, x^{2i}.
\end{equation}

\noindent
This polynomial has degree $2k$ in the coordinate $x$; hence, $k=1$ corresponds to a quadratic approximation, $k=2$ to a quartic approximation, and so on.  
Below, in Table \ref{Tl0} we present the relative error $\Delta E$ in the ground-state energy as a function of the number of variational parameters employed in the ansatz.

\begin{table}[h!]
\centering
\renewcommand{\arraystretch}{1.25}
\setlength{\tabcolsep}{8pt}
\begin{tabular}{c c l}
\toprule
\textbf{Degree of prefactor} $Q_k^{(0)}$ & \textbf{Relative error $\Delta E$} & \textbf{Optimized parameters} \\
\midrule

$2$  &
$2.6727\times 10^{-2}$ &
$\displaystyle a_{1}=-0.3126809708$ \\[4pt]

$4$  &
$9.7362\times 10^{-4}$ &
$\displaystyle
\begin{aligned}
a_{1} &= -0.4531994490,\\
a_{2} &= 0.06722709332
\end{aligned}$ \\[6pt]

$6$  &
$2.3945\times 10^{-5}$ &
$\displaystyle
\begin{aligned}
a_{1} &= -0.4912011963,\\
a_{2} &= 0.1080204448,\\
a_{3} &= -0.01013727919
\end{aligned}$ \\[8pt]

$8$  &
$4.4247\times 10^{-7}$ &
$\displaystyle
\begin{aligned}
a_{1} &= -0.4986287422,\\
a_{2} &= 0.12125882396,\\
a_{3} &= -0.01728711954,\\
a_{4} &= 0.001160882879
\end{aligned}$ \\[10pt]

$10$ &
$6.5691\times 10^{-9}$ &
$\displaystyle
\begin{aligned}
a_{1} &= -0.4998134059,\\
a_{2} &= 0.12432545252,\\
a_{3} &= -0.01995369865,\\
a_{4} &= 0.002079244055,\\
a_{5} &= -0.000107134049
\end{aligned}$ \\[12pt]

$12$ &
$8.1665\times 10^{-11}$ &
$\displaystyle
\begin{aligned}
a_{1} &= -0.4999771682,\\
a_{2} &= 0.12489540008,\\
a_{3} &= -0.02065622957,\\
a_{4} &= 0.002461369670,\\
a_{5} &= -0.000200353830,\\
a_{6} &= 8.287020711\times 10^{-6}
\end{aligned}$ \\[12pt]

\bottomrule
\end{tabular}
\caption{Relative energy error for the variational ground state at $\lambda=0$.}
\label{Tl0}
\end{table}

The corresponding Shannon entropies in position and momentum space are presented in Table \ref{Tl0S}.

The Shannon entropy in position space is defined as
\begin{equation}
S_x
= - \int_{-\infty}^{\infty} \rho(x)\, \ln \rho(x)\, dx ,
\qquad
\rho(x) = |\psi(x)|^{2} .
\end{equation}

\noindent
Similarly, the Shannon entropy in momentum space is defined as
\begin{equation}
S_p
= - \int_{-\infty}^{\infty} \tilde{\rho}(p)\, \ln \tilde{\rho}(p)\, dp ,
\qquad
\tilde{\rho}(p) = |\phi(p)|^{2},
\end{equation}
where $\phi(p)$ is the momentum--space wavefunction obtained via the Fourier transform
\begin{equation}
\phi(p)
= \frac{1}{\sqrt{2\pi}}
\int_{-\infty}^{\infty}
\psi(x)\, e^{-\, i p x}\, dx .
\end{equation}

\begin{table}[h!]
\centering
\renewcommand{\arraystretch}{1.25}
\setlength{\tabcolsep}{10pt}
\begin{tabular}{c c c c c}
\toprule
\textbf{Degree of $Q_k^{(0)}$} &
$\boldsymbol{S_x^{(\rm var)}}$ &
$\boldsymbol{\displaystyle \frac{|S_x^{(\rm var)}-S_x^{(\text{exact})}|}{S_x^{(\text{exact})}}}$ &
$\boldsymbol{S_p^{(N)}}$ &
$\boldsymbol{\displaystyle \frac{|S_p^{(\rm var)}-S_p^{(\text{exact})}|}{S_p^{(\text{exact})}}}$ \\
\midrule

2 &
$0.8226566228$ &
$3.989180864\times10^{-2}$ &
$1.3778410103$ &
$9.809501700\times10^{-4}$ \\[4pt]

4 &
$0.7932084673$ &
$2.667412873\times10^{-3}$ &
$1.3795780274$ &
$2.784935788\times10^{-4}$ \\[4pt]

6 &
$0.7911855391$ &
$1.102993113\times10^{-4}$ &
$1.3792446341$ &
$3.676299851\times10^{-5}$ \\[4pt]

8 &
$0.7911010621$ &
$3.317926110\times10^{-6}$ &
$1.3791985068$ &
$3.317926110\times10^{-6}$ \\[4pt]

10 &
$0.7910983603$ &
$9.955151240\times10^{-8}$ &
$1.3791938477$ &
$6.024819014\times10^{-8}$ \\[4pt]

12 &
$0.7910982837$ &
$2.715741793\times10^{-9}$ &
$1.3791939566$ &
$1.868646829\times10^{-8}$ \\[4pt]

\bottomrule
\end{tabular}
\caption{Position-space Shannon entropy $S_x$ and momentum-space Shannon
entropy $S_p$ obtained with the variational ansatz of degree $2\,k$ at $\lambda=0$
(parameters from Table~\ref{Tl0}). Exact values:
$S_x^{(\text{exact})} = 0.79109828152\ldots$ and
$S_p^{(\text{exact})} = 1.37919393077\ldots$.}
\label{Tl0S}
\end{table}

\clearpage

Now, for $\lambda=1$ the exact QES ground state ($n=0$) solution is given by 
\[
\psi_0 \ = \ \left(1+(\sqrt{3}-1) \,x^2\right)\,e^{-\frac{1}{4}x^4-\frac{1}{2}x^2} \ ,
\]
with exact energy
\[
E_0 \ = \ \frac{3}{2}-\sqrt{3} \ .
\]
In Table \ref{Tl1} we present the relative error $\Delta E$ in the ground-state energy as a function of the number of variational parameters employed in the ansatz.

\begin{table}[h!]
\centering
\renewcommand{\arraystretch}{1.25}
\setlength{\tabcolsep}{8pt}
\begin{tabular}{c c l}
\toprule
\textbf{Degree of prefactor} $Q_k^{(0)}$ & \textbf{Relative error $\Delta E$} & \textbf{Optimized parameters} \\
\midrule

$2$ &
$4.6563\times 10^{-2}$ &
$\displaystyle a_{1} = -0.05860507075$ \\[4pt]

$4$ &
$7.6744\times 10^{-3}$ &
$\displaystyle
\begin{aligned}
a_{1} &= 0.10253495387,\\
a_{2} &= -0.07314452001
\end{aligned}$ \\[6pt]

$6$ &
$4.4991\times 10^{-4}$ &
$\displaystyle
\begin{aligned}
a_{1} &= 0.19677960722,\\
a_{2} &= -0.17203725937,\\
a_{3} &= 0.02459448236
\end{aligned}$ \\[8pt]

$8$ &
$1.5361\times 10^{-5}$ &
$\displaystyle
\begin{aligned}
a_{1} &= 0.22479924945,\\
a_{2} &= -0.22122664714,\\
a_{3} &= 0.05132969065,\\
a_{4} &= -0.004372210324
\end{aligned}$ \\[10pt]

$10$ &
$3.6657\times 10^{-7}$ &
$\displaystyle
\begin{aligned}
a_{1} &= 0.23082022174,\\
a_{2} &= -0.23659628008,\\
a_{3} &= 0.06478777917,\\
a_{4} &= -0.009040695644,\\
a_{5} &= 0.0005476974990
\end{aligned}$ \\[12pt]

$12$ &
$6.7071\times 10^{-9}$ &
$\displaystyle
\begin{aligned}
a_{1} &= 0.23186998035,\\
a_{2} &= -0.24020220308,\\
a_{3} &= 0.06926098046,\\
a_{4} &= -0.01148943090,\\
a_{5} &= 0.001148088913,\\
a_{6} &= -0.00005358401893
\end{aligned}$ \\[12pt]

\bottomrule
\end{tabular}
\caption{Relative energy error for the variational ground state at $\lambda=1$.}
\label{Tl1}
\end{table}

The associated Shannon entropies in position and momentum space are presented in Table \ref{Tl1S}.

\begin{table}[h!]
\centering
\renewcommand{\arraystretch}{1.25}
\setlength{\tabcolsep}{10pt}
\begin{tabular}{c c c c c}
\toprule
\textbf{Degree of $Q_k^{(0)}$} &
$\boldsymbol{S_x^{(\rm var)}}$ &
$\boldsymbol{\displaystyle \frac{|S_x^{(\rm var)}-S_x^{(\text{exact})}|}{S_x^{(\text{exact})}}}$ &
$\boldsymbol{S_p^{(\rm var)}}$ &
$\boldsymbol{\displaystyle \frac{|S_p^{(\rm var)}-S_p^{(\text{exact})}|}{S_p^{(\text{exact})}}}$ \\
\midrule

2 &
$0.9884178551$ &
$4.9276\times 10^{-3}$ &
$1.2319666665$ &
$2.4055\times 10^{-2}$ \\[4pt]

4 &
$0.9952220079$ &
$1.9127\times 10^{-3}$ &
$1.2594248236$ &
$2.3540\times 10^{-3}$ \\[4pt]

6 &
$0.9935384436$ &
$2.1810\times 10^{-4}$ &
$1.2619202139$ &
$3.7476\times 10^{-4}$ \\[4pt]

8 &
$0.9933313443$ &
$7.6604\times 10^{-7}$ &
$1.2624182767$ &
$1.9855\times 10^{-5}$ \\[4pt]

10 &
$0.9933218112$ &
$2.3122\times 10^{-7}$ &
$1.2623903778$ &
$2.2811\times 10^{-6}$ \\[4pt]

12 &
$0.9933215821$ &
$3.1360\times 10^{-10}$ &
$1.2623935474$ &
$2.2437\times 10^{-7}$ \\[4pt]

\bottomrule
\end{tabular}
\caption{Position-space Shannon entropy $S_x$ and momentum-space Shannon
entropy $S_p$ obtained with the variational ansatz of degree $2\,k$ at $\lambda=1$
(parameters from Table~\ref{Tl1}). Exact values:
$S_x^{(\text{exact})}=0.993321580789...$ and
$S_p^{(\text{exact})}=1.262393264036...$\ .}
\label{Tl1S}
\end{table}

\vspace{3.0cm}

Therefore, the convergence of the variational energies mirrors that of the Shannon entropies, indicating that the trial wavefunctions closely reproduce the exact ground state. Since entropies are highly sensitive to the detailed shape of the probability density, their agreement with the exact values confirms the overall accuracy and reliability of the variational approximation.

\noindent
The present study is focused on the qualitative evolution of phase-space and entropic structures as functions of the control parameter $\lambda$, rather than on high-precision benchmarking. All numerical results are stable to at least four significant digits, and further increases in numerical accuracy do not alter the qualitative behavior of the entropy and information-theoretic profiles, nor their visual representation.

\end{document}